\documentclass[11pt]{article}
\usepackage{jheppub}
\usepackage{amsmath,amssymb,amsfonts,graphicx}
\usepackage{braket}
\usepackage{setspace}
\usepackage{subcaption}

\usepackage[left=1in,right=1in,top=1in,bottom=1.2in]{geometry}
\usepackage{xcolor}

\title{Magnetic Braneworlds: Cosmology and Wormholes}
\author[1]{Stefano Antonini,}
\author[1]{Luis Gabriel C. Bariuan}

\affiliation[1]{Center for Theoretical Physics and Department of Physics, University of California, Berkeley, California 94720, U.S.A.}

\emailAdd{santonini@berkeley.edu}
\emailAdd{lbariuan@berkeley.edu}

\abstract{We construct 4D flat Big Bang-Big Crunch cosmologies and Anti-de Sitter (AdS) planar eternally traversable wormholes using braneworlds embedded in asymptotically AdS${}_5$ spacetimes. The background geometries are the AdS${}_5$ magnetic black brane and the magnetically charged AdS${}_5$ soliton, respectively. The two setups arise from different analytic continuations of the same saddle of the gravitational Euclidean path integral, in which the braneworld takes the form of a Maldacena-Maoz Euclidean wormhole. We show the existence of a holographic dual description of this setup in terms of a microscopic Euclidean boundary conformal field theory (BCFT) on a strip. By analyzing the BCFT Euclidean path integral, we show that the braneworld cosmology is encoded in a pure excited state of a CFT dual to a black brane microstate, whereas the braneworld wormhole is encoded in the ground state of the BCFT. The latter confines in the IR, and we study its confining properties using holography. We also comment on the properties of bulk reconstruction in the two Lorentzian pictures and their relationship via double analytic continuation. 
This work can be interpreted as an explicit, doubly-holographic realization of the relationship between cosmology, traversable wormholes, and confinement in holography, first proposed in arXiv:2102.05057, arXiv:2203.11220. 
}

\date{\today}

\begin{document}

\maketitle
\newpage
\parskip=10pt

\section{Introduction}
\label{sec:intro}

How to describe cosmology remains one of the most important open questions in holography. The holographic duality \cite{tHooft:1993dmi,Susskind:1994vu} is a powerful framework to describe quantum gravity and it is best understood in the context of the Anti-de Sitter/Conformal Field Theory (AdS/CFT) correspondence \cite{Maldacena:1997re,Witten:1998qj,Gubser:1998bc,Aharony:1999ti}. The structure of the AdS/CFT correspondence seems prima facie to rely on the features of asymptotically AdS spacetimes, which have a negative cosmological constant $\Lambda<0$---which asymptotically is the only relevant source in the Einsten's equations---and have a timelike conformal boundary where the dual theory is defined. These are all specific features distinguishing AdS from cosmological models of our universe, which typically do not have any asymptotic boundary, are filled with energy density (almost) homogeneously and isotropically, and, at least in the simplest possible models in accordance with observations, have a positive cosmological constant $\Lambda>0$ \cite{Weinberg:2008zzc,Baumann:2022mni}.

However, there is evidence that holography is an intrinsic feature of quantum gravity, extending well beyond the AdS realm. In fact, the idea that gravity stores information holographically can be dated back to the work of Bekenstein \cite{Bekenstein:1972tm,Bekenstein:1973ur,Bekenstein:1974ax} and Hawking \cite{Hawking:1974rv,Hawking:1975vcx}, and was understood before \cite{tHooft:1993dmi,Susskind:1994vu} its explicit realization in AdS/CFT. The holographic description of different backgrounds, mainly de Sitter \cite{Alishahiha:2004md,Coleman:2021nor,Batra:2024kjl,Susskind:2021omt,Shaghoulian:2022fop,Araujo-Regado:2022gvw,Chandrasekaran:2022cip,Witten:2023xze,Cotler:2023eza,Cotler:2024xzz}\footnote{See also \cite{McFadden:2009fg} for a different approach to the holographic description of $\Lambda>0$ cosmology.} and asymptotically flat \cite{Strominger:2017zoo,Pasterski:2021raf}, have been under investigation for the last two decades. In particular, de Sitter holography, unlike AdS/CFT, has the advantage of naturally describing a universe undergoing an accelerated expansion.\footnote{Although our universe is currently undergoing an accelerated expansion and possibly contains a positive cosmological constant (which would imply the acceleration lasts forever and our universe approaches de Sitter in the far future, modulo the conjectured metastability of de Sitter leading to vacuum decay to a neighboring AdS vacuum in the string theory landscape), it is not exactly de Sitter. A similar reasoning applies to the inflationary era, during which the universe was nearly de Sitter. Therefore, de Sitter holography should be regarded as one step forward from AdS/CFT towards a holographic description of $\Lambda>0$ cosmology, rather than the final goal.} \footnote{More recently, using general quantum information-theoretic properties of holography \cite{Ryu2006a,Ryu2006b,Hubeny:2007xt,Engelhardt:2014gca,Swingle:2009bg,VanRaamsdonk:2010pw}, steps towards a more generic holographic duality independent of the global properties of the spacetime were taken \cite{Bousso:2022hlz,Bousso:2023sya}.} Although these approaches are promising and could lead to breakthroughs in the holographic description of cosmology in the next few years, they are not yet as well understood as AdS/CFT and their application to a concrete description of a cosmological universe seems premature.

An alternative line of research \cite{Maldacena:2004rf,McInnes:2004nx,Cooper:2018cmb,Antonini:2019qkt,VanRaamsdonk:2020tlr,VanRaamsdonk:2021qgv,Antonini:2022blk,Antonini:2022ptt,Antonini:2022fna,Waddell:2022fbn,Sahu:2023fbx,Antonini:2023hdh,Betzios:2024oli}\footnote{See also \cite{Usatyuk:2024mzs} for a lower-dimensional construction.} is aimed at describing $\Lambda<0$ cosmology in holography. Broadly speaking, the reason for this endeavor is that quantum gravity with a negative cosmological constant is better understood thanks to AdS/CFT and we can hope to obtain a well-controlled holographic description of the cosmological universe, as we will see. The resulting cosmology will typically be a Big Bang-Big Crunch universe which, after an expanding phase, starts recollapsing due to the effect of the negative cosmological constant. This is clearly unlike the $\Lambda>0$ $\Lambda$CDM standard model of cosmology, which is the simplest model matching current observational data. However, other mechanisms could lead to phases of accelerated expansion \cite{Peebles:1987ek,Ratra:1987rm}, with the resulting models able to in principle match current cosmological observations \cite{Antonini:2022ptt,Antonini:2022fna,VanRaamsdonk:2023ion}. 

Regardless of the phenomenological viability of these models, the holographic description of $\Lambda<0$ cosmologies has great theoretical value. Having a negative cosmological constant does not immediately imply a simple, ordinary AdS/CFT description. In fact, these $\Lambda<0$ Big Bang-Big Crunch cosmological universes are not asymptotically AdS \cite{Antonini:2022blk}. Nonetheless, as we will see, a holographic description using AdS/CFT can be obtained with some effort. Therefore, these constructions provide us with a more ``ordinary'' holographic setting where to learn general lessons about how spacetimes without an asymptotic AdS boundary are encoded in a dual microscopic quantum mechanical theory. Hopefully, these lessons can then be applied to the description of phenomenologically relevant cosmological universes.

Motivated by these ideas, in this paper we study (3+1)-dimensional, $\Lambda<0$, symmetric Big Bang-Big Crunch braneworld cosmologies, which we build by embedding a codimension-1 end-of-the-world (ETW) brane in a (4+1)-dimensional magnetic black brane (MBB) background \cite{DHoker:2009mmn}. This setup is associated with a dominant Euclidean saddle of the gravitational path integral and has a well-defined holographic dual description in terms of an excited pure state of a microscopic 4D CFT. Interestingly, the Euclidean saddle also admits a different analytic continuation to a (4+1)-dimensional magnetically charged AdS${}_5$ soliton cut off by a static ETW brane. In this case, the braneworld physics\footnote{Here and in the rest of the paper, when we talk about ``braneworld physics'' or ``braneworld picture'' we mean the effective 4D description of physics on the ETW brane.} is given by a planar eternally traversable wormhole connecting two distinct asymptotic AdS${}_4$ boundaries. 

The existence of this alternative analytic continuation is of its own interest both from a bulk and from a boundary dual viewpoint.  Our braneworld construction is the first explicit realization of the ideas presented in \cite{VanRaamsdonk:2021qgv,Antonini:2022blk,Antonini:2022ptt} to build higher dimensional eternally traversable wormholes connecting two distinct asymptotic AdS${}_4$ boundaries. For a different approach to the construction of higher dimensional AdS traversable wormholes see \cite{Bintanja:2021xfs}.\footnote{Other eternally traversable wormhole constructions include lower dimensional examples \cite{Gao:2016bin,Maldacena:2018lmt}, 4D wormholes connecting two regions of the same asymptotically flat spacetime \cite{Maldacena:2018gjk,Fu:2019vco}, 4D braneworld wormholes connecting two different regions of the same asymptotic AdS${}_4$ boundary \cite{Maldacena:2020sxe}, and charged wormholes with AdS${}_{d-p} \times S^{p}$ interior topologies \cite{Deshpande:2022zfm}. See Section \ref{subsec:otherconstr} for a discussion of the relationship between our construction and those presented in \cite{Maldacena:2018gjk,Maldacena:2020sxe}. We also remark that the construction of our wormhole involves embedding a braneworld in a higher dimensional spacetime, and is different in this sense from the constructions in \cite{Gao:2016bin,Maldacena:2018gjk,Maldacena:2018lmt,Fu:2019vco}. However, an effective 4D description of the braneworld wormhole not involving the 5D spacetime does exist, see Sections \ref{sec:relation} and \ref{subsec:gravitylocalization}.} These wormholes admit a dual description in terms of the vacuum state of a microscopic theory which exhibits a mass gap and IR confinement, as we will discuss. Additionally, our construction realizes in an explicit and controlled setup the ideas advanced in \cite{VanRaamsdonk:2021qgv,Antonini:2022blk} connecting cosmology, traversable wormholes, and confining properties of a holographic microscopic theory. Before summarizing the main results of our paper, we will review these ideas and explain their connection to our construction.

\subsection{Relation to previous work and motivation}
\label{sec:relation}

First of all, let us review the ideas presented in \cite{VanRaamsdonk:2021qgv,Antonini:2022blk}, which are one basic motivation for the present work.

The realization that certain time symmetric $\Lambda<0$ Big Bang-Big Crunch cosmologies are obtained by analytic continuation of reflection-symmetric Euclidean wormholes connecting two AdS asymptotic boundaries dates back to the work of Maldacena and Maoz \cite{Maldacena:2004rf}. The metric of these Euclidean wormholes generically takes the form
\begin{equation}
    ds^2=dw^2+a^2(w)ds^2_{\Sigma_d}
\end{equation}
where $ds^2_{\Sigma_d}$ is the metric on the $d$-dimensional manifold $\Sigma_d$ and $a(w)$ is symmetric $a(w)=a(-w)$ and diverges exponentially for $w=\pm \infty$. This behavior is typical of asymptotically AdS spacetimes, as it becomes clear by recasting the metric in the conformal form (which will be useful in the rest of the paper)
\begin{equation}
    ds^2=a^2(\tau)\left(d\tau^2+ds^2_{\Sigma_d}\right)
    \label{eq:maldamaozmetric}
\end{equation}
in which $a(\tau)\approx 1/(\tau_\infty\pm\tau)$ for $\tau\to\mp\tau_\infty$, with $\tau_\infty$ a finite number.

These wormholes analytically continue under $w=i\lambda$ to Friedmann-Lema\^itre-Robertson-Walker (FLRW) metrics for homogeneous and isotropic cosmological universes. The cosmological scale factor is simply given by $a(\lambda)$ after substituting $w\to i\lambda$ in the wormhole scale factor. $a(\lambda)$ is guaranteed to be real by reflection symmetry and it goes to zero at the Big Bang and Big Crunch for $\lambda=\pm\lambda_0$. The simplest explicit example is given by a flat FLRW universe sourced by a negative cosmological constant and a radiation term, giving rise to a scale factor $a(\lambda)=A\sqrt{\cos(B\lambda)}$, where $A,B$ are constants simply related to $\Lambda$ and the radiation density parameter. Analytically continuing to Euclidean signature, the scale factor becomes $a(w)=A\sqrt{\cosh(Bw)}$, which indeed diverges exponentially at the two asymptotic AdS boundaries $w\to\pm\infty$.

\begin{figure}[h]
    \centering
    \includegraphics[width=0.9\linewidth]{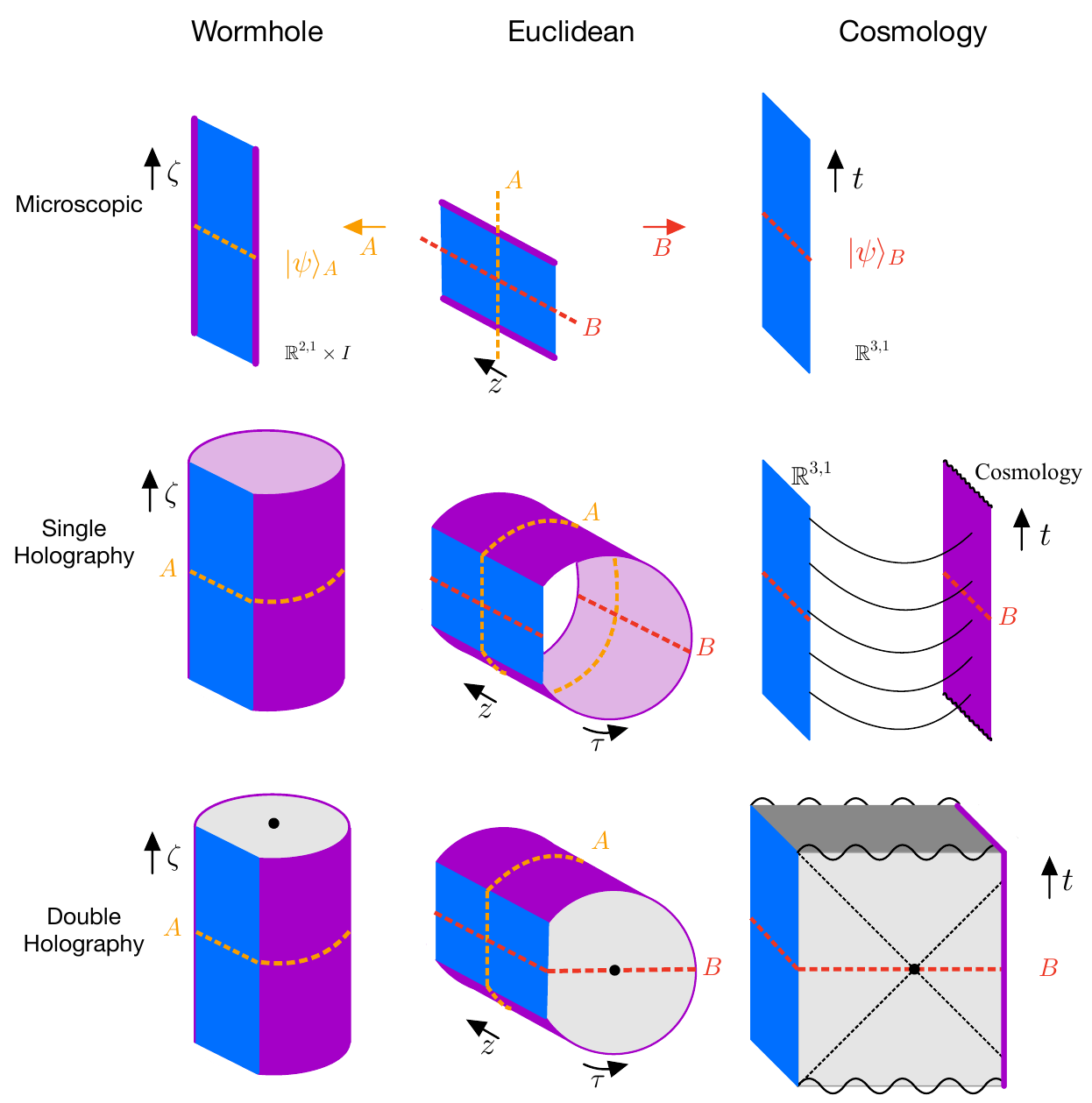}
    \caption{{\setstretch{1.0}\small{Recap of all the constructions discussed in this paper. The yellow slices labeled by $A$ are the reflection-symmetric slices associated with the wormhole picture, whereas the red slices labeled by $B$ are the reflection-symmetric slices associated with the cosmology picture. In the first line we have the microscopic, UV complete description, with boundaries indicated by a purple line. From left to right: the vacuum state of the BCFT on $\mathbb{R}^{2,1}\times I$, the Euclidean BCFT on $\mathbb{R}^{3}\times I$, and the excited state of the CFT on $\mathbb{R}^{3,1}$. In the second line we have the singly holographic/braneworld description. From left to right: the eternally traversable wormhole (in purple) with boundaries coupled by non-gravitational DOF (in blue), the Maldacena-Maoz Euclidean wormhole with boundaries coupled by non-gravitational DOF, and the Big Bang-Big Crunch cosmology entangled to the bath on $\mathbb{R}^{3,1}$. In the third line we have the doubly holographic description. The black dot indicates $r=r_H$. From left to right: the magnetically charged AdS${}_5$ soliton spacetime cut off by the ETW brane (in purple), the 5D Euclidean saddle cut off by the ETW brane, and the double-sided magnetic black brane cut off by the ETW brane.}}}
    \label{fig:slicingduality}
\end{figure}

In \cite{VanRaamsdonk:2020tlr,VanRaamsdonk:2021qgv,Antonini:2022blk,Antonini:2022ptt} (see also \cite{Betzios:2021fnm}) it was proposed that Maldacena-Maoz-like wormholes in $d+1=4$ dimensions are saddles of a Euclidean gravitational path integral dual to the path integral of a microscopic Euclidean theory given by two 3D holographic CFTs on $\Sigma_3$ coupled via a relevant interaction to 4D, in general non-holographic auxiliary degrees of freedom (DOF) on $\Sigma_3\times I$. $I$ here is a finite interval connecting the two 3D CFTs. See the top line, center panel of Figure \ref{fig:slicingduality}. The holographic 3D theories are taken to have many more degrees of freedom than the 4D auxiliary theory, such that an effective 4D holographic description (the Euclidean wormhole) exists. Following \cite{VanRaamsdonk:2020tlr,VanRaamsdonk:2021qgv,Antonini:2022blk,Antonini:2022ptt}, let us focus our attention on the case $\Sigma_3=\mathbb{R}^3$.
Given this microscopic description, the bulk effective field theory (EFT) is given by a 4D Maldacena-Maoz-like Euclidean wormhole---which can be understood to be dual to the two 3D holographic CFTs---with the two asymptotic boundaries coupled by auxiliary, non gravitational DOF on $\mathbb{R}^3\times I$, which account for the non-holographic microscopic 4D DOF.\footnote{\label{footnote:DOF}This is an intuitive, simplified description of the relationship between microscopic and bulk EFT DOF. In fact, the wormhole physics does not encode only the holographic 3D microscopic boundary DOF, but also the coupling of these with the 4D microscopic DOF. In particular, matter fields associated with both 3D and 4D microscopic DOF live on the wormhole background and are coupled  by appropriate boundary conditions to the non-gravitational bath \cite{Antonini:2022blk}.} \footnote{This can be interpreted in analogy with the (1+1)-dimensional holographic models of black hole evaporation studied in \cite{Almheiri:2019psf,Almheiri:2019qdq}. In that case, the microscopic theory was given by a holographic (0+1)-dimensional theory coupled to a non-holographic (1+1)-dimensional CFT. In the bulk setup, a (1+1)-dimensional black hole replaces the holographic DOF and is coupled to a non-gravitational bath at the asymptotic AdS boundary.} See the center line, center panel of Figure \ref{fig:slicingduality}. Notice that the presence of these auxiliary non-gravitational DOF coupling the two boundaries is fundamental to avoid a factorization puzzle \cite{Maldacena:2004rf,VanRaamsdonk:2020tlr}.

In this Euclidean bulk EFT, the reflection-symmetric slice $w=0$ where the initial state of the corresponding Lorentzian cosmology is prepared also intersect the non-gravitational DOF (the ``bath''). The Lorentzian description therefore includes a Big Bang-Big Crunch cosmology \textit{and} a non-gravitational bath on $\mathbb{R}^{3,1}$, with bulk fields in the cosmology entangled to bulk fields in the bath, see the center line, right panel of Figure \ref{fig:slicingduality}. This has interesting implications for the encoding of the cosmological universe in the dual microscopic theory and is related to the construction of cosmological islands \cite{Hartman:2020khs,Bousso:2022gth,Antonini:2022blk}, see Section \ref{sec:islands}.

What is the state of the microscopic theory dual to the Lorentzian cosmology (plus the non-gravitational bath)? From the associated slicing of the microscopic Euclidean path integral, we see that the interval $I$ is taken to be the Euclidean time direction. This Euclidean path integral therefore prepares an excited pure state of the 4D non-holographic DOF only on $\mathbb{R}^{3,1}$, see top line, right panel of Figure \ref{fig:slicingduality}. Note that the 3D holographic theories are not present at all in the Lorentzian description, but rather they are in the Euclidean past and future, representing boundary conditions for the Euclidean path integral preparing the microscopic state.

It was observed in \cite{VanRaamsdonk:2020tlr,VanRaamsdonk:2021qgv,Antonini:2022blk,Antonini:2022ptt} that, given the translational symmetry in the $\mathbb{R}^3$ directions, the bulk Euclidean saddle admits a different analytic continuation also yielding a real Lorentzian theory. This is obtained by identifying one of the $\mathbb{R}^3$ directions, let us label it by $z$, with the Euclidean time. The resulting bulk effective field theory is static and it is given by a planar eternally traversable wormhole (namely the Lorentzian version of a Maldacena-Maoz-like wormhole), with the two boundaries coupled by non-gravitational auxiliary DOF, see the center line, left panel of Figure \ref{fig:slicingduality}. Once again, the presence of these additional DOF is fundamental, because a traversable wormhole can only exist if the holographic theories living on the two boundaries are coupled to each other \cite{Antonini:2022blk,Maldacena:2018lmt}. 

The microscopic dual description of the traversable wormhole is given by the vacuum state\footnote{The $z$ direction, which is chosen to be the Euclidean time in this picture, is non-compact. Therefore, the microscopic path integral naturally prepares the vacuum state of the theory.} of the 3D-4D-3D theory on $\mathbb{R}^{2,1}\times I$, see top-left panel of Figure \ref{fig:slicingduality}.
Note that in the infrared (IR), namely at scales much larger than the interval $I$, this 3D-4D-3D theory on $\mathbb{R}^{2,1}\times I$ flows to a 3D theory on $\mathbb{R}^{2,1}$. Because the center of the traversable wormhole is at a finite proper distance away from any other bulk point in the wormhole, it acts as a cut off in the emergent holographic direction. In the spirit of the holographic renormalization group (RG) flow \cite{deBoer:1999tgo}, this suggests the existence of a IR energy cutoff in the dual microscopic theory, below which physics is trivial. This fact was interpreted in \cite{VanRaamsdonk:2020tlr,VanRaamsdonk:2021qgv,Antonini:2022blk,Antonini:2022ptt} as evidence that the 3D and 4D theories and their coupling must be such that the IR theory on $\mathbb{R}^{2,1}$ confines and possesses a mass gap (hence the choice of relevant coupling between the 3D CFTs and the 4D DOF). A similar argument for confinement in holography was first introduced by Witten \cite{Witten:1998zw} (see Section \ref{sec:MBBdual}).

The existence of these two possible slicings of the same Euclidean path integral, leading to two different states of two different microscopic theories and, holographically, to two different bulk setups, was dubbed ``slicing duality'' in \cite{Antonini:2022blk}. It was argued \cite{VanRaamsdonk:2021qgv,Antonini:2022blk,Antonini:2022opp} that observables in the two Lorentzian bulk setups are related to each other by double analytic continuation, and that the slicing duality could be used to reconstruct observables in the cosmology from vacuum observables in the confining theory dual to the Lorentzian wormhole. Because direct reconstruction of cosmological observables from the dual highly excited pure state is understood to be very complex, the slicing duality has the potential to provide a new, much simpler dictionary in this context. We will come back to these points in Sections \ref{sec:islands} and \ref{sec:slicing}.

\subsubsection*{Double holography and braneworlds}

In order to build an explicit realization of these ideas in a concrete setting, it is useful to consider the special case in which the 4D microscopic auxiliary degrees of freedom that couple the two 3D CFTs are also holographic. Let us focus initially on the Euclidean setup. The microscopic theory can then be understood to be a 4D holographic boundary conformal field theory (BCFT) \cite{Cardy:1989ir,Affleck:1991tk,Cardy:2004hm} on $\mathbb{R}^3\times I$. According to the AdS/BCFT prescription \cite{Takayanagi:2011zk,Fujita:2011fp}, the bulk description is now given by a 5D Euclidean bulk geometry cut off by a constant tension ETW brane,\footnote{Depending on the setup, there could be different saddles of the 5D gravitational path integral, with some of them having two disconnected ETW branes each one intersecting the boundary at one of the two endpoints of the interval \cite{Cooper:2018cmb}. In our setup, this alternative phase does not exist, and the gravitational ensemble is dominated by the connected phase of our interest (see Appendix \ref{app:actioncomparison}).} which intuitively is an extension into the bulk of the boundary of the BCFT. See the bottom center panel of Figure \ref{fig:slicingduality}. The tension parameter $T$ is related to the boundary entropy of the Cardy state $\ket{B}$ in the dual BCFT \cite{Takayanagi:2011zk,Fujita:2011fp}, and is therefore uniquely determined by the specific choice of conformal boundary conditions at the endpoints of the interval $I$. 

Under specific conditions that we will explain in Section \ref{subsec:gravitylocalization}, an effective 4D description of gravity localized on the brane via the Karch-Randall-Sundrum mechanism \cite{Randall:1999ee,Randall:1999vf,Karch:2000ct} is possible. When this is the case, there exists an effective 4D description of physics on the ETW brane (the ``braneworld'') without referring to the higher dimensional 5D bulk setup. The gravitational sector of the induced effective field theory on the brane is given generically by Einstein gravity corrected by specific higher derivative terms \cite{Bueno:2022log}. The higher derivative corrections are suppressed in the large tension limit in which gravity localization holds. Because, as we will see, this limit represents the regime of our interest, the gravitational effective field theory in our braneworld setup is simply captured by an Einstein-Hilbert action to good approximation. A detailed study of the higher derivative corrections arising in our specific setup is an interesting endeavour that we leave for future work.

Several studies have explored properties of braneworld cosmologies. For example, \cite{Huey:2001ae, Vinet:2004bk, Cline:2003ak, Majumdar:2005ba,Anber:2008qu,Schmidt:2009sg, Bilic:2018cyh} explore their phenomonological properties, whereas \cite{Emparan:2022ijy, Panella:2023lsi, Aguilar-Gutierrez:2023tic} explored the possibility to study various de Sitter ($\Lambda>0$) setups using braneworlds. On the other hand, as we have discussed, our attention will be focused on $\Lambda<0$ braneworlds that admit a dual microscopic description via ordinary AdS/CFT. In our setup, the braneworld will be coupled to non-gravitational auxiliary DOF accounting for the physics of the 5D bulk.\footnote{Similarly to what we explained in Footnote \ref{footnote:DOF}, this is just an intuitive description. In reality, the physics of the 5D bulk affects both the auxiliary DOF and the 4D gravitational effective theory on the ETW brane.} This braneworld description takes the form of the 4D Euclidean Maldacena-Maoz wormhole reviewed above. The 5D setup with ETW branes can then be seen as a doubly holographic realization of the ideas introduced in \cite{VanRaamsdonk:2020tlr,VanRaamsdonk:2021qgv,Antonini:2022blk,Antonini:2022ptt}, in a similar way as doubly holographic setups for black hole evaporation were built in \cite{Almheiri:2019hni}. The advantage is that this doubly holographic realization can be built explicitly in setups with a well-understood holographic dual description.

\begin{figure}
    \centering
    \includegraphics[width=0.8\linewidth]{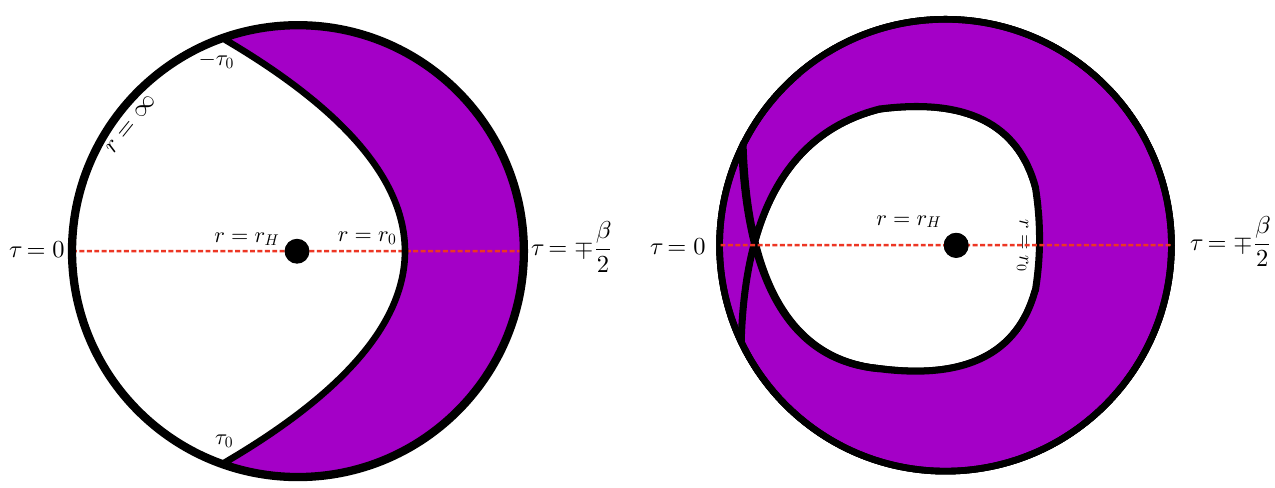}
    \caption{Left: Euclidean black hole cut off by a codimension-1 ETW brane with trajectory in the $r-\tau$ directions. Right: in the AdS-Schwarzschild case \cite{Cooper:2018cmb} the brane overlaps itself for large tension (which is needed for gravity localization), signaling an ill-defined bulk setup and no holographic dual description in this regime. In our magnetic setup, gravity localization can be achieved without the brane overlapping.}
    \label{fig:overlapping}
\end{figure}

The first construction of this AdS/BCFT setup was given in \cite{Cooper:2018cmb} (see also \cite{Kourkoulou:2017zaj,Almheiri:2018ijj} for earlier lower-dimensional realizations of similar constructions), where the 5D bulk spacetime was taken to be a Euclidean AdS-Schwarzschild black hole cutoff by an ETW brane, see Figure \ref{fig:overlapping}. The analytic continuation corresponding to the cosmology picture yields a Lorentzian double-sided black hole with one of the two asymptotic boundaries completely cut off by an ETW brane, similar to the bottom-right panel of Figure \ref{fig:slicingduality}. This is a single-sided black hole microstate, with the corresponding pure excited state of the dual microscopic theory living on the only remaining asymptotic boundary (the left one in Figure \ref{fig:slicingduality}). An observer comoving with the brane experiences the expansion and contraction of a Big Bang-Big Crunch FLRW universe. 

Notice that, in the planar limit, the same Euclidean saddle admits also an alternative analytic continuation yielding a real Lorentzian spacetime: the static AdS${}_5$ soliton spacetime \cite{Horowitz:1998ha,Witten:1998zw} cut off by a static ETW brane, see bottom-left panel of Figure \ref{fig:slicingduality}. The braneworld description is now that one of an eternal traversable wormhole connecting two asymptotic AdS${}_4$ boundaries, realizing the Lorentzian wormhole picture of the singly holographic description reviewed above \cite{VanRaamsdonk:2021qgv,Antonini:2022blk,Antonini:2022ptt}. AdS${}_5$ solitons are understood to be dual to negative energy ground states of the dual holographic theory on $\mathbb{R}^{2,1}\times S^1$ \cite{Horowitz:1998ha}, which is expected to confine in the IR at length scales much larger than the size of the $S^1$ \cite{Witten:1998zw}. In our case, the soliton cut off by the ETW brane is then dual to the ground state of the BCFT, which is also expected to confine at length scales much larger than the interval $I$.

One important issue with the AdS-Schwarzschild construction of \cite{Cooper:2018cmb} is that gravity localization (and therefore an effective 4D braneworld description) cannot be achieved in a well-defined bulk setup with a well-defined dual description. In fact, if we try to increase the tension $T$ of the brane enough to obtain gravity localization\footnote{As we will discuss in Section \ref{subsec:gravitylocalization}, increasing the tension allows us to ``push'' the brane far into the asymptotic AdS region, which is necessary for gravity localization to hold. From a microscopic point of view, this corresponds to increasing the number of boundary degrees of freedom of the BCFT.}, the brane overlaps itself in the Euclidean setup, completely cutting off the asymptotic boundary, see Figure \ref{fig:overlapping}. Therefore, we lose control over the 5D bulk EFT and no well-defined microscopic dual description exists.
This problem was addressed in \cite{Antonini:2019qkt} by replacing the AdS-Schwarzschild spacetime by a near-extremal AdS-Reissner-Nordstr\"om (AdS-RN) black hole. In this case, it was shown that the tension can be taken arbitrarily close to the critical value $T_{crit}=1/L$ (where $L$ is the 5D AdS radius), and the brane does not overlap if the black hole charge is close enough to its extremal value. However, in the AdS-RN setup there is no corresponding well-defined soliton solution because, under double analytic continuation, the electric field in AdS-RN becomes an imaginary magnetic field in the soliton.

In this paper, we consider a magnetically charged setup which allows for gravity localization on the brane while retaining a well-defined dual theory. Furthermore, both analytic continuations (to a Lorentzian magnetic black brane (MBB) and to a magnetically charged AdS${}_5$ soliton) of the Euclidean saddle are real and well-defined, giving rise to a braneworld cosmology and a braneworld traversable wormhole, respectively. Our setup is therefore a first explicit and complete doubly-holographic realization of the ideas advanced in \cite{VanRaamsdonk:2021qgv,Antonini:2022blk,Antonini:2022ptt}. It realizes in a concrete setting the embedding of a $\Lambda<0$ cosmological universe in a holographic setup, it relates this to the physics of traversable wormholes in AdS, and allows us to study explicitly the features of the dual holographic theory.

\subsection{Summary of results and outline}

Let us briefly summarize the main constructions and results of this paper:

\begin{itemize}
    \item We start by studying the (4+1)-dimensional MBB Lorentzian spacetime first constructed in \cite{DHoker:2009mmn} (obtained numerically) and we observe that it arises from analytic continuation of a Euclidean saddle of the gravitational path integral. We also observe the existence of an alternative analytic continuation of the same Euclidean saddle yielding a static, magnetically charged AdS${}_5$ soliton spacetime. Unlike the AdS-RN case, this soliton solution is well-defined and has a real magnetic field. This is thanks to the breaking of rotational symmetry due to the presence of a magnetic field in 5D. The drawback of this symmetry breaking is that the spacetime metric cannot be obtained analytically.
    \item The holographic theory on $\mathbb{R}^{2,1}\times S^1$ dual to the magnetic soliton spacetime is confining on scales much larger than the $S^1$. Following \cite{Witten:1998zw}, we compute the discrete spectrum of massive particles (glueballs \cite{Csaki:1998qr}) by means of a bulk calculation and show the existence of a mass gap.
    \item Embedding an ETW brane in the Euclidean 5D spacetime, for a sufficiently large magnetic field the brane tension can be increased enough to allow for gravity localization on the brane without the latter overlapping with itself. This guarantees a well-defined 5D bulk solution and dual theory (given by a 4D BCFT), and a well-defined 4D effective braneworld description on the brane. The Euclidean braneworld takes the form of a Maldacena-Maoz wormhole. 
    \item One analytic continuation of this Euclidean saddle leads to the double-sided MBB with one boundary completely cut off by an ETW brane. This MBB microstate is dual to an excited state of a 4D CFT living on the only remaining asymptotic boundary (which is $\mathbb{R}^{3,1}$). The 4D braneworld description is given by a Big Bang-Big Crunch homogeneous but anisotropic (Bianchi I \cite{Taub:1950ez,Ellis:1968vb,Montani:2009hju}) cosmology. In the region where a good 4D braneworld description exists, the anisotropies are very small and the braneworld cosmology approximately takes the form of a flat FLRW universe. In the braneworld cosmology setting, gravity localization only holds for (a large) part of the brane trajectory around the time-symmetric point.
    \item At early times (but where gravity localization still holds and an effective 4D braneworld description still exists), the braneworld cosmology undergoes a phase of accelerated expansion, likely driven by a negative (from the point of view of the braneworld) energy density proportional to the magnetic field. The expansion then decelerates and eventually stops when the negative cosmological constant $\Lambda_4<0$ takes over. The recollapse is time-symmetric.
    \item A different analytic continuation of the same Euclidean saddle leads to the magnetically charged AdS${}_5$ soliton with part of the boundary cut off by an ETW brane. This is dual to the ground state of a BCFT on $\mathbb{R}^{2,1}\times I$, which confines on scales much larger than the size of the interval $I$. The 4D braneworld description is given by a nearly Poincar\'e symmetric eternally traversable wormhole connecting two distinct AdS${}_4$ boundaries.
    In the braneworld wormhole setting, gravity localization holds everywhere on the brane.
    The confined spectrum of the dual theory can be computed, following \cite{Antonini:2022opp}, directly from the 4D braneworld effective theory.
    \item The Euclidean saddle giving rise to the results above, which contains a connected ETW brane, is the only known saddle contributing to the gravitational ensemble. The existence of a natural alternative Euclidean saddle containing two disconnected ETW branes is ruled out in Appendix \ref{app:actioncomparison}.
\end{itemize}

The paper is organized as follows. In Section \ref{sec:MBB} we study in depth the properties of the AdS${}_5$ magnetic black brane (MBB) solution, its double analytic continuation given by the magnetically charged AdS${}_5$ soliton, and their holographic dual description. Many interesting features of these constructions are discussed, and a derivation of the mass spectrum of confining particles (glueballs) of the theory dual to the soliton is given. Although knowledge of these properties and results will deepen the understanding of the following sections, the reader eager to learn the main upshot related to braneworld cosmology and braneworld wormholes can find the MBB and AdS${}_5$ soliton background metrics in equations \eqref{eq:MBBmetric} and \eqref{eq:MBBsolitonmetric}, with the metric functions obtained numerically plotted in Figure \ref{fig:fgh}. This knowledge should be enough for a surface-level understanding of Sections \ref{sec:euclidean}, \ref{sec:cosmology}, and \ref{sec:wormhole}. In Section \ref{sec:euclidean} we embed an ETW brane in the Euclidean geometry and show that well-defined solutions supporting gravity localization and admitting a holographic dual description can be obtained. We then show that the braneworld description is given by a Maldacena-Maoz-like wormhole. In Section \ref{sec:cosmology} we study the braneworld cosmology embedded in the Lorentzian MBB solution and its dual description, and comment on the properties of bulk reconstruction in this setting. In Section \ref{sec:wormhole} we study the Lorentzian traversable braneworld wormhole embedded in the magnetically charged AdS${}_5$ soliton, study the confining properties of the dual theory, comment on the slicing duality, and explain the relationship of our solution with other constructions of 4D eternally traversable wormholes. Finally, in Section \ref{sec:discussion} we give our final remarks. The appendices contain technical details related to the main sections of the paper. In particular, Appendix \ref{app:actioncomparison} shows that a naively possible alternative saddle of the gravitational Euclidean path integral does not exist in our setup. The Euclidean saddle examined in the main text is therefore the only (known) saddle of the Euclidean path integral and it dominates the gravitational ensemble.

In order to facilitate the reading, we include at the beginning of each section a quick overview of the most relevant results and the organization of that section.

\subsubsection*{Glossary}

To avoid confusion, we recap here the different descriptions depending on how many layers of holography and which analytic continuation we consider (depicted in Figure \ref{fig:slicingduality}), and we specify the terms we will use in the rest of the paper to refer to each one of them.
\begin{itemize}
    \item We refer to the microscopic UV complete description depicted in the top row of Figure \ref{fig:slicingduality} interchangeably as: microscopic description/theory/DOF, 4D BCFT. This is what in the double holography literature \cite{Almheiri:2019hni} is typically called the ``fundamental description''.
    \item We refer to the intermediate 4D bulk EFT depicted in the central row of Figure \ref{fig:slicingduality} given by the wormhole/cosmology and non-gravitational auxiliary DOF interchangeably as: effective 4D EFT/description, singly holographic setup, braneworld description.
    \item We refer to the 5D bulk setup including an ETW brane depicted in the bottom row of Figure \ref{fig:slicingduality} interchangeably as: 5D bulk theory/description, doubly holographic setup, higher dimensional setup.
\end{itemize}
Depending on which description we are analyzing (microscopic, singly, or doubly holographic, respectively), we refer to the Euclidean and the two Lorentzian pictures as:
\begin{itemize}
    \item Left column of Figure \ref{fig:slicingduality}, generically referred to as ``wormhole picture''. Top to bottom: vacuum state of the confining theory on $\mathbb{R}^{2,1}\times I$; AdS${}_4$ eternally traversable wormhole or braneworld wormhole; magnetically charged AdS${}_5$ soliton cut off by ETW brane.
    \item Center column of Figure \ref{fig:slicingduality}, generically referred to as ``Euclidean picture''. Top to bottom: Euclidean BCFT on $\mathbb{R}^3\times I$; Maldacena-Maoz/Euclidean wormhole; 5D Euclidean bulk geometry.
    \item Right column of Figure \ref{fig:slicingduality}, generically referred to as ``cosmology picture''. Top to bottom: highly excited pure state of CFT on $\mathbb{R}^{3,1}$; braneworld cosmology; MBB cut off by ETW brane.
\end{itemize}

\section{$\text{AdS}_{5}$ Magnetic Black Brane and soliton}
\label{sec:MBB}

In this section we introduce the asymptotically AdS${}_5$ background spacetimes of our interest: the AdS${}_5$ magnetic black brane (MBB), the magnetically charged AdS${}_5$ soliton, and the Euclidean saddle they arise from. The MBB is a solution of Einstein-Maxwell equations in five dimensions with a negative cosmological constant $\Lambda<0$ and a constant magnetic field $B$ tangent to the asymptotic boundary. First introduced by D'Hoker and Kraus \cite{DHoker:2009mmn} (see also \cite{DHoker:2009ixq,Martinez-y-Romero:2017awl,Avila:2018sqf}), it is a one-parameter family of solutions at finite temperature $\mathcal{T}$ with a spatially flat asymptotic boundary, parametrized by the dimensionless ratio $\mathcal{T}/\sqrt{B}$. 

In order to get acquainted with the properties of planar solutions of Einstein-Maxwell equations, we first consider in Section \ref{sec:planaradsrn} the planar limit of the electrically charged AdS-Reissner-Nordstr\"om (AdS-RN) black hole, pointing out the similarities and differences with the MBB solution of our interest. Working out the AdS-RN case, on which we have full analytic control, is particularly useful to guide our intuition when treating the MBB solution, which can only be found numerically. We then numerically construct the MBB solution in Section \ref{sec:MBBsol}. We use a different choice of coordinates with respect to \cite{DHoker:2009mmn}, which will turn out to be favorable to study ETW branes embedded in the MBB background. In Section \ref{sec:MBBsoliton} we consider a double analytic continuation of the MBB solution yielding a magnetically charged AdS${}_5$ soliton spacetime and comment on its properties and role in the context of AdS/CFT. Both the MBB and the soliton arise from analytic continuation of the same Euclidean saddle, which we describe. Finally, in Section \ref{sec:MBBdual} we comment on the microscopic dual description of the MBB and the soliton and, following \cite{Witten:1998zw}, provide evidence for the existence of a mass gap and a spectrum of confined particles in the holographic theory dual to the magnetic soliton.

\subsection{Planar limit of AdS-Reissner-Nordstr\"om}
\label{sec:planaradsrn}

As a warm up, let us consider the planar limit of the electrically charged AdS-RN black hole in five dimensions. This can be obtained by generalizing the procedure employed in \cite{Witten:1998zw} to derive the planar limit of the AdS-Schwarzschild black hole. The metric of the 5D AdS-RN black hole is given by
\begin{equation}
    ds^2=-f(r)dt^2+\frac{dr^2}{f(r)}+r^2d\Omega_3^2
\end{equation}
where
\begin{equation}
    f(r)=1+\frac{r^2}{L^2}-\frac{2M}{r^2}+\frac{Q^2}{r^4},
\end{equation}
$M$ is the mass parameter (proportional to the ADM mass), and $Q$ is the charge parameter. This metric has two horizons at $r=r_\pm$ and its inverse temperature, given by $\beta=4\pi/f'(r_+)$, diverges in the extremal limit for which $r_-=r_+$. The electromagnetic field strength has $F_{tr}=-F_{rt}\neq 0$, with all other components vanishing.

First, we can rescale the radial coordinate as $r=(2M/L^2)^{\frac{1}{4}}\rho$ and the time coordinate as $t=(L^2/2M)^{\frac{1}{4}}\tilde{t}$, obtaining the metric
\begin{equation}
    ds^2=-\tilde{f}(\rho)d\tilde{t}^2+\frac{d\rho^2}{\tilde{f}(\rho)}+\rho^2 \sqrt{\frac{2M}{L^2}} d\Omega_3^2,
    \label{eq:rnmetricrescaled}
\end{equation}
where we defined
\begin{equation}
    \tilde{f}(\rho)=\sqrt{\frac{L^2}{2M}}f(\rho)=\sqrt{\frac{L^2}{2M}}+\left(\frac{\rho^2}{L^2}-\frac{L^2}{\rho^2}+\frac{\alpha L^4}{\rho^4}\right), \quad \quad \alpha\equiv \frac{Q^2}{L(2M)^{\frac{3}{2}}}.
    \label{eq:frho}
\end{equation}
We can then take the limit of infinite mass and infinite charge, keeping the ratio $\alpha$ fixed. In this limit, we can drop the first term on the right-hand side of equation \eqref{eq:frho}. From \eqref{eq:rnmetricrescaled} we also see that the size of the $S^3$ diverges as $M\to\infty$ yielding a 3-plane, $\sqrt{(2M/L^2)}d\Omega_3^2\to dx^2+dy^2+dz^2$. We are therefore left with the planar AdS-RN metric
\begin{equation}
    ds^2=-\left(\frac{\rho^2}{L^2}-\frac{L^2}{\rho^2}+\frac{\alpha L^4}{\rho^4}\right)dt^2+\frac{d\rho^2}{\left(\frac{\rho^2}{L^2}-\frac{L^2}{\rho^2}+\frac{\alpha L^4}{\rho^4}\right)}+\rho^2(dx^2+dy^2+dz^2),
    \label{eq:planaradsrn}
\end{equation}
which in the zero charge limit $\alpha=0$ reduces to the planar AdS-Schwarzschild solution \cite{Witten:1998zw}. 

A few comments are in order. First, notice that taking the planar limit we reduced a two-parameter ($M,Q$) family of solutions to a one-parameter ($\alpha$) family of solutions. This can be readily understood in terms of the conformal symmetry of the dual boundary theory. Initially, there are three length scales in the boundary theory\footnote{The dual boundary theory has a finite chemical potential $\mu$, which is proportional to the charge parameter $Q$ introduced above, see e.g \cite{Guo:2015swu,Andrade:2013rra}.}: the inverse temperature $\beta$ (i.e. the periodicity of the Euclidean time circle), the size of the spatial $S^3$, and the length scale associated with the chemical potential $l_\mu\propto \mu^{-1}$. Since conformal invariance guarantees that only adimensional parameters are physically meaningful, these three length scales can be combined into two physical parameters, e.g. the ratios $\beta/R$ and $l_\mu/R$ with $R$ the radius of the spatial $S^3$. In the planar limit $R\to\infty$ we are clearly left with only two length scales, $\beta$ and $l_\mu$, and therefore with a single dimensionless parameter given by their ratio, which can be expressed in terms of bulk parameters as a combination of $Q$, $M$, and $L$, namely the $\alpha$ parameter of equation \eqref{eq:frho}. This fact remains true for the magnetic black branes we will study in the next subsections, which are indeed a one-parameter family of solutions as we will discuss.

Second, the planar solution still admits a well-defined extremal limit for which $l_\mu/\beta=0$. By setting $\tilde{f}(\rho)=\tilde{f}'(\rho)=0$ after taking the planar limit, the extremal value of the free parameter can be found to be $\alpha_e=2/(3\sqrt{3})$. This is contrast with the MBB solution which, as we will see, is not well defined in the strictly extremal limit.

Finally, the Euclidean geometry associated with the planar AdS-RN black hole admits a different analytic continuation yielding a real Lorentzian geometry, where one of the planar directions (e.g. $z$) becomes timelike. Notice that the $\tau=it$ direction is now compact and it pinches off at $r=r_+$. This static cigar geometry is an AdS${}_5$ soliton spacetime. AdS${}_5$ solitons are understood to be dual to negative energy ground states of CFTs living on manifolds with one compact direction \cite{Horowitz:1998ha}. In this case, the (3+1)-dimensional dual theory would live on a spatial $\mathbb{R}^2\times S^1$ manifold. As we will explain in Section \ref{sec:MBBdual}, similar theories flow in the IR to (2+1)-dimensional theories on spatial $\mathbb{R}^2$ which are expected to be confining \cite{Witten:1998zw}. However, in the electrically charged setup studied in this subsection, this alternative analytic continuation does not yield a well-defined Lorentzian theory. In fact, it is straightforward to show that the electric field $F_{tr}$ becomes an imaginary magnetic field in the bulk theory, and, correspondingly, the chemical potential becomes imaginary in the dual boundary theory.\footnote{One possible way to make sense of this theory is to Wick rotate also the charge $Q$, see e.g. \cite{Anabalon:2021tua}. Note that in this case the Euclidean saddles associated to the two Lorentzian spacetimes are different, because the size of the $S^1$ at infinity (i.e. the periodicity $\beta$) changes under analytic continuation of $Q$. The physical motivation and implications of this procedure are not entirely clear and we will not explore this possibility in the present paper.} 

On the other hand, in the magnetically charged black branes studied in the next subsections, the soliton solution and the corresponding dual confining theory will be well-defined.

\subsection{AdS${}_5$ magnetic black brane}
\label{sec:MBBsol}

With the properties of the planar AdS-RN solution in mind, let us now construct the MBB solution of our interest following \cite{DHoker:2009mmn}. The Lorentzian Einstein-Maxwell action in five dimensions is given by
\begin{equation}
    S=\frac{1}{16\pi G}\int d^5x \sqrt{-\bar{g}} (R-2\Lambda-F_{\mu\nu}F^{\mu\nu})+S_{\partial}
    \label{eq:bulkaction}
\end{equation}
where $G$ is the 5D Newton's constant, $\Lambda=-6/L^2$ is the negative cosmological constant, $F_{\mu\nu}=\partial_\mu A_\nu-\partial_\nu A_\nu$ is the electromagnetic field strength (with $A_\mu$ the electromagnetic potential), and $S_\partial$ includes boundary terms such as the Gibbons-Hawking-York term, the necessary electromagnetic boundary terms \cite{Hawking:1995ap}, and other counterterms \cite{DHoker:2009ixq,Ammon:2017ded}.\footnote{A Chern-Simons term can also be added, and our action then becomes the action for the bosonic sector of minimal gauged supergravity in 5D \cite{DHoker:2009mmn}. We do not include this term here, which would not give a contribution to our solution.} We denoted the determinant of the 5D metric by $\bar{g}$ to distinguish it from the metric function $g$ appearing in the ansatz \eqref{eq:MBBmetric}. With these conventions, the Einstein-Maxwell equations are\footnote{These equations need to be supplemented by the Bianchi identity.}
\begin{equation}
\begin{aligned}
    &R_{\mu\nu}-\frac{1}{2}(R-2\Lambda)\bar{g}_{\mu\nu}=2T_{\mu\nu}\\
    &\nabla_\mu F^{\mu\nu}=0
    \end{aligned}
    \label{eqn:einsteinmaxwelleqn}
\end{equation}
where $T_{\mu\nu}=\bar{g}^{\rho\sigma}F_{\mu\rho}F_{\nu\sigma}-\frac{1}{4}\bar{g}_{\mu\nu}F^{\rho\sigma}F_{\rho\sigma}$. We are interested in finding an asymptotically AdS, planar solution sourced by a constant magnetic field $B$ tangent to the planar directions, which we label by $x,y,z$ in analogy with \eqref{eq:planaradsrn}. We can then simply take $F_{xy}=-F_{yx}=B$,\footnote{A simple choice of potential yielding this field strength is $A=Bx dy$.} with all other components of the field strength vanishing, and plug it into the Einstein-Maxwell equations. Notice that a magnetic field breaks rotational symmetry in five dimensions by distinguishing the $x-y$ plane from the perpendicular $z$ direction. The most general ansatz for the spacetime metric is therefore
\begin{equation}
    ds^2=-f(r)dt^2+\frac{dr^2}{f(r)}+g(r)(dx^2+dy^2)+h(r)dz^2
    \label{eq:MBBmetric}
\end{equation}
with the AdS asymptotics requiring $f(r)\sim g(r)\sim h(r)\sim r^2$ as $r\to\infty$. The planar directions $x,y,z$ can be taken to be compactified into a torus $\mathbb{T}^3$, but we will be interested in the non-compact case $\mathbb{R}^3$. 
The absence of rotational symmetry is the reason why an analytic solution of Einstein's equations is hard (if possible at all) to find.

By defining $g(r)\equiv \exp(2V(r))$, $h(r)\equiv\exp(2W(r))$, Einstein's equations simplify significantly, taking the form
\begin{equation}
    \begin{aligned}
     &f''+f'(2V'+W')=8+\frac{4}{3}B^2e^{-4V}\\
     &2V''+W''+2(V')^2+(W')^2=0\\
     &f(V''-W'')+[f'+f(2V'+W')](V'-W')=-2B^2e^{-4V}\\
     &f'(2V'+W')+2fV'(V'+2W')=12-2B^2e^{-4V}
    \end{aligned}
    \label{eq:einsteineq}
\end{equation}
where a prime denotes a derivative with respect to $r$. Notice that, naturally, $f(r)=g(r)=h(r)=r^2$ is a solution for $r\to\infty$. Moreover, an exact solution of Einstein's equations is given by the product of the BTZ black hole and $\mathbb{R}^2$, namely $f(r)=3(r^2-r_H^2)$, $g(r)=B/\sqrt{3}$, $h(r)=3r^2$ \cite{DHoker:2009mmn}, where $r_H$ is the black hole horizon radius. This is not asymptotically AdS${}_5$, but it suggests that the solution of our interest, i.e. the MBB, has a horizon and it interpolates between a near-horizon region given by BTZ$\times \mathbb{R}^2$ and the AdS${}_5$ asymptotics.\footnote{This suggests that the near-horizon dynamics is captured by an emergent (1+1)-dimensional CFT in the $t-z$ directions \cite{DHoker:2009mmn}.} Although we could not construct this solution analytically, it is straightforward to find numerically.

The boundary conditions to impose to find a numerical solution are (see Appendix \ref{app:MBBbc} for a detailed derivation, we set here and in the following $L=1$ and shift the $r$ coordinated such that $r_H=1$)
\begin{equation}
    f(1)=V(1)=W(1)=0, \quad f'(1)=\frac{8}{3}(3-b^2), \quad V'(1)=\frac{1}{2},\quad W'(1)=\frac{6+b^2}{4(3-b^2)}.
    \label{eq:MBBbc}
\end{equation}
With these boundary conditions (which are completely general), $g$ and $h$ behave asymptotically as $g(r)\sim v(b)r^2$, $h(r)\sim w(b)r^2$, with $v(b)$ and $w(b)$ constants depending only on the value of the magnetic field parameter $b$. After obtaining our solution, we will therefore need to numerically find the value of $v(b),w(b)$ and appropriately rescale the $x$, $y$ coordinates to obtain the correct asymptotically AdS${}_5$ solution. The physical magnetic field is then given by $B=b/v(b)$. We remark that the difference between our boundary conditions in \eqref{eq:MBBbc} and those used in \cite{DHoker:2009mmn} is non-physical and only corresponds to a different choice of coordinates. Our choice is more convenient to numerically study gravity localization on ETW branes embedded in this background. See Appendix \ref{app:MBBbc} for details.

\begin{figure}[h]
    \centering
    \includegraphics[width=0.5\textwidth]{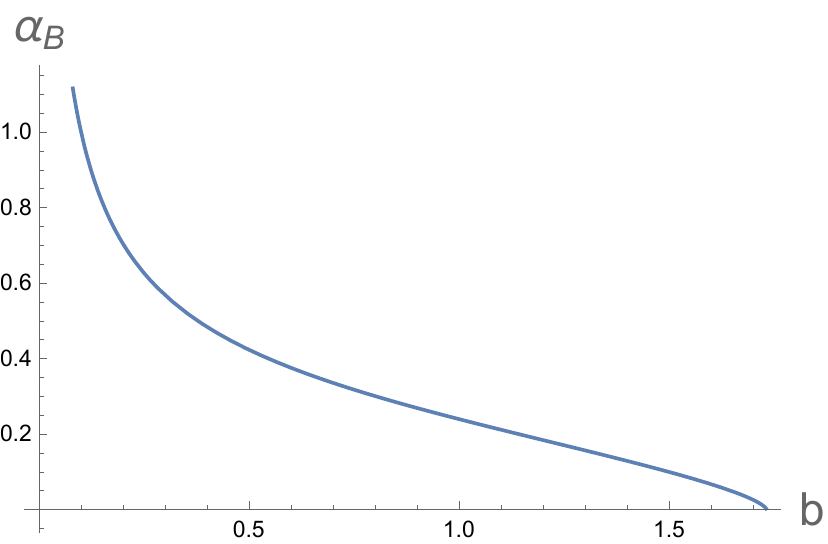}
    \caption{Dimensionless parameter $\alpha_B$ (defined in equation \eqref{eq:alphab}) as a function of the magnetic field parameter $b$. $\alpha_B$ diverges for $b\to 0$, signaling a large temperature phase dominated by the planar limit of AdS-Schwarzschild, and it vanishes for $b\to\sqrt{3}$, where the MBB solution ceases to exist. At exactly zero temperature, the gravitational thermodynamic ensemble is dominated by a different (unique for all values of $b$) solution interpolating between AdS${}_3\times \mathbb{R}^2$ deep into the bulk and AdS${}_5$ asymptotically \cite{DHoker:2009mmn}.}
    \label{fig:alphab}
\end{figure}

We can now solve the three dynamical equations in the system \eqref{eq:einsteineq} with the boundary conditions \eqref{eq:MBBbc}.\footnote{For the sake of numerical implementation, the boundary conditions \eqref{eq:MBBbc} must be perturbed slightly away from the black hole horizon.} Similar to the AdS-RN case studied in Section \ref{sec:planaradsrn}, this defines a one-parameter family of solutions parametrized by $b$ in our numerical solution. The only dimensionless physical parameter is given by
\begin{equation}
    \alpha_B\equiv \frac{\mathcal{T}}{\sqrt{B}}=\frac{2(3-b^2)\sqrt{v(b)}}{3\pi\sqrt{b}}.
    \label{eq:alphab}
\end{equation}
Smooth solutions can be found in the range $0\leq |b|<\sqrt{3}$.\footnote{Without loss of generality, we will focus on $b\geq 0$ in the rest of the paper. The solutions for $b<0$, which differ only by the direction of the magnetic field, are identical.} For $b=0$, two solutions are simply Poincar\'e-AdS${}_5$ and the planar limit of AdS-Schwarzschild, with the latter dominating the gravitational thermodynamic ensemble, because the planar limit is equivalent to a large temperature limit \cite{Witten:1998zw}.\footnote{This is readily understood because the ratio of the size of the thermal $S^1$ to the size of the spatial $S^3$ at the boundary vanishes in the planar limit. It is also made manifest by the divergence of the $\alpha_B$ parameter defined in equation \eqref{eq:alphab}.} On the other hand, because $\alpha_B\to 0$ for $b\to\sqrt{3}$ (see Figure \ref{fig:alphab}), it is tempting to interpret $b=\sqrt{3}$ as an extremal solution, similar to the AdS-RN case. However, from \eqref{eq:MBBbc} we also see that $W'(1)\to\infty$ in the same limit, signaling that no smooth extremal solution exists and the MBB solution simply ceases to exist in the $b\to\sqrt{3}$ limit.\footnote{Our choice of coordinates makes this breakdown of the MBB solution for $b\to\sqrt{3}$ analytically manifest in the blowup of the boundary condition for $W'$. With different coordinate choices the breakdown is not immediately manifest, but nonetheless numerical solutions with the correct AdS${}_5$ asymptotics can again only be found for $b<\sqrt{3}$ \cite{DHoker:2009mmn}. Notice also that $v(b)$ and $w(b)$ blow up for $b\to\sqrt{3}$ (with the combination $(3-b^2)\sqrt{v(b)}\to 0$, guaranteeing $\alpha_B\to 0$), which is another sign of the breakdown of the MBB solution.} Another indication of the difference between the $b\to\sqrt{3}$ limit and the extremal limit of the AdS-RN solution is that the entropy density in the MBB case vanishes in the $b\to\sqrt{3}$ limit \cite{DHoker:2009mmn}, whereas the extremal AdS-RN solution has a finite entropy. It was shown in \cite{DHoker:2009mmn} that at exactly zero-temperature a different solution exists, interpolating between Poincar\'e-AdS${}_3\times\mathbb{R}^2$ deep into the bulk and AdS${}_5$ asymptotically. This solution exhibits Poincar\'e invariance in the $t-z$ plane, i.e. $f(r)=h(r).$\footnote{\label{poincarefootnote}The absence of a Poincar\'e symmetry in the $x,y$ directions is due to the presence of the magnetic field, which breaks rotational invariance in the planar directions.} Similar to the planar AdS-Schwarzschild case, in this limit there are no dimensionless parameters and the solution is therefore unique, independent of the value of the magnetic field. More details of this zero temperature solution and why it does not contribute to the Euclidean path integral in our braneworld setup are reported in Appendix \ref{app:actioncomparison}. We refer to \cite{DHoker:2009mmn} for details of the numerical construction of this solution.

\begin{figure}[h]
    \centering
    \includegraphics[width=0.65\textwidth]{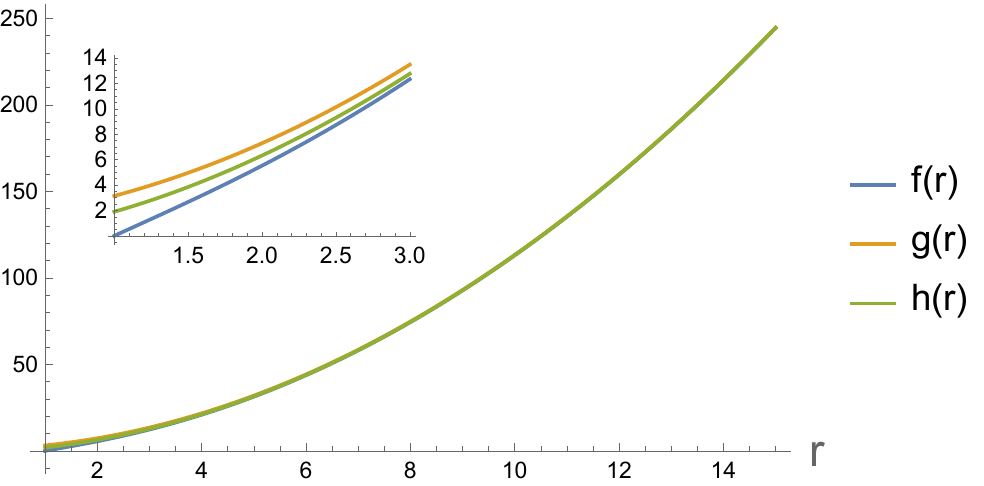}
    \caption{Metric functions $f(r)$, $g(r)$, $h(r)$ as a function of the radial coordinate for $b=1$ (i.e. $\alpha_B=0.239494$). In the BTZ$\times\mathbb{R}^2$ near-horizon region (see the inset) the three functions are distinct, but they become indistinguishable for larger values of $r$, as expected for an asymptotically AdS${}_5$ solution.}
    \label{fig:fgh}
\end{figure}

From now on, we will focus on the range $0<b<\sqrt{3}$ where the MBB solution is well-defined. As an exemplificatory numerical solution, we report in Figure \ref{fig:fgh} the metric functions $f(r)$, $g(r)$, $h(r)$ for $b=1$ (with $g$ and $h$ appropriately rescaled by $v(b)$ and $w(b)$, respectively). It is worth mentioning that our numerical results show $f(r)$ and $h(r)$ becoming indistinguishable for any value of $r$ in the $b\to\sqrt{3}$ limit. This is in accordance with the expectation that we should recover Poincar\'e invariance in the $t-z$ plane in the zero temperature limit.

The BTZ$\times\mathbb{R}^2$ structure of the near-horizon region of the MBB suggests that the causal structure of the MBB is similar to AdS-Schwarzschild, namely presenting a single horizon and a spacelike singularity. This expectations is confirmed by the extension of our numerical solution inside the black hole horizon (see also \cite{Avila:2018sqf,HosseiniMansoori:2018gdu}). The maximally extended Penrose diagram for the MBB is then depicted on the right of Figure \ref{fig:MBBreflectionsymmetries}, where we also depict one of the non-compact directions transverse to the $r-t$ directions.

\subsection{Magnetically charged AdS${}_5$ soliton}
\label{sec:MBBsoliton}

\begin{figure}
    \centering
    \includegraphics[width=0.7\linewidth]{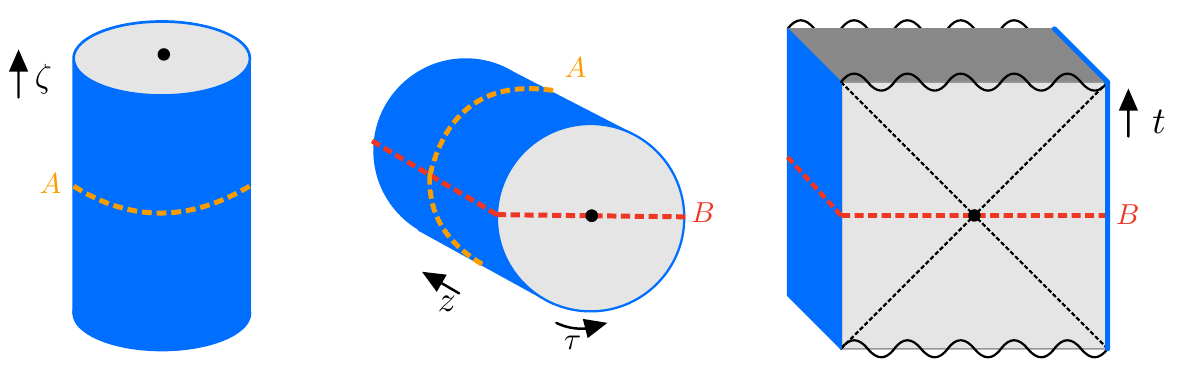}
    \caption{Center: Euclidean saddle associated with the MBB and magnetic AdS${}_5$ soliton solutions. The black dot indicates $r=1$, where the $\tau$ circle pinches off. We are suppressing two non-compact directions ($x$ and $y$). Right: Identifying $\tau$ with the Euclidean time, the semiclassical gravitational Euclidean path integral prepares the initial state for bulk fields in the Lorentzian maximally extended MBB solution on the $\tau=0,\pm\beta/2$ slice represented by the red dotted line, which is left invariant by the analytic continuation. The causal structure of the MBB is similar to AdS-Schwarzschild. Left: Identifying $z$ with the Euclidean time, the initial state for Lorentzian time evolution is prepared on the $z=0$ slice, identified by the yellow dotted line. The corresponding Lorentzian geometry is the static AdS${}_5$ soliton \eqref{eq:MBBsolitonmetric}. The black dot indicates $r=1$, where the $\tau$ circle pinches off. }
    \label{fig:MBBreflectionsymmetries}
\end{figure}

Let us now consider the Euclidean metric associated with the MBB metric \eqref{eq:MBBmetric}:
\begin{equation}
    ds^2=f(r)d\tau^2+\frac{dr^2}{f(r)}+g(r)(dx^2+dy^2)+h(r)dz^2.
    \label{eq:euclideanMBB}
\end{equation}
The Euclidean time $\tau$ is periodic with periodicity $\beta=4\pi/f'(1)$. The Euclidean geometry is then given, as usual for Euclidean black holes, by a cigar geometry in the $\tau-r$ plane with the $\tau$ circle pinching off at $r=1$, see Figure \ref{fig:MBBreflectionsymmetries}. The transverse directions $x,y,z$ are all non-compact. This Euclidean geometry is the dominant saddle of the Euclidean gravitational path integral.\footnote{It is the only known saddle given our boundary conditions for $\beta<\infty$. If other saddles exist, it is also expected to be the dominant one because, as we shall see, it gives rise to the AdS${}_5$ soliton Lorentzian solution, which is understood to be dual to the ground state of the dual CFT \cite{Horowitz:1998ha}. For $\beta=\infty$ there exists a different saddle \cite{DHoker:2009mmn}, but this clearly does not contribute to the finite $\beta$ thermodynamic ensemble.}

This Euclidean saddle has multiple reflection symmetries, see Figure \ref{fig:MBBreflectionsymmetries}. The first one is the one around the $\tau=0,\pm\beta/2$ slice (depicted in red). The others are around any slice identified by a fixed value of one of the non-compact directions, e.g. $x=0$, or $y=0$, or $z=0$ (depicted in yellow). By identifying any of these coordinates ($\tau$, $x$, $y$, or $z$) with the Euclidean time, the semiclassical gravitational Euclidean path integral in the saddle point approximation can be interpreted as preparing the initial state for bulk fields on any such reflection symmetric slices. These initial states can then be evolved in real time yielding real Lorentzian geometries, which can be easily obtained by analytically continuing the corresponding coordinate in the Euclidean metric \eqref{eq:euclideanMBB}.

Choosing $\tau=it$ to be the Euclidean time leads to the maximally extended Lorentzian MBB studied in the previous subsection, with Penrose diagram given on the right of Figure \ref{fig:MBBreflectionsymmetries} and a real magnetic field $F=Bdx\wedge dy$. Choosing $x=i\chi$ or $y=i\upsilon$ to be the Euclidean time yields a real Lorentzian geometry, but an imaginary electric field, e.g. $F=iBd\chi\wedge dy$. This is a problem similar to that seen for the double analytic continuation of the planar AdS-RN case discussed at the end of Section \ref{sec:planaradsrn}, and the resulting Lorentzian theory is not well-defined.  Interestingly, choosing $z=i\zeta$ to be the Euclidean time yields a real geometry
\begin{equation}
    ds^2=f(r)d\tau^2+\frac{dr^2}{f(r)}+g(r)(dx^2+dy^2)-h(r)d\zeta^2
    \label{eq:MBBsolitonmetric}
\end{equation}
\textit{and} a real magnetic field $F=Bdx\wedge dy$ (which is left invariant by both this and the MBB analytic continuations). This is the magnetically charged AdS${}_5$ soliton. The existence of this well-defined double analytic continuation of the MBB solution is a consequence of the breaking of rotational symmetry introduced by the magnetic field in five dimensions. 

\begin{figure}
    \centering
    \includegraphics[width=0.4\linewidth]{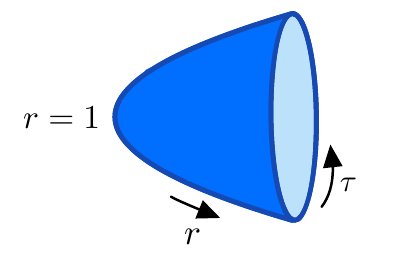}
    \caption{Time slice of the magnetically charged AdS${}_5$ soliton. Only the $\tau$ and $r$ directions are represented, with the non compact $x$, $y$ directions suppressed. The $S^1$ in the $\tau$ direction pinches off in the bulk at $r=1$, a sign of the presence of a mass gap and a confined spectrum of massive particles in the dual theory \cite{Witten:1998zw}, see Section \ref{sec:MBBdual}.}
    \label{fig:solitonslice}
\end{figure}

Note that each time slice of the spacetime \eqref{eq:MBBsolitonmetric} is given by a cigar geometry in the $\tau-r$ plane times two non-compact directions (see Figure \ref{fig:solitonslice}), and the spacetime is static. This comes as no surprise, because the soliton should be regarded as a ground state solution of the semiclassical Einstein's equations, and representing the ground state of the dual holographic CFT. In fact, the $z$ direction is non-compact and a Euclidean path integral performed over a manifold with a non-compact Euclidean time direction prepares the vacuum state.

The asymptotic AdS boundary of the soliton spacetime has spatial topology $S^1\times\mathbb{R}^2$, with the $S^1$ in the $\tau$ direction. Note that the $S^1$ pinches off in the bulk at $r=1$, where $f(1)=0$. This feature is associated with a mass gap and IR confinement in the dual theory, as we will explore in the next subsection \cite{Witten:1998zw}. AdS soliton spacetimes similar to \eqref{eq:MBBsolitonmetric} have been studied and shown to have negative energy.\footnote{The zero energy configuration is taken here to be pure AdS with one direction periodically identified \cite{Horowitz:1998ha}.} There is evidence that they represent the negative energy ground state of the dual theory with one compact $S^1$ spatial direction. In the dual theory, the negative energy is interpreted as a Casimir energy arising from the supersymmetry-breaking antiperiodic boundary conditions for fermions on the $S^1$ \cite{Horowitz:1998ha}.

Before turning to the properties of the holographic dual description of the MBB and of the soliton, we would like to point out that these two solutions are associated with the same saddle of the gravitational Euclidean path integral. Therefore, although the two geometries and the state of quantum fields on top of them (when they are present) are radically different, observables in the two setups are expected to be related by double analytic continuation. In particular, observables in the MBB on the codimension-2 slice $t=z=0$, which is left invariant by the double analytic continuation, are equal to the corresponding observables in the soliton spacetime on the codimension-2 slice $\tau=\zeta=0$. As we have pointed out in Section \ref{sec:intro}, this ``slicing duality'' has interesting implications for the holographic description of cosmology (see also \cite{Antonini:2022blk,Antonini:2022ptt,Antonini:2022opp} and Section \ref{sec:wormhole}).

\subsection{The dual description: $\mathcal{N}=4$ SYM and confinement}
\label{sec:MBBdual}

Now that we have explored in some detail the properties of the bulk spacetimes of our interest, we can describe their holographic dual description. In general, the boundary value of a bulk gauge field acts as an external source turned on for a global $U(1)$ conserved current in the dual theory. In \cite{DHoker:2009mmn} this conserved current $J^\alpha$ was understood to be associated with a $U(1)$ subgroup in the Cartan subalgebra of the $SU(4)$ R-symmetry of $\mathcal{N}=4$ Super-Yang-Mills (SYM):
\begin{equation}
    S=S_{SYM}+\int d^4x J^\alpha A_\alpha^{ext},
\end{equation}
where $A_\alpha^{ext}$ is the boundary value of the bulk gauge field.
Evidence for the duality between the MBB solution and $\mathcal{N}=4$ SYM with such an external magnetic field was provided in \cite{DHoker:2009mmn,DHoker:2009ixq,Ammon:2017ded}. Notice that the coupling of the external source to the $U(1)$ subgroup of the R-symmetry partially breaks supersymmetry \cite{DHoker:2016ncv}.

In the following we will explore some generic features of the boundary theory which will apply to any holographic theory dual to our bulk setup, regardless of whether it takes the exact form described above. For simplicity, we will refer to it as $\mathcal{N}=4$ SYM${}_B$ (with the subscript $B$ indicating the presence of the external magnetic field), but our results are completely general.

\begin{figure}
    \centering
    \includegraphics[width=0.7\linewidth]{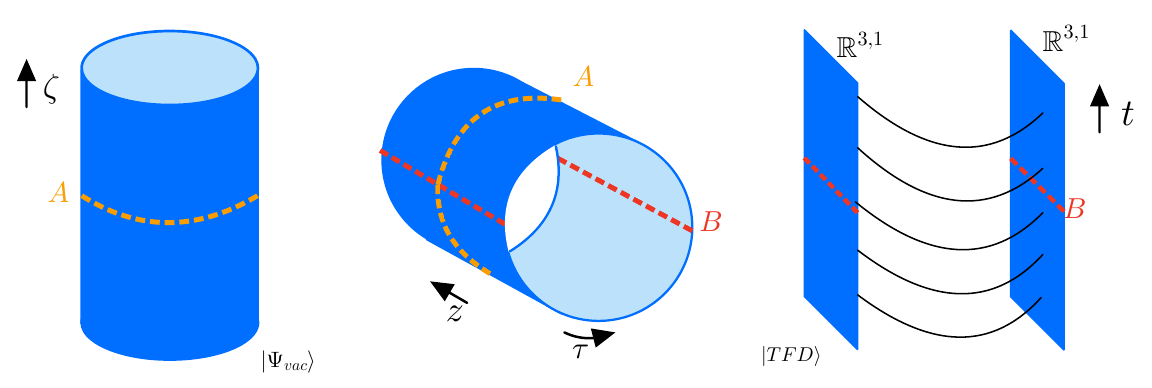}
    \caption{Holographic theory dual to our 5D bulk geometries. This could for example be $\mathcal{N}=4$ SYM${}_B$. Center: the Euclidean theory lives on $S^1\times \mathbb{R}^3$ (two non-compact directions are suppressed). Right: the Lorentzian theory dual to the MBB is given by two copies of $\mathcal{N}=4$ SYM${}_B$ on $\mathbb{R}^{3,1}$ in the thermofield double state. Left: the Lorentzian theory dual to the magnetic AdS${}_5$ soliton is given by $\mathcal{N}=4$ SYM${}_B$ on $S^1\times \mathbb{R}^{2,1}$ in the vacuum state. This theory flows in the IR to a (2+1)-dimensional confining theory on $\mathbb{R}^{2,1}$.}
    \label{fig:microscopicslicings}
\end{figure}

Let us start by noticing that the Euclidean boundary theory lives on a manifold $S^1\times \mathbb{R}^3$. The analytic continuation corresponding to the MBB solution is the one identifying the $S^1$ as the Euclidean time direction. It is then immediate to conclude that the state dual to the MBB and prepared by the boundary Euclidean path integral is the thermofield double (TFD) state of two copies of $\mathcal{N}=4$ SYM${}_B$ on $\mathbb{R}^{3,1}$ (see Figure \ref{fig:microscopicslicings}, right panel):\footnote{With an abuse of notation, we are using here the familiar discrete sum definition of the TFD, even though $\mathcal{N}=4$ SYM${}_B$ on spatial $\mathbb{R}^3$ has a continuous energy spectrum and the sum should be replaced by an integral.}
\begin{equation}
    \ket{TFD}=\frac{1}{\sqrt{Z(\beta\sqrt{B})}}\sum_{n}e^{-\frac{\beta E_n}{2}}\ket{E_n,B}_L\ket{E_n,B}_R,
    \label{eq:TFD}
\end{equation}
where we restricted the sum to a fixed magnetic charge subsector and we emphasized, as we have pointed out in Sections \ref{sec:planaradsrn} and \ref{sec:MBBsol}, that the finite temperature partition function only depends on the dimensionless quantity $\beta\sqrt{B}$, as dictated by conformal symmetry.

On the other hand, the analytic continuation corresponding to the magnetic AdS${}_5$ soliton is the one where one of the non-compact directions (the one breaking rotational symmetry, which we labeled by $z$) is identified with the Euclidean time. The state dual to the AdS${}_5$ soliton is then the negative energy ground state of $\mathcal{N}=4$ SYM${}_B$ on $S^1\times \mathbb{R}^{2,1}$, see left panel of Figure \ref{fig:microscopicslicings}. The negative energy arises from the antiperiodic boundary conditions for fermions around the $S^1$ \cite{Witten:1998zw,Horowitz:1998ha}. This theory flows in the IR (on length scales much larger than the size of the $S^1$) to a (2+1)-dimensional theory on $\mathbb{R}^{2,1}$. This lower-dimensional theory is essentially a pure $SU(N)$ gauge theory\footnote{This is because the long distance dynamics in the $\mathbb{R}^{2,1}$ directions of $\mathcal{N}=4$ SYM on $S^1\times \mathbb{R}^{2,1}$ is that one of (2+1)-dimensional $SU(N)$ gauge fields, because all other fields become massive from the (2+1)-dimensional point of view due to the boundary conditions along the $S^1$ \cite{Witten:1998zw}.} and is expected to display a mass gap and a discrete mass spectrum of confining particles \cite{Witten:1998zw,VanRaamsdonk:2021qgv}.\footnote{On intermediate length scales much larger than the magnetic field scale $1/\sqrt{B}$ but smaller than the size of the $S^1$, a 2D CFT emerges in the $\tau-\zeta$ directions \cite{DHoker:2009mmn,Maldacena:2020sxe}. On length scales much larger than the size of the $S^1$ this conformal structure is also lost and the remaining theory on $\mathbb{R}^{2,1}$ is confining.}

From the bulk spacetime point of view, the gapped nature (in the (2+1)-dimensional sense explained above) of the dual theory is manifested in the $S^1$ pinching off in the bulk at $r=1$. In fact, this feature acts as a radial cutoff at small $r$ in the bulk at a finite proper distance from any point in the bulk spacetime. By means of the holographic renormalization group \cite{deBoer:1999tgo}, we can identify this bulk radial cutoff with a IR cutoff in the dual boundary theory. This intuitively implies that, in the boundary theory, physics is trivial at arbitrarily long wavelengths. In other words, the dual theory is not a CFT in the IR, but rather a confining 
theory with a mass gap.\footnote{This is different from the pure AdS spacetime with spatially $S^3$ boundary, where the center of AdS $r=0$ is also a finite proper distance away from any bulk point. In that case, the IR cutoff is due to the compact nature of the $S^3$. In our case, the dual theory lives on a non-compact manifold, and the IR cutoff cannot be interpreted in terms of a geometric feature of the boundary manifold. Our case should be compared instead to the Poincar\'e-AdS case, in which the boundary is spatially $\mathbb{R}^3$ and $r=0$ is infinitely far away in proper distance from any bulk point, corresponding to no IR cutoff in the dual boundary theory.}
We can make this argument precise using a bulk calculation similar to the one carried out in \cite{Witten:1998zw,Csaki:1998qr}. The goal is to show that bulk fields on the background \eqref{eq:MBBsolitonmetric} give rise to a mass gap and a discrete tower of massive particles (glueballs \cite{Csaki:1998qr}) in the dual IR theory on $\mathbb{R}^{2,1}$.\footnote{Another way to investigate confinement is by computing the expectation values of spatial Wilson loops and check whether they satisfy an area law \cite{Witten:1998zw}. Recent works \cite{Betzios:2019rds,Betzios:2023obs} presented this confinement criterion in dual theories of Euclidean wormholes by calculating correlation functions of local and non-local operators in these backgrounds. We will not pursue this direction in the present paper.}

\subsubsection{Mass gap and confinement in the dual theory}

Let us focus our attention on a bulk scalar field $\phi$ with mass $\mu$ satisfying the Klein-Gordon equation $(\nabla^\mu\nabla_\mu-\mu^2)\phi=0$, which can be conveniently rewritten as
\begin{equation}
    \left[\frac{1}{\sqrt{-g}}\partial_\mu\left(\sqrt{-g}g^{\mu\nu}\partial_\nu\right)-\mu^2\right]\phi=0.
    \label{eq:KG}
\end{equation}
Exploiting the translational symmetry in the $\zeta$, $x$, $y$, $\tau$ directions, we can decompose the scalar field as
\begin{equation}
    \phi=\psi(r)e^{-i\omega\zeta+ik_xx+ik_yy+ik_\tau \tau},
\end{equation}
where $k_\tau=2\pi n/\beta$ due to the periodicity $\beta=4\pi/f'(1)$ of the $\tau$ coordinate. Using the metric \eqref{eq:MBBsolitonmetric}, the Klein-Gordon equation \eqref{eq:KG} takes the form
\begin{equation}
    \mathcal{D}_r\psi(r)=\left(\frac{k_\tau^2}{f(r)}+\frac{k_x^2+k_y^2}{g(r)}-\frac{\omega^2}{h(r)}\right)\psi(r),
\end{equation}
where the differential operator $\mathcal{D}_r$ is given by\footnote{Notice that by defining $\bar{\psi}(r)=\sqrt{f(r)g(r)\sqrt{h(r)}}\psi(r)$ we could eliminate the linear derivative term in equation \eqref{eq:diffop} at the expense of introducing a complicated potential term $V(r)$. However, we find that keeping the differential operator in the form \eqref{eq:diffop} simplifies our numerical analysis.} \footnote{For the numerical analysis it is convenient to rewrite the operator \eqref{eq:diffop} in terms of the $V,W$ functions appearing in the Einstein's equations \eqref{eq:einsteineq}.}
\begin{equation}   \mathcal{D}_r=f(r)\partial_r^2+\left[f'(r)+\frac{f(r)g'(r)}{g(r)}+\frac{f(r)h'(r)}{2h(r)}\right]\partial_r-\mu^2.
    \label{eq:diffop}
\end{equation}
When solving the Klein-Gordon equation \eqref{eq:KG} we must impose smoothness at the tip of the cigar, i.e. $\psi'(1)=0$, as well as require a normalizable solution at $r=\infty$. Asymptotically, where $f(r)\sim g(r)\sim h(r)\sim r^2$, there are two linearly independent solutions, one behaving as $r^\mu$ and the other one behaving as $r^{-4-\mu}$. We must therefore pick the latter (for which clearly $\psi(\infty)=\psi'(\infty)=0$). For any given value of $k_\tau, k_x, k_y$ this set of boundary conditions can only be satisfied by a discrete set of solutions associated to a discrete set of values of $\omega$. In particular, let us label the $\omega$ eigenvalues associated with $k_\tau=k_x=k_y=0$ by $\omega=M_n$. The first five $M_n$ eigenvalues found numerically for $b=1$ and a scalar field of mass $\mu=1$ are
\begin{equation}
    M_n=\{5.10, 8.81, 12.39, 15.92, 19.44\}.
    \label{eq:solitonspectrum}
\end{equation}
Notice that the eigenvalues are strictly positive, $M_n>0$.

Let us now think of the bulk fields as (2+1)-dimensional fields living on the spatial plane identified by $x,y$. The (2+1)-dimensional dispersion relation arising from our analysis above of the scalar field takes the form
\begin{equation}    \omega_{bulk}=\sqrt{M_n^2+F(k_\tau)+G(k_x^2+k_y^2)}
\end{equation}
with $F(0)=G(0)=0$. Notice that $F$ and $G$\footnote{$G$ only depends on the combination $k_x^2+k_y^2$ because of rotational symmetry in the $x-y$ plane.} are complicated functions in general, and in particular $F(k_\tau)\neq k_\tau^2$ and $G(k_x^2+k_y^2)\neq k_x^2+k_y^2$ because the $\zeta-\tau$ and $\zeta-x-y$ Poincar\'e invariance is broken in the bulk, as it is clear from the soliton metric \eqref{eq:MBBsolitonmetric}.\footnote{We thank Gary Horowitz for a useful discussion on this point.} In this (2+1)-dimensional sense, a scalar field on our AdS${}_5$ soliton spacetime gives therefore rise to discrete towers of massive particles with mass $\sqrt{M_n^2+F(k_\tau)}$.

As we approach the asymptotic boundary $r\to\infty$, Poincar\'e symmetry is restored because $f(r)\sim g(r)\sim h(r)\sim r^2$. Therefore, in the boundary theory the dispersion relation reduces to
\begin{equation}    \omega_{\partial}=\sqrt{M_n^2+k_\tau^2+k_x^2+k_y^2}.
\label{eq:boundarydisp}
\end{equation}
In terms of the IR boundary theory on $\mathbb{R}^{2,1}$, this is associated to towers of massive particles with mass $m_n=\sqrt{M_n^2+k_\tau^2}$. The strict positivity of the $M_n$ eigenvalues (see equation \eqref{eq:solitonspectrum}) guarantees the existence of a mass gap in the dual IR theory on $\mathbb{R}^{2,1}$.

We emphasize that this spectrum of massive particles is not simply the discrete spectrum one would obtain by Kaluza-Klein (KK) reduction of a theory on $S^1\times \mathbb{R}^{2,1}$ to a theory on $\mathbb{R}^{2,1}$. In that case, the mass spectrum would simply be given by $m_n^{KK}=k_\tau=2\pi n/\beta$, and the KK zero mode implies there would be no mass gap. Our mass spectrum differs from the KK one due to the presence of the $M_n$ eigenvalues, which arise from the presence of the radial cutoff in the bulk, i.e. the IR cutoff in the boundary theory. The infinite, discrete tower of $M_n$'s should be therefore interpreted as a confining spectrum, i.e. as the masses of glueballs \cite{Csaki:1998qr}. Notice that for each $M_n$ a KK tower of massive particles exists associated with different discrete values of $k_\tau$. The dispersion relation \eqref{eq:boundarydisp} therefore really describes a tower of towers of massive particles, with one tower associated to confinement and the other one to KK reduction.

\begin{figure}[h]
    \centering
    \includegraphics[width=0.5\textwidth]{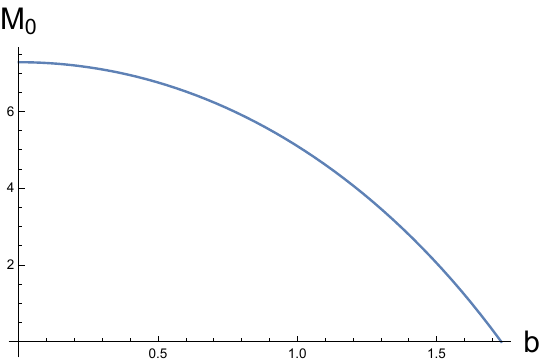}
    \caption{Mass gap $M_0$ as a function of the magnetic field parameter $b$ for a scalar field of mass $\mu=1$. As $b\to\sqrt{3}$ the gap closes, signaling deconfinement. The gap between the different $M_n$ eigenvalues also vanishes in this limit.}
    \label{fig:losinggap}
\end{figure}

Finally, we would like to remark that the mass gap reduces to zero and the $M_n$ eigenvalues become continuous in the limit $b\to\sqrt{3}$, see Figure \ref{fig:losinggap}.\footnote{We remind that this limit cannot be taken strictly, as we have explained in Section \ref{sec:MBBsol}.} This result has a simple interpretation both in the bulk and in the boundary. From the bulk point of view, the radial cutoff at $r=1$ (i.e. the tip of the cigar) becomes infinitely far away in proper radial distance from any bulk point. This is because $f'(1)\to 0$ in this limit, implying that $f(r)$ develops a double pole at $r=1$ which causes the proper distance $l_r\propto\int_1^{r^*}dr/\sqrt{f(r)}$ to diverge logarithmically.
Therefore the radial cutoff responsible for the confining spectrum is removed, the gap closes $M_0\to 0$, and the spacing between different $M_n$'s also vanishes. Analogously, from the boundary theory perspective, the corresponding IR cutoff determining the confining scale is taken to zero. In fact, as we have explained in Section \ref{sec:MBBsol}, the $b\to \sqrt{3}$ limit corresponds to the $\beta\to\infty$ limit (in the MBB picture this is the zero temperature limit). In this limit, the dual theory becomes $\mathcal{N}=4$ SYM${}_B$ on $\mathbb{R}^{3,1}$ in its zero energy vacuum state, which does not flow in the IR to a theory on $\mathbb{R}^{2,1}$ and is known to not be confining, see e.g. \cite{Witten:1998zw}. 

Now that we have analyzed in full detail the MBB solution, the magnetically charged AdS${}_5$ soliton, and the properties of the holographic dual theories, we are ready to consider ETW branes embedded in these spacetimes, which will allow us to study braneworld cosmologies, braneworld traversable wormholes, and their holographic descriptions.

\section{Euclidean braneworld analysis}
\label{sec:euclidean}

In the previous section, we constructed the asymptotically AdS${}_5$ MBB background spacetime and the associated Euclidean saddle \eqref{eq:euclideanMBB}, whose metric we report here:
\begin{equation}
ds^2 = f(r)d\tau^2 + \frac{1}{f(r)}dr^2 + g(r)(dx^2+dy^2)+h(r) dz^2,
\label{eqn:euclidean5dmetric}
\end{equation}
where $\tau$ is periodically identified with period $\beta$. We will take the $\tau$ coordinate in the range $\tau\in [-\beta/2,\beta/2]$.
As we explained in the previous section, this Euclidean solution is dual to $\mathcal{N}=4$ SYM${}_B$ with an external magnetic field on $\mathbb{R}^{3} \times S^1$.

\begin{figure}
    \centering
    \includegraphics[width=0.5\linewidth]{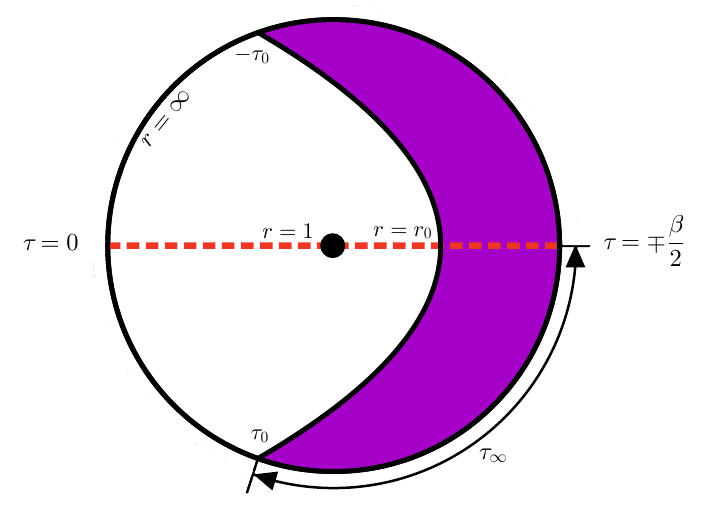}
    \caption{Brane trajectory in the $\tau-r$ direction. The region shaded in purple is cut off by the ETW brane. The transverse $x$, $y$, and $z$ directions are suppressed.}
    \label{fig:eucltrajectory}
\end{figure}

Following \cite{Cooper:2018cmb,Antonini:2019qkt}, let us now place a dynamical codimension-1 End-of-the-World (ETW) brane with a constant tension $T$ to excise a portion of the asymptotic region of this Euclidean spacetime. Neumann boundary conditions for bulk fields, including the metric,\footnote{Technically, the boundary condition for the metric is Robin, see Footnote \ref{footnote:robin}.} are imposed at the brane, allowing it to dynamically evolve in the bulk. We are interested in a brane whose trajectory lies in the $\tau-r$ directions and cuts off a portion of the asymptotic boundary identified by an interval in the $\tau$ direction, see Figure \ref{fig:eucltrajectory}. The brane extends infinitely in the $x,y,z$ coordinates.

In the  Euclidean picture, the brane evolves from the asymptotic boundary $r \to \infty$ to a minimum radius $r_{0}$, then expands back to $r \to \infty$. We are interested in brane trajectories preserving the reflection symmetry of the Euclidean saddle around the $\tau=0,\pm\beta/2$ slice.\footnote{By construction the reflection symmetry around $z=0$ is also preserved.} We will therefore take the minimum radius $r_0$ of the trajectory to occur for $\tau=\pm\beta/2$, and study symmetric trajectories around this point. We label by $\tau_{\infty}$ the amount of $\tau$ coordinate employed by the brane to reach the asymptotic boundary from $r_0$. The trajectory therefore excises a portion of the asymptotic boundary of size $2\tau_{\infty}$, leaving us with a strip-shaped boundary of width
\begin{equation}
    l\equiv 2\tau_0=\beta-2\tau_{\infty},
    \label{eq:preptime}
\end{equation}
where we defined the boundary width parameter $\tau_0$.\footnote{In \cite{Cooper:2018cmb,Antonini:2019qkt} $\tau_0$ was labeled ``preparation time'', because $\tau$ was identified with the Euclidean time and $\tau_0$ corresponds to the amount of Euclidean time evolution applied to a Cardy state $\ket{B}$ to prepare the state dual to the double-sided black hole (or MBB in our case) with the second asymptotic region cut off by the ETW brane, see Section \ref{sec:cosmodual}. Here we avoid this terminology, because we will also be interested in a different analytic continuation in which $z$ is identified with the Euclidean time and $2\tau_0$ is simply the width of the boundary strip.} See Figure \ref{fig:eucltrajectory}. Clearly, the only dimensionless parameter in the boundary theory is now $\tau_0\sqrt{B}$, where the boundary width parameter $\tau_0$ took the place of $\beta$, the length of the $S^1$ at the boundary.

Notice that the AdS boundary is now a manifold with a boundary and it is topologically a strip $\mathbb{R}^3\times I$, where $I=[-\tau_0,\tau_0]$ is an interval in the $\tau$ direction. The boundary theory dual to this bulk setup is therefore a Euclidean BCFT \cite{Cardy:1989ir,Affleck:1991tk,Cardy:2004hm}, as customary in the context of the AdS/BCFT correspondence \cite{Takayanagi:2011zk,Fujita:2011fp}. In particular, following \cite{DHoker:2009ixq} and Section \ref{sec:MBBdual}, the Euclidean dual theory is given by $\mathcal{N}=4$ SYM${}_B$ on $\mathbb{R}^3\times I$, with appropriate conformal boundary conditions imposed at the boundaries of the interval. The choice of conformal boundary conditions determines the tension of the brane in the bulk setup \cite{Takayanagi:2011zk,Fujita:2011fp}. In order for this dual theory to exist, the brane must not excise all of the asymptotic boundary, i.e. it must not overlap with itself. In other words, the boundary width parameter $\tau_0$ must be positive. 

On the other hand, since our goal is to describe braneworld cosmologies and braneworld wormholes, namely effective (3+1)-dimensional theories localized on the ETW brane, it is crucial that an effective (3+1)-dimensional description of gravity localized on the brane exists. This can be achieved in our setup via the Karch-Randall-Sundrum gravity localization mechanism\cite{Randall:1999ee,Randall:1999vf,Karch:2000ct}. Gravity localization will hold only locally and for portions of the brane trajectory lying deep into the asymptotic AdS region \cite{Antonini:2019qkt,Clarkson:2005mg}. Since the minimum radius $r_0$ becomes a maximum radius in the braneworld cosmology setting studied in Section \ref{sec:cosmology}, effective 4D gravity for part of the evolution of the braneworld cosmology is possible only if $r_0$ is large enough to be deep into the asymptotic AdS region (i.e. the metric \eqref{eqn:euclidean5dmetric} approximately reduces to Euclidean AdS${}_5$ for $r\sim r_0$).

From this discussion we understand that two conditions are necessary for our setup to realize braneworld cosmologies and braneworld wormholes with well-defined holographic dual descriptions: 
\begin{enumerate}
    \item $\tau_{0}>0$ to ensure that a well-defined dual description exists;
    \item $r_0$ lies deep into the AdS region to guarantee Karch-Randall gravity localization works for at least a portion of the evolution of the braneworld.
\end{enumerate}
It was shown in \cite{Cooper:2018cmb} that satisfying these two conditions simultaneously in a Euclidean AdS-Schwarzschild setup is impossible. Contrarily, the AdS-RN setup of \cite{Antonini:2019qkt} achieved this goal by considering a near-extremal black hole and a near-critical brane tension $T\to 1/L$. The drawback is that there is no well-defined solitonic solution associated with the AdS-RN Euclidean saddle as we discussed in Section \ref{sec:planaradsrn}, and so no braneworld wormhole can be defined. 

In this section we will study the embedding of an ETW brane in the Euclidean background \eqref{eqn:euclidean5dmetric} and show that, similar to the AdS-RN setup, a regime of parameters exist (namely, $b\lesssim \sqrt{3}$, $T\lesssim 1/L$) for which both conditions can be satisfied simultaneously. In this case, as we have discussed, a solitonic solution also exists, which will allow us to build a braneworld traversable wormhole in Section \ref{sec:wormhole}. In particular, in Sections~\ref{subsec:branetrajectory} and \ref{subsec:preparationtime} we analyze the brane trajectory and numerically obtain $r_{0}$ and $\tau_{0}$ across different values of the magnetic field parameter $b$ and tension $T$. In Section~\ref{subsec:gravitylocalization}, we study at what values of the radial coordinate $r$ the metric \eqref{eqn:euclidean5dmetric} approximately takes the form of a pure AdS metric (i.e. at what values of $r$ we are deep within the AdS asymptotic region) for various values of $b$. We then show that a range of parameters $b,T$ exists such that $r_0$ is well within such region while keeping $\tau_0>0$. In Section \ref{sec:maldamaoz} we discuss the 4D effective braneworld description of the Euclidean setup, which is given by a Maldacena-Maoz-like wormhole with the two AdS${}_4$ asymptotic boundaries coupled by non-gravitational auxiliary degrees of freedom. 

We remark that, by analogy with \cite{Cooper:2018cmb}, one could expect the existence of an alternative saddle built by cutting off the zero temperature solution of \cite{DHoker:2009mmn} with two disconnected ETW branes. As we show in Appendix \ref{app:actioncomparison}, this alternative phase does not exist in our setup, and the Euclidean saddle studied in the present section is the only known saddle contributing to the Euclidean path integral. It therefore dominates the gravitational ensemble.

\subsection{Brane Trajectory}
\label{subsec:branetrajectory}

The Euclidean ETW brane action is given by
\begin{equation}
    S_{ETW}=-\frac{1}{8\pi G}\int_{ETW}d^4x\sqrt{h}(K-3T)+S_{ETW}^{em}
    \label{eq:braneaction}
\end{equation}
where $h$ is the determinant of the induced metric on the brane, $K$ is the trace of the extrinsic curvature, and $S_{ETW}^{em}$ is a necessary electromagnetic term needed for the well-posedness of the variational problem \cite{Hawking:1995ap,Antonini:2019qkt}, which will not affect the brane's equations of motion.

We can describe the motion of the brane by varying the full action of the bulk (equation \eqref{eq:bulkaction}) and brane  and imposing Robin boundary conditions, \footnote{\label{footnote:robin}In most of the literature (e.g. \cite{Takayanagi:2011zk,Fujita:2011fp,Cooper:2018cmb, Antonini:2019qkt}), it is common to regard this boundary condition as ``Neumann" rather than ``Robin". However, in principle, the boundary condition imposed is between the induced metric and its derivative along the brane. Thus, it is more precise to characterize this condition as the ``Robin" boundary condition which we adopt in the rest of the work.

}
\begin{equation}
    K_{ab} - K h_{ab}=-3 T h_{ab}
    \label{eqn:robinbraneeqn}
\end{equation}
where $K_{ab}$ is the extrinsic curvature and $K$ its trace. The tension can take values $-1<T<1$ for a $\Lambda<0$ brane and we will focus on the positive case $T>0$ (we remind that $L=1$). We relegate the explicit derivation of the brane equation to Appendix \ref{app:braneeqn}. Parametrizing the brane trajectory by $\tau$, the final result we obtain in terms of the $f,U,V$ metric functions defined in Section \ref{sec:MBB} is:

\begin{equation}
    \frac{dr}{d\tau} = \pm \frac{f \sqrt{4 fV'W' + 4f V'^2 + fW'^2 -9T^2}}{3T}
    \label{eqn:braneeuclidean}
\end{equation}
where a prime denotes a derivative with respect to $r$.   The $+$ and $-$ sign denotes the contracting (top half of Figure \ref{fig:eucltrajectory}) and expanding (bottom half of Figure \ref{fig:eucltrajectory}) phase, respectively. 

We define the minimum radius $r_{0}$ (reached at the reflection-symmetric slice $\tau=\pm\beta/2$) of the brane's evolution as the largest $r$ such that $dr/d\tau =0$ in equation~\eqref{eqn:braneeuclidean} is satisfied.

More explicitly, 
\begin{equation}
    r_{0} = \max{\{ r | 4f(r) V'(r) W'(r) +4 f(r) V'(r)^2 +f(r) W'(r)^2 -9T^2=0 \} }.
\end{equation}

\begin{figure}
\centering
    \includegraphics[width=0.7\linewidth]{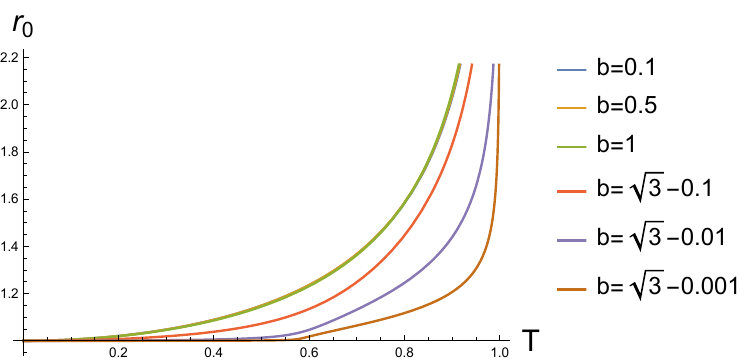}
\caption{The minimum radius $r_{0}$ of the ETW brane as a function of the brane tension $T$ across different values of the magnetic field parameter $b$. For any fixed value of $b$, $r_0$ diverges for $T\to T_{crit}=1$.
}
\label{fig:minradius}
\end{figure}

We numerically evaluate the minimum radius $r_{0}$ across different values of the magnetic field parameter $b$ and tension $T$ and plot the results in Figure~\ref{fig:minradius}. As expected, $r_{0}$ increases as $T$ increases for any value of $b$, and for $T=0$ we have $r_0=r_H=1$, i.e. the brane passes through the tip of the cigar and we are left with exactly half of the Euclidean geometry. In principle, as $T$ is taken to criticality $T_{\text{crit}}=1$, $r_{0}$ approaches infinity for any fixed value of $b$, which corresponds to the brane approaching the asymptotic boundary. 

Ultimately, the numerical evaluation of our Euclidean MBB scenario shows that, given high numerical precision, we can attain a larger value of $r_{0}$ by simply tuning the tension closer to criticality. This will play an important role when considering the range of parameters that allow gravity localization to be satisfied in the Euclidean, cosmology, and wormhole pictures.

\subsection{Boundary width parameter}
\label{subsec:preparationtime}
An essential property we need to consider for a well-defined state is the positivity of the boundary width parameter $\tau_{0}$, defined in equation \eqref{eq:preptime}. Only if $\tau_0>0$ the brane does not overlap and a portion of the asymptotic boundary where to define the dual BCFT is retained. This condition generically competes with the large tension condition seen above needed to support gravity localization, as demonstrated in both the AdS-Schwarzschild and the AdS-RN scenarios \cite{Cooper:2018cmb, Antonini:2019qkt}. More specifically, a tension too close to criticality $T_{\text{crit}}=1$ leads to a vanishing and eventually negative boundary width parameter, signaling an ill-defined solution with an overlapping brane cutting off the entire asymptotic boundary.

$\tau_0$ can be computed in our bulk solution by numerically evaluating the interval $2\tau_{\infty}$ in $\tau$ coordinate taken by the brane to undergo its trajectory. This can then be subtracted from the $\tau$ periodicity $\beta =4\pi/f'(r)|_{r=r_{H}}= 3\pi/(2(3-b^2))$ to find twice the boundary width parameter. Thus, $\tau_{0}$ has the explicit form
\begin{equation}
    \tau_{0}= \frac{3\pi}{4(3-b^2)}-\int_{r_{0}}^{\infty} dr \frac{3T}{f \sqrt{4 fV'W' + 4 f V'^2 + fW'^2 -9T^2}}.
\end{equation}
\begin{figure}
    \centering
    \begin{subfigure}{0.6\linewidth}
    \centering
    \includegraphics[width=\linewidth]{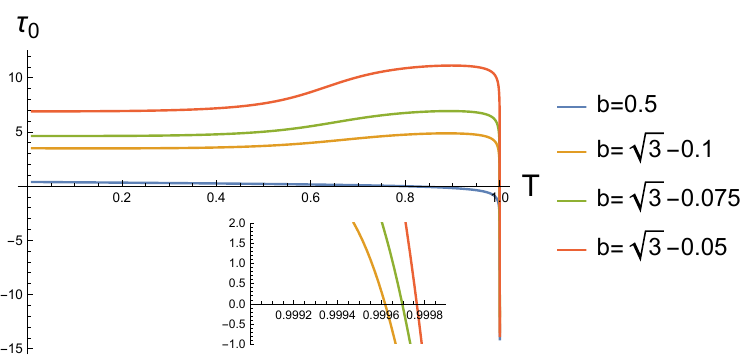}
   \caption{}
    \end{subfigure}
    \begin{subfigure}{0.6\linewidth}
    \centering
     \includegraphics[width=\linewidth]{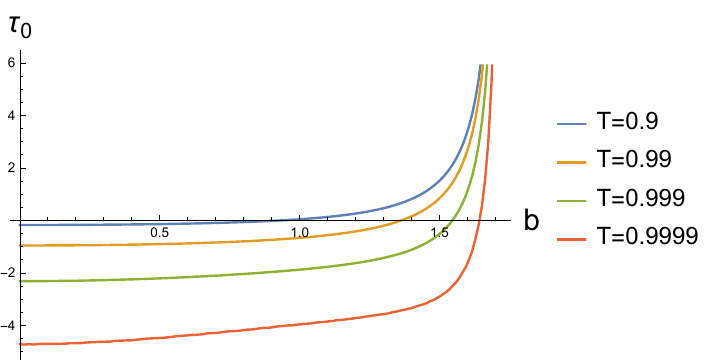}
    \caption{}
    \end{subfigure}
    \caption{(a) The boundary width parameter $\tau_{0}$ as a function of the brane tension $T$ across different magnetic fields $b$. For any fixed $b$, $\tau_0$ becomes negative if $T$ is tuned too close to criticality. The inset shows how the value of $T$ for which $\tau_0$ becomes negative increases as $b$ is tuned closer to $\sqrt{3}$. (b) $\tau_{0}$ as a function of the magnetic field $b$ evaluated across different brane tensions $T$. For any fixed $T$, $\tau_0$ can be made positive by tuning $b$ sufficiently close to $\sqrt{3}$.}
    \label{fig:preptime}
\end{figure}
We numerically evaluate $\tau_{0}$ and plot the results in Figure~\ref{fig:preptime}. There are several features to notice. As the tension approaches criticality $T \to T_{\text{crit}}$, $\tau_{0}$ vanishes and eventually becomes negative for any value of $b$. However, for the purposes of constructing a valid solution that provides a large enough $r_{0}$ such that gravity localization is possible, we only need to take $T$ close to criticality but strictly smaller than $T_{\text{crit}}$. The second feature to notice is that taking $b\to \sqrt{3}$ while keeping $T$ fixed increases $\tau_{0}$. In particular, $\tau_0$ can always be made positive for any value of $T$ by tuning $b$ close enough to $\sqrt{3}$. Intuitively, this is because taking $b\to\sqrt{3}$ corresponds to taking $\beta\to\infty$ (see Section \ref{sec:MBB}), whereas $\tau_{\infty}$ remains finite in this limit for fixed $T$.
This suggests that to allow for solutions which can support gravity localization we should tune $b$ close to $\sqrt{3}$ to provide a larger range of valid brane tensions with $\tau_0>0$. This behavior is completely analogous to the AdS-RN case studied in \cite{Antonini:2019qkt}.

We remark that within our family of solutions parameterized by $b$ and $T$, there exist values of the parameters that allow for a positive boundary width parameter and locally localized gravity, as we will see in the next subsection.

\subsection{Gravity Localization}
\label{subsec:gravitylocalization}
The second property we need to examine is whether observers living on the ETW brane see 4-dimensional gravity.  Gravity localization on a brane was initially explored by \cite{Karch:2000ct, Randall:1999ee, Randall:1999vf}. To review, Randall and Sundrum \cite{Randall:1999ee,Randall:1999vf} demonstrated that a 4D Minkowski brane embedded in a warped AdS$_{5}$ geometry supports 4-dimensional gravity: a stable 4D massless graviton exists on the brane. A massive tower of Kaluza-Klein (KK) modes is also present, and the 4D graviton is the zero mode. Karch and Randall \cite{Karch:2000ct} extended this result to AdS${}_4$ branes embedded in AdS${}_5$, showing that in this case the 4D graviton acquires a small mass, leading to localized gravity on shorter length scales. In the case of more general asymptotically AdS${}_5$ spacetimes, local gravity localization can still be achieved as long as the brane sits in a region which is nearly pure AdS${}_5$. For instance, for a black hole background the brane needs to sit far from the black hole horizon and deep into the AdS asymptotic region \cite{Clarkson:2005mg,Antonini:2019qkt}. In this case the 4D graviton is also massive and it has a finite lifetime, i.e. it is a quasi-bound mode. However, if we only consider relatively short spatial and time scales, gravity is effectively localized on the brane and the graviton is nearly massless. Thus, 4D observers living on the brane would experience 4D gravity. 

Therefore, the question we need to answer is: in the Lorentzian braneworld cosmology and braneworld wormhole pictures studied in Sections \ref{sec:cosmology} and \ref{sec:wormhole}, can the brane sit, at least for part of its trajectory, in the asymptotic AdS region of the respective Lorentzian spacetimes (the MBB and the soliton)? The answer to this question resides in the value of $r_0$ studied in Section \ref{subsec:branetrajectory}. If $r_0$ is large enough to be in a region where the spacetime looks nearly AdS${}_5$, then the Lorentzian brane trajectories will support local gravity localization. In the braneworld cosmology picture, as we will see, $r_0$ becomes a maximum radius and the brane collapses into the MBB horizon. Therefore, gravity localization is achieved only for part of the trajectory around the time-symmetric slice, when the brane is far away from the horizon. On the other hand, $r_0$ remains a minimum radius in the braneworld wormhole picture, implying that gravity localization will hold for the whole trajectory of the brane.

As we have seen, $r_0$ can be made arbitrarily large by tuning the tension close to criticality.\footnote{This is in line with the idea that a 4D effective description of physics on the brane exists whenever the number of DOF associated with the boundary of the microscopic BCFT is large compared to the number of DOF associated with the bulk of the BCFT \cite{Karch:2000ct,VanRaamsdonk:2021qgv,Antonini:2022blk}, and therefore the boundary entropy is large.} The difficulty resides in also retaining a positive value of $\tau_0$, warranty of a well-defined solution and dual description.
In the AdS-Schwarzschild setup of \cite{Cooper:2018cmb}, even in the planar limit, valid solutions that allow for gravity localization do not exist because, increasing $T$, $\tau_{0}$ becomes negative well before  $r_{0}$ reaches the asymptotic AdS region. In the AdS-RN setup, \cite{Antonini:2019qkt} explicitly demonstrated gravity localization is possible for values of the electric charge parameter sufficiently close to extremality. Because, as we have seen in the previous subsection, $\tau_0$ can be made positive for any value of $T$ by tuning $b\to\sqrt{3}$, the same result should be attainable in our magnetically charged setup.

To verify whether our setup supports valid ($\tau_0>0$) solutions with gravity localization, we will now check for which value $r=\bar{r}$ of the radial coordinate the metric \eqref{eqn:euclidean5dmetric} is approximately pure AdS${}_5$ given a certain value of the parameter $b$, and then make sure that, for the same value of $b$, it is possible to choose a value of $T$ such that $r_0>\bar{r}$ and $\tau_0>0$.\footnote{Note that we cannot simply find the asymptotic AdS region by numerically checking the range of $r$ for which $f(r)\approx g(r)\approx h(r)\approx r^2$. This is because we are always allowed (and in fact made use of in Section \ref{sec:MBB}) a shift in the radial coordinate $r\to r+c$, which implies $f(r)\approx (r+c)^2$ might hold in the asymptotic AdS region. Although eventually $f(r)\approx r^2$ for large enough $r$, for large values of $c$ we could be deep into the AdS asymptotic region with $f(r)\neq r^2$.} 

\begin{figure}
    \centering
    \includegraphics[width=0.48\linewidth]{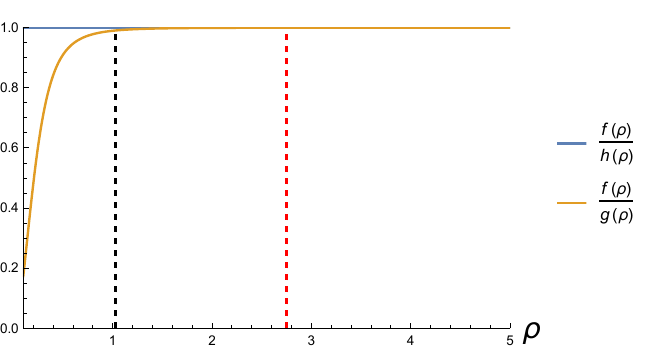}
    \includegraphics[width=0.48\linewidth]{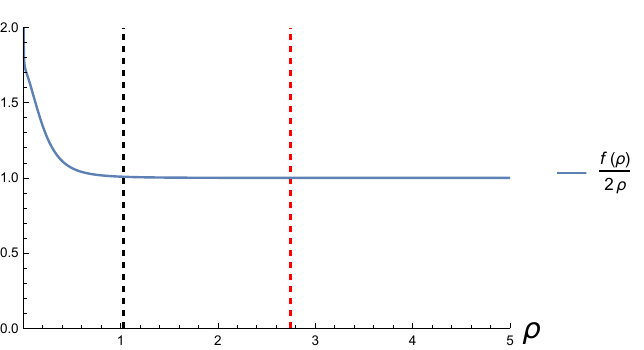}
    \caption{Numerical evaluation of the two conditions defining the asymptotic AdS region for $b=\sqrt{3}-0.001$ and $T=0.9999$. The (red) dashed line denotes $\rho_{0}=\rho(r_0)$, namely the value of $\rho$ corresponding to the minimum radius of the brane. Notice that for these values of the parameters, $\tau_0>0$ also holds. The (black) dashed line denotes $\bar{\rho}$ such that for all $\rho<\bar{\rho}$ at least one of the conditions for gravity localization is violated by more than $1\%$. The brane minimum radius is clearly deep into the asymptotic AdS region, implying that our solutions support gravity localization.}
    \label{fig:gravitylocalization}
\end{figure}

Beginning with the metric \eqref{eqn:euclidean5dmetric}, we perform a coordinate transformation such that $\rho^2=f(r)$, which recasts the metric into the form
\begin{equation}
    ds^2 =\rho^2 d\tau^2 +\frac{4 d \rho^2}{(f'(r)|_{r=r(\rho)})^{2}}+g(r(\rho))(dx^2+dy^2) + h(r(\rho))dz^2.
\end{equation}
As we have seen in Section \ref{sec:MBB}, $f(r)\approx g(r)\approx h(r)$ for large enough $r>\bar{r}$. Assuming we are in this range of values of $r$, the metric reduces to
\begin{equation}
    ds^2 \approx \rho^2 (d\tau^2+dx^2+dy^2+dz^2) + \frac{4 d\rho^2}{(f'(r)|_{r=r(\rho)})^{2}}
\end{equation}
Finally, if $f'(r)|_{r=r(\rho)}=2\rho$, then the metric takes the form of Euclidean AdS$_{5}$:
\begin{equation}
    ds^2 \approx\rho^{2} (d\tau^2+dx^2+dy^2+dz^2) + \frac{d\rho^2}{\rho^2}.
    \label{eqn:ads5form}
\end{equation}
The range of $r$ associated with the asymptotic AdS region is therefore $r>\bar{r}$, where $\bar{r}$ is the value of $r$ for which (1) $f(r)\approx g(r)\approx h(r)$ starts to be satisfied and (2) $f'(r)|_{r=r(\rho)}\approx 2\rho$ starts to hold. The numerical plots in Figure \ref{fig:gravitylocalization} show that, for an appropriate choice of $b$ and $T$ (we chose here $b=\sqrt{3}-0.001$ and $T=0.9999$), these conditions are satisfied for values of $r$ considerably smaller than $r_0$ while $\tau_0>0$ holds. Therefore, at least part of the brane trajectory sits in the asymptotic AdS region in all Lorentzian pictures. The results of \cite{Antonini:2019qkt} then imply that local gravity localization can be achieved. We leave a more thorough study of gravity localization and the 4D quasi-bound mode in our setup to future work.

\subsection{Maldacena-Maoz wormhole on the brane}
\label{sec:maldamaoz}

Before moving on to the study of Lorentzian continuations of our Euclidean setup, we would like to emphasize the connection between our Euclidean braneworld solution and the setup discussed in \cite{VanRaamsdonk:2021qgv,Antonini:2022blk,Antonini:2022ptt} and reviewed in Section \ref{sec:intro}. In particular, the gravity localization analysis of the previous subsection suggests that an effective 4D bulk description on the Euclidean brane should exist, which does not rely on the 5D Euclidean geometry. We can therefore ``remove one layer of holography''. This should be understood in complete analogy with the relationship between singly and doubly holographic models of black hole evaporation \cite{Almheiri:2019hni}.

\begin{figure}
    \centering
    \includegraphics[width=0.55\linewidth]{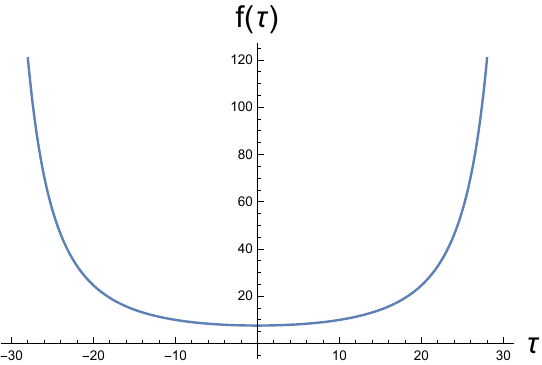}
    \caption{Wormhole conformal factor $f(\tau)$ for the (Euclidean and Lorentzian) wormhole solution given $b=\sqrt{3}-0.001$ and $T=0.9999$. $f(\tau)$ has double poles at the two asymptotic AdS${}_4$ boundaries at $\tau=\pm\tau_\infty$. In the doubly holographic setup, this is where the ETW brane meets the asymptotic AdS${}_5$ boundary.}
    \label{fig:wormholescalefactor}
\end{figure}

The metric induced on the brane takes the form
\begin{equation}
    ds^2=\left(f(\tau)+\frac{(r'(\tau))^2}{f(\tau)}\right)d\tau^2+g(\tau)(dx^2+dy^2)+h(\tau)dz^2
    \label{eq:braneeuclmetric}
\end{equation}
where we defined $f(\tau)\equiv f(r(\tau))$ (and similarly for $g$ and $h$), with $r(\tau)$ describing the brane trajectory.
In the asymptotically AdS region, where the brane sits in the solutions supporting gravity localization, this reduces to the nearly isotropic form\footnote{In regions where gravity localization is possible, the $(r'(\tau))^2/f(\tau)$ can be neglected. We verified this numerically for the examples we displayed. Moreover, it can be seen analytically by examining equation~\eqref{eqn:braneeuclidean}. }
\begin{equation}
    ds^2\approx f(\tau)(d\tau^2+dx^2+dy^2+dz^2).
    \label{eq:branemaldamaoz}
\end{equation}
$f(\tau)$ has a minimum at $\tau=0$ and two double poles at $\tau=\pm\tau_\infty$,\footnote{It is easy to show that $f(\tau)$ has double poles by solving equation \eqref{eqn:braneeuclidean} in the asymptotic region where $f(r)\approx g(r)\approx h(r)\approx r^2$, see Section \ref{subsec:wormholesol}.} where the brane meets the asymptotic boundary, see Figure \ref{fig:wormholescalefactor}. The geometry of the brane is therefore that of a Euclidean wormhole connecting two distinct asymptotically AdS${}_4$ boundaries, corresponding to the two 3D boundaries of the dual 4D BCFT. Note that in the convention used so far, the turnaround point of the brane was at $\tau=\pm\beta/2$, but for the braneworld wormhole description we choose to set $\tau=0$ there for simplicity of notation. 

This wormhole is similar to those studied by Maldacena and Maoz \cite{Maldacena:2004rf}, with the metric taking the same form \eqref{eq:maldamaozmetric}. Notice that in our setup the two boundaries connected by the wormhole are also coupled by non-gravitational auxiliary degrees of freedom living on $\mathbb{R}^3\times I$, see center line, center panel of Figure \ref{fig:slicingduality}. These can be understood to account for the coupling of the brane to the 5D bulk physics, and are a low-energy effective version of the bulk\footnote{By bulk here we mean the region of the microscopic 4D BCFT away from its boundaries.} degrees of freedom of the microscopic 4D holographic BCFT.\footnote{These auxiliary non-gravitational degrees of freedom are the analog of the bath degrees of freedom in the singly holographic description of the black hole evaporation setting \cite{Almheiri:2019hni}.} This is in fact exactly the setup considered in \cite{VanRaamsdonk:2021qgv,Antonini:2022blk,Antonini:2022ptt}, of which our higher dimensional braneworld setup can be interpreted as a doubly holographic realization.

\section{Lorentzian Braneworld Cosmology}
\label{sec:cosmology}

\begin{figure}
    \centering
    \includegraphics[width=0.7\linewidth]{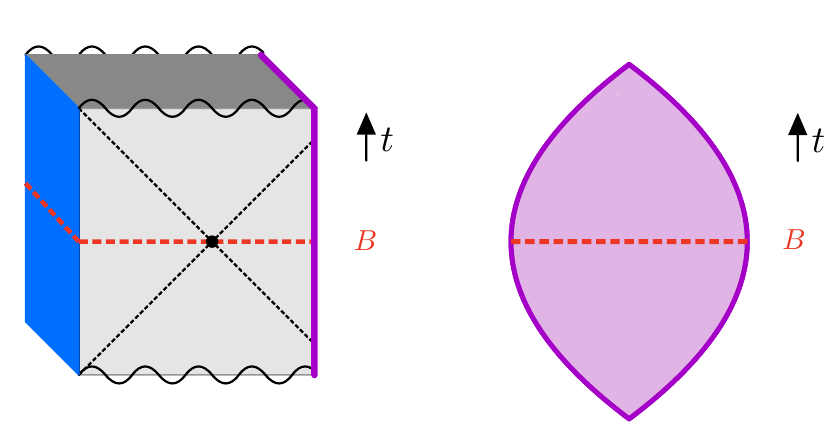}
    \caption{Left: MBB spacetime with an ETW brane (purple) cutting off the right asymptotic boundary. Two non-compact directions ($x$ and $y$) are suppressed.  Right: The effective 4D description on the brane is that of a Big Bang-Big Crunch Bianchi I cosmology. The time-symmetric slice is denoted by the red dashed line. Gravity is effectively 4D for a large portion of the cosmological evolution around the time-symmetric point.}
    \label{fig:branecosmo}
\end{figure}

In the previous section, we embedded an ETW brane in the Euclidean geometry that excises a portion of the asymptotic boundary. As we have discussed, our Euclidean solution including the ETW brane possesses two reflection symmetries, and therefore admits two well-defined analytic continuations. In this section, we consider the MBB analytic continuation, i.e. we identify $\tau$ with the Euclidean time direction and obtain the metric \eqref{eq:MBBmetric}. This leads to a double-sided MBB bulk spacetime with an ETW brane cutting off the second asymptotic boundary, see Figure \ref{fig:branecosmo}. This is therefore a one-sided black brane microstate, with the dual holographic theory living on the only asymptotic AdS boundary (the left one in Figure \ref{fig:branecosmo}).

From the point of view of an observer comoving with the brane, the expansion and contraction of the brane looks like the evolution of a Big Bang-Big Crunch cosmology. As we will see, this is a special case of a Bianchi type I universe \cite{Taub:1950ez,Ellis:1968vb,Montani:2009hju}. As we have examined in Section \ref{subsec:gravitylocalization}, there exist values of $b$ and $T$ for which gravity is locally localized on the brane for part of its trajectory. In particular, gravity is effectively 4D on the brane when the brane is deep into the AdS${}_5$ asymptotic region, but not when it sits near or inside the MBB horizon. Therefore, the effective 4D description in terms of a cosmology is only sensible for part of the evolution of the crunching universe, namely around the time-symmetric point. As we will see, this includes a large portion of the cosmological evolution in cosmic time. Notice that the past and future singularities of the braneworld cosmology correspond to the past and future singularities of the higher dimensional MBB. However, as we have explained, in the regions near the singularities there is no effective 4D cosmological description.

In Section~\ref{subsec:cosmology}, we derive the Bianchi I metric of the braneworld cosmology and the equation of motion for the cosmological evolution. The cosmic scale factors can be numerically evaluated in the region outside the MBB horizon. Because this contains the region where gravity localization holds and the effective 4D description exists, this is enough for the study of our braneworld cosmology. In Section~\ref{sec:cosmoaccel}, we study the behavior of the scale factors of this cosmology, observe the presence of a phase of accelerated expansion, and comment on the possible source of this accelerated expansion. Finally, in Section \ref{sec:cosmodual} we will comment on the holographic dual description of this bulk setup, which is given by a specific pure excited state of a theory on $\mathbb{R}^{3,1}$, and on the properties of bulk reconstruction, pointing out the connection with entanglement islands in cosmology \cite{Hartman:2020khs,Bousso:2022gth}.

\subsection{Big Bang/Big Crunch Cosmology}
\label{subsec:cosmology}
To describe cosmological evolution, we derive the brane equation of motion in Lorentzian signature. The first step is to analytically continue $\tau=it$ and define the brane proper time as 
\begin{equation}
    d \lambda = \sqrt {f(r(t)) - \frac{1}{f(r(t))} \left ( \frac{dr}{dt} \right )^{2}} dt
    \label{eqn:dlambda}
\end{equation}
where $r(t)$ denotes the brane trajectory.
The induced metric on the brane then reads
\begin{equation}
    ds^2_{ETW} = -d\lambda^2 + g(\lambda) (dx^2+dy^2) + h(\lambda) dz^2
\end{equation}
where $g(\lambda)\equiv g(r(\lambda))$, and similarly for $h$.
This is the metric of a homogeneous but anisotropic cosmology with scale factors $g(\lambda)$ and $h(\lambda)$, namely a special case of a Bianchi I universe \cite{Taub:1950ez,Ellis:1968vb,Montani:2009hju} in which two directions ($x$ and $y$) share the same scale factor. Here $\lambda$ plays the role of the cosmic time in the braneworld cosmology. Although generically $g(\lambda) \neq h(\lambda)$ and the cosmology is anisotropic, we will see that it remains relatively isotropic $g(\lambda) \approx h(\lambda)$ for a large portion of its evolution around the turning point, where the brane is the furthest from the MBB horizon. In fact, $g\approx h$ was one of the necessary conditions for gravity localization to hold, see Section \ref{subsec:gravitylocalization}. This implies that whenever gravity is localized on the brane and therefore an effective 4D cosmological description exists, the 4D universe is nearly-isotropic and approximately takes the form of a flat Big Bang-Big Crunch FLRW universe.

\begin{figure}
    \centering
    \begin{subfigure}{0.47\textwidth}
        \centering
        \includegraphics[width=\linewidth]{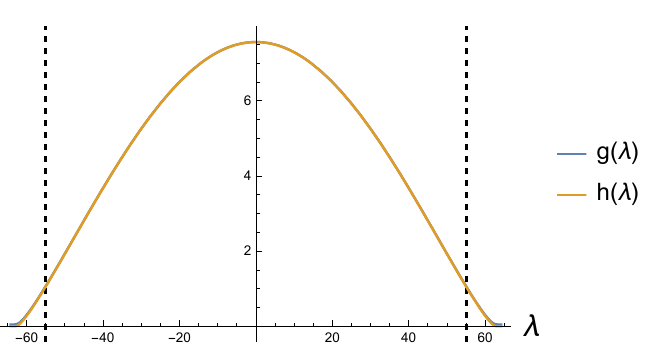}
        \caption{}
    \end{subfigure}
    \begin{subfigure}{.47\textwidth}
        \centering
        \includegraphics[width=\linewidth]{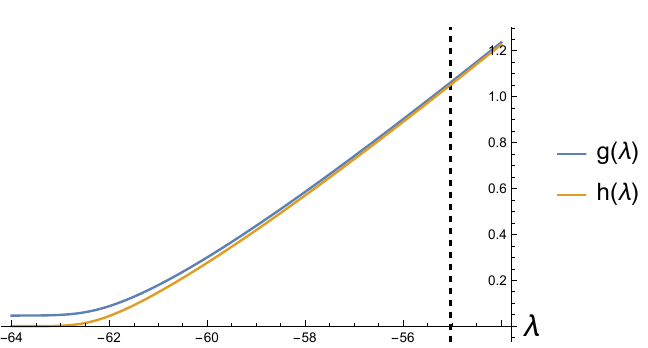}
        \caption{}
    \end{subfigure}
    \caption{(a) The scale factors $g(\lambda)$ and $h(\lambda)$ as a function of the brane comoving time $\lambda$ (i.e. the cosmic time). We chose $b=\sqrt{3}-0.001$ and $T=0.9999$ for the free parameters. Note that for a large portion of the evolution, $g(\lambda)$ and $h(\lambda)$ are practically equal to each other. Thus, the cosmology is approximately isotropic. (b) Detail of the scale factors at early times for the expanding phase of the cosmological evolution. At early times, where the cosmology becomes more anisotropic, we also lose gravity localization and a 4D effective description of the cosmology ceases to make sense.
    The (black) dashed lines denote $\lambda=\pm \bar{\lambda}\equiv \pm |\lambda(\bar{r})|$ such that for all $|\lambda|>\bar{\lambda}$ at least one of the conditions for gravity localization is violated by more than $1\%$.
    } 
    \label{fig:cosmoscalefactor}
\end{figure}

Notice that the analytic continuation $\tau=it$ has the effect of flipping the sign of the terms under the square root in the brane equation of motion~\eqref{eqn:braneeuclidean}. 
This implies that the minimum radius $r_0$ of the brane in Euclidean signature becomes a maximum radius in the Lorentzian picture.
Combining the analytic continuation of equation \eqref{eqn:braneeuclidean} with the definition of the comoving time $\eqref{eqn:dlambda}$, the brane equation of motion in Lorentzian signature can be cast in the form

\begin{equation}
    \left ( \frac{dr}{d\lambda} \right )^{2} = \frac{9 T^2}{4V'W' +4V'^2 + W'^2}-f.
\end{equation}
We can integrate this equation with initial condition $r(\lambda=0)=r_{0}$ to find the brane trajectory $r(\lambda)$ as a function of the brane comoving time. We can then evaluate the scale factors $g(\lambda)$ and $h(\lambda)$ numerically for $r>1$. We display the results in Figure~\ref{fig:cosmoscalefactor}, where we selected $b=\sqrt{3}-0.001$ and $T=0.9999$. We remark that this choice of parameters satisfies the conditions needed for gravity localization around the turning point of the brane evolution. Evidently, the cosmology is nearly isotropic for a large part of its trajectory. Additionally, gravity localization holds and an effective 4D cosmological description exists for a large fraction of the total brane proper time outside the horizon. 

To quantify the portion of the trajectory for which gravity localization holds (and the cosmology is nearly isotropic), we can evaluate the total proper time as the brane moves from the horizon $r_{H}=1$ to $r_{0}$ and eventually back to $r_{H}=1$:
\begin{equation}
    \lambda_{\text{tot}} = 2 \int_{1}^{r_{0}}\sqrt{\frac{4V'W'+4V'^2+W'^{2}}{9T^2-4fV'W'-4fV'^2-fW'^2}}dr
    \label{eq:totalproptime}
\end{equation}
where the factor of two accounts for the time symmetry of the brane evolution. We can then evaluate the total proper time for which the conditions for gravity localization hold by simply substituting the lower extreme of the integral \eqref{eq:totalproptime} with $\bar{r}$ defined in Section \ref{subsec:gravitylocalization}. By choosing $\bar{r}$ such that the conditions for gravity localization are satisfied to more than $1\%$ accuracy, for our choice of parameters $b=\sqrt{3}-0.001$, $T=0.9999$ we obtain that gravity localization holds for $85.8\%$ of the total brane proper time of the trajectory outside the horizon. 

A 4D effective cosmological description on the brane is therefore well-defined for a large portion of the braneworld cosmology evolution. We leave a more accurate quantification of gravity localization to future work.

\subsection{Accelerating cosmology at early times}
\label{sec:cosmoaccel}
Examining Figure~\ref{fig:cosmoscalefactor} reveals that a phase of accelerated cosmic expansion takes place near the beginning of the braneworld cosmology evolution.\footnote{Naturally, this corresponds to a phase of decelerated contraction near the end of the braneworld cosmology evolution, because the full evolution is time-symmetric.} In particular, notice that gravity localization still holds during part of this accelerated phase. After the accelerated phase, the universe undergoes a decelerated expansion and eventually it starts to symmetrically recollapse when the effective negative 4D cosmological constant, given by (we restore here the AdS radius $L$)
\begin{equation}
    \Lambda_4=T^2-\frac{1}{L^2}<0,
\end{equation}
becomes dominant.

\begin{figure}
    \centering
    \includegraphics[width=0.55\linewidth]{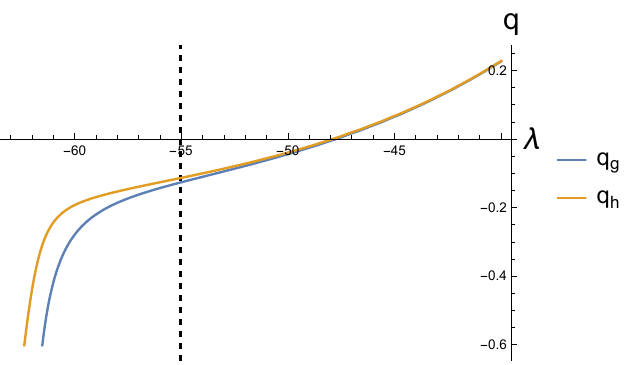}
    \caption{The deceleration parameters $q_{g,h}$ for $g(\lambda)$, and $h(\lambda)$ evaluated in the accelerated expansion phase. The (black) dashed line denotes the  cosmic time $\lambda$ where gravity localization starts to be possible (i.e. the conditions for gravity localization are satisfied to better than $1\%$ accuracy). Notice that $q<0$, i.e. the cosmological expansion is accelerated, in a region where gravity localization likely holds and 4D braneworld cosmology is well-defined.}
    \label{fig:cosmoaccel}
\end{figure}

In order to quantify the acceleration, we can evaluate the cosmological deceleration parameters $q_g$ and $q_h$ defined as 
\begin{equation}
    q_g=-\frac{\ddot{g}g}{\dot{g}^2}, \quad \quad q_h=-\frac{\ddot{h}h}{\dot{h}^2}
\end{equation}
where the derivatives are taken with respect to the cosmic (i.e. comoving) time $\lambda$. In Figure~\ref{fig:cosmoaccel}, we plot the results for $q_{g}$ and $q_{h}$ for $b=\sqrt{3}-0.001$ and $T=0.9999$.  We remark that $q_g,q_h<0$ (signaling an accelerated expansion) for an interval of time where gravity localization still holds and therefore the 4D effective description of the cosmology makes sense. Therefore, our braneworld cosmology undergoes a phase of accelerated expansion.

What drives this accelerated expansion? In \cite{Cooper:2018cmb,Antonini:2019qkt}, where the form of the 5D bulk metric is known analytically, the Friedmann equation governing the scale factor evolution was obtained explicitly, allowing an identification of the energy densities driving the cosmological evolution. For the AdS-RN case, for instance, the Friedmann equation read
\begin{equation}
    \left(\frac{\dot{a}}{a}\right)^2=\Lambda_4-\frac{1}{a^2}+\frac{2\mu}{a^4}-\frac{Q^2}{a^6},
    \label{eq:friedmann}
\end{equation}
where $a$ is the scale factor and we can identify a (negative) cosmological constant $\Lambda_4=T^2-1/L^2$, a curvature term, a radiation term proportional to mass of the AdS-RN black hole, and a repulsive term proportional to the black hole charge (known in the cosmology literature as stiff matter with negative energy density \cite{Chavanis:2014lra}). The last term, which arises from the repulsive nature of the AdS-RN timelike singularity and is a negative energy source from the point of view of the braneworld cosmology, is responsible for an acceleration at early times. This eventually leads to a Big Bounce, but this accelerated phase takes place in a region where gravity localization does not hold, and the AdS-RN solution itself is unstable (namely inside the Cauchy horizon) \cite{Antonini:2019qkt}.

The absence of an analytic form for the MBB metric and therefore for the scale factors' equations of motion prevents us from identifying what kind of energy density sources the accelerated expansion in our setup. By analogy with the AdS-RN case, it seems plausible to assume that, from the 4D cosmology viewpoint, such an energy density is negative and is proportional to the magnetic field. This is also consistent with the species driving the acceleration being dominant at early times, see Appendix \ref{app:accel}. We leave a more thorough investigation of this possibility to future work. 
In our numerical setting, all we can do is constrain the equation of state of the species driving the accelerated expansion. We report this analysis in Appendix \ref{app:accel}.

\subsection{Holographic dual description}
\label{sec:cosmodual}

We can now turn to the question of what is the microscopic state dual to this MBB bulk spacetime cut off by the ETW brane. First of all, notice that the dual theory lives on the $\mathbb{R}^{3,1}$ asymptotic boundary of the MBB microstate. In fact, the microscopic state is defined by slicing the microscopic Euclidean path integral in the middle of the interval $I$, see the top line, right panel of Figure \ref{fig:slicingduality}. The boundary of the Euclidean BCFT is in the Euclidean past. The conformal boundary conditions imposed there should be seen as initial conditions for the Euclidean path integral preparing the microscopic state, which is given by
\begin{equation}
    \ket{\psi}_B=e^{-\tau_0H}\ket{B}
\end{equation}
where $\ket{B}$ is the Cardy state defined at the boundary of the BCFT at $\tau=-\tau_0$, and $\tau_0$ plays here the role of the preparation time. The state $\ket{\psi}_B$ is therefore an excited pure state of the dual theory ($\mathcal{N}=4$ SYM${}_B$) on $\mathbb{R}^{3,1}$, obtained by Euclidean time evolution of a boundary state. It can be also understood to arise from the TFD \eqref{eq:TFD} dual to the double-sided MBB spacetime after a projective measurement of the entire right boundary is performed in an appropriate basis \cite{Kourkoulou:2017zaj,Cooper:2018cmb,Antonini:2022sfm,Antonini:2023aza}. This construction is analogous to the one described in \cite{Cooper:2018cmb,Antonini:2019qkt} and can be interpreted as a higher dimensional realization of the Kourkoulou-Maldacena setup \cite{Kourkoulou:2017zaj}, see also \cite{Almheiri:2018ijj,Antonini:2021xar}.

\subsubsection{Hartman-Maldacena surfaces and cosmological islands}
\label{sec:islands}

We remark that probing the cosmological physics from the dual microscopic theory is difficult, because the brane sits behind the MBB horizon. In fact, the braneworld cosmology is inside a Python's lunch \cite{Brown:2019rox,Engelhardt:2021mue,Engelhardt:2021qjs}, with the non-minimal quantum extremal surface being the bifurcate horizon and the minimal one the empty set. Thus, for a CFT observer living on the left asymptotic boundary, reconstructing observables within the ETW brane is exponentially complex. 

\begin{figure}
    \centering
    \includegraphics[width=0.7\linewidth]{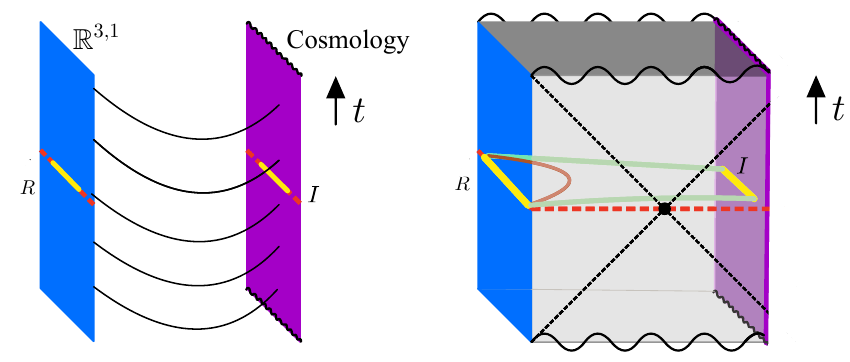}
    \caption{Left: in the effective (singly-holographic) 4D theory, the entanglement entropy of a large region $R$ of the microscopic theory receives contributions from the same region $R$ in the non-gravitational bath and from an entanglement island $I$ in the cosmology. Both $R$ and $I$ are depicted in light yellow. Right: in our 5D (doubly-holographic) setup, the RT surface associated with subregions of the microscopic theory has a connected phase (in brown) and a disconnected phase given by a Hartman-Maldacena-like surface ending on the brane (in green). The latter becomes dominant for large $R$. The endpoints on the brane of this surface identify the boundary of the island $I$.}
    \label{fig:islands}
\end{figure}

Some known observables are nonetheless expected to be able to probe the cosmology. One example is the entanglement entropy of large subregions of the dual theory. In fact, for large enough subregions the Ryu-Takayanagi surface computing the holographic entanglement entropy transitions from a connected to a disconnected phase, extending behind the horizon and ending on the ETW brane, see Figure \ref{fig:islands}. These surfaces, studied in \cite{Cooper:2018cmb}, are similar to Hartman-Maldacena surfaces \cite{Hartman:2013qma}. In this case, the entanglement entropy should display a time evolution matching the brane's evolution in the bulk. The scale factor behavior is therefore encoded in the entanglement entropy of these large subregions.

Interestingly, these Hartman-Maldacena-like surfaces ending on the brane are strictly related to the existence of entanglement islands in cosmology \cite{Hartman:2020khs,Bousso:2022gth}.\footnote{Similar setups of entanglement islands in the Karch-Randall braneworld setup via replica wormholes was explored in a recent work by \cite{Geng:2024xpj}.} To understand that, we should realize that the 4D (singly holographic) effective description of this Lorentzian setup involves the Big Bang-Big Crunch cosmology \textit{and} a non gravitational bath on $\mathbb{R}^{3,1}$ disconnected from but entangled with the cosmology \cite{VanRaamsdonk:2021qgv,Antonini:2022blk,Antonini:2022ptt}, as explained in Section \ref{sec:intro}, see the center line, right panel of Figure \ref{fig:slicingduality}. This bath once again accounts for the physics of the 5D bulk and is a low-energy EFT version of the microscopic dual theory on $\mathbb{R}^{3,1}$. This is exactly the setup in which cosmological islands arise. In particular, for a flat cosmology such as ours, it was shown in \cite{Hartman:2020khs} that a contribution coming from an island in the cosmological universe is present when computing the holographic entanglement entropy of large regions of the microscopic theory.

In our 5D setup, which can be seen as a doubly-holographic realization of the construction in \cite{Hartman:2020khs}, the island is identified by the region comprised between the two intersection points of the Hartman-Maldacena-like surface with the brane, see Figure \ref{fig:islands}. This is completely analogous to what happens in doubly holographic models of black hole evaporation \cite{Almheiri:2019hni}, where islands can be understood in terms of classical RT surfaces ending on an ETW brane. We leave a thorough study of the quantitative agreement between the doubly holographic calculation and the cosmological island calculation in our setup to future study.

Although it is possible to reconstruct some cosmological observables such as the scale factor using entanglement entropy or holographic complexity \cite{Cooper:2018cmb}, these boundary quantities are very complex observables from a microscopic theory point of view. Additionally, probing other cosmological observables---such as correlation functions of fields in the cosmology---from the dual theory is also an exponentially complex task, and the dictionary is unknown and expected to be state-dependent \cite{Papadodimas:2012aq,Papadodimas:2013jku,Akers:2021fut,Akers:2022qdl}. However, the slicing duality introduced in Section \ref{sec:intro} could provide a solution to this issue. In fact, one can relate braneworld cosmological observables to observables in a traversable braneworld wormhole, which will be the subject of the next section. Those observables can then be reconstructed from the confining theory dual to the wormhole using an ordinary holographic dictionary such as HKLL.

\section{Eternally Traversable Lorentzian Wormhole on the Brane}
\label{sec:wormhole}

\begin{figure}
    \centering
    \includegraphics[width=0.7\linewidth]{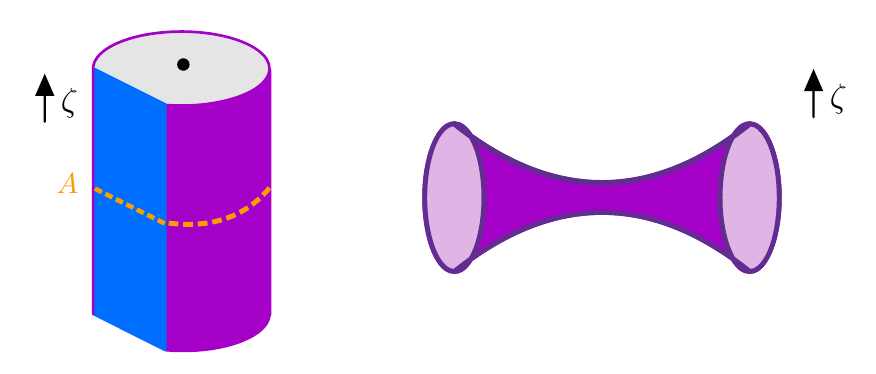}
    \caption{Left: AdS${}_5$ soliton geometry cut off by an ETW brane (in purple). The black dot denotes $r=1$, where the compact direction $\tau$ pinches off. The orange dashed line denotes the time reflection symmetric slice on which the bulk (vacuum) state is prepared by the Euclidean path integral. Right: a time-slice of the Lorentzian eternally traversable wormhole on the brane, connecting two asymptotic AdS${}_4$ boundaries. }
    \label{fig:solitonwormhole}
\end{figure}

In this section, we focus on the other possible analytic continuation of our Euclidean saddle, where the $z$ coordinate is identified with the Euclidean time. This yields the magnetically charged AdS${}_5$ soliton \eqref{eq:MBBsolitonmetric} with part of the asymptotic boundary cut off by the ETW brane, see Figure \ref{fig:solitonwormhole}. Notice that in this setup the brane is static, since both $r$ and $\tau$ are spatial directions. Its trajectory is exactly the same as in the Euclidean analysis of Section \ref{sec:euclidean}. The only difference is that the planar $z$ direction is now analytically continued and plays the role of the Lorentzian time $\zeta=-iz$.

The geometry induced on the brane is that one of a traversable planar wormhole connecting two distinct AdS$_{4}$ boundaries.\footnote{See \cite{Maldacena:2020sxe} for a different construction of braneworld traversable wormholes starting from the zero temperature 5D solution constructed in \cite{DHoker:2009mmn}. Notice that the authors of \cite{Maldacena:2020sxe} take a different approach, embedding the ETW brane in a Lorentzian geometry analogous to our MBB rather than to the soliton solution. See Section~\ref{subsec:otherconstr} for additional comments on this point.} In fact, it is a Lorentzian version of the Maldacena-Maoz-like wormhole seen in Section \ref{sec:maldamaoz}. Given that the induced geometry on the brane, just like the 5D geometry, is static, and given that the proper distance between any two bulk points in this wormhole is finite, we can conclude that this is an eternally traversable wormhole. 

We remark that this 4D effective description of the brane as a wormhole is valid for the whole brane trajectory. In fact, we have seen that for an appropriate choice of the $b,T$ parameters (e.g. the ones employed in the numerical analysis of the previous section), the radius of the brane at the turnaround point $r_{0}$ lies in a region where gravity can be localized on the brane. For radii $r>r_0$ gravity localization only works better because the brane is farther into the asymptotic AdS region, approaching the setup of \cite{Karch:2000ct}. Since in this Lorentzian soliton picture, just like in the Euclidean case, $r_0$ is the minimum radius of the brane, a 4D local effective description of gravity on the brane exists everywhere, and the braneworld traversable wormhole description is sensible everywhere. We also emphasize that in the 4D effective description, the two boundaries of the wormhole are also coupled by 4D non-gravitational auxiliary degrees of freedom on $\mathbb{R}^{2,1}\times I$, just like in the Euclidean case discussed in Section \ref{sec:maldamaoz}. This is something to be expected: a traversable wormhole connecting two AdS boundaries can only exist if the two asymptotic boundaries (where the holographic theories live) are coupled \cite{Maldacena:2018lmt}.

Another interesting feature is that, due to the slicing duality of the Euclidean saddle, cosmological observables in the braneworld cosmology picture of the previous section can be analytically continued to observables in the braneworld wormhole of this section. The absence of a horizon in the soliton spacetime (and, equivalently, in the 4D effective description of the brane as a traversable wormhole) implies that reconstruction of these observables is simple, and can be performed using the tools needed for causal wedge reconstruction, such as the HKLL dictionary. This simplifies greatly the reconstruction of braneworld cosmology observables with respect to a direct reconstruction in the dual theory described in Section \ref{sec:cosmodual}.

In Section~\ref{subsec:wormholesol} we derive the traversable wormhole geometry by examining the induced metric on the brane and show via a numerical evaluation that the wormhole connects two AdS asymptotic boundaries. In Section~\ref{subsec:nogotheorem} we explore traversability of the wormhole by evaluating the amount of boundary time needed for a light ray shot from one boundary to reach the other one and showing that it is finite. We also find that in the limit $b\to\sqrt{3}$ (corresponding to the $\beta\to\infty$ limit), the wormhole becomes non-traversable. We comment how this result allows us to reconcile our solution with known ``no-go" arguments for Poincar\'e-symmetric traversable wormholes \cite{Freivogel:2019lej}. In Section~\ref{subsec:wormholeconfinement}, we discuss the dual theory, which is given by the vacuum state of a BCFT on $\mathbb{R}^{2,1}\times I$, and demonstrate that it is confining by calculating the mass gap using the 4D braneworld wormhole description via the procedure described in \cite{Antonini:2022opp}. We compare and find qualitative agreement with the behavior of the mass gap obtained in the 5D soliton calculation in Section~\ref{sec:MBBdual}. In Section \ref{sec:slicing} we comment about bulk reconstruction and the slicing duality between the braneworld cosmology and the braneworld wormhole pictures. Finally, in Section~\ref{subsec:otherconstr} we explain the relationship between our braneworld wormholes and other previously constructed eternally traversable wormhole solutions in four dimensions \cite{Maldacena:2018gjk,Maldacena:2020sxe}.

\subsection{Braneworld Wormhole Solution}
\label{subsec:wormholesol}

The metric induced on the brane embedded in the magnetically charged AdS${}_5$ soliton \eqref{eq:MBBsolitonmetric} is given by
\begin{equation}
    ds^{2} = \left ( f(\tau)  + \frac{(r'(\tau))^{2}}{f(\tau)} \right ) d\tau^{2} +g(\tau) (dx^2+dy^2)-h(\tau)d\zeta^{2}
    \label{eqn:wormholetraversable}
\end{equation}
and it can be obtained by simply analytically continuing the Euclidean induced metric \eqref{eq:braneeuclmetric}.

As we have seen in Section \ref{sec:maldamaoz}, in regions where gravity localization is possible, we can recast the metric into an approximately conformally flat form 
\begin{equation}
    ds^2 \approx f(\tau)(d\tau^{2} +dx^2 +dy^2-d\zeta^{2}).
    \label{eqn:wormholeconformalform}
\end{equation}
A numerical evaluation of $f(\tau)$ for $b=\sqrt{3}-0.001$ and $T=0.9999$ was reported in Figure~\ref{fig:wormholescalefactor}. $f(\tau)$ has a positive minimum (given by $f(r_0)$) which we set to be at $\tau=0$,\footnote{Like for the Euclidean Maldacena-Maoz wormhole, we choose to set $\tau=0$ at the turnaround point in the 4D braneworld wormhole description, as opposed to $\tau=\pm\beta/2$ in the doubly holographic description.} and it diverges for finite values of $\tau$, which we label $\tau=\pm\tau_\infty$. Naturally, this divergence corresponds to the intersection between the brane and the asymptotic boundary of the AdS${}_5$ soliton spacetime, and $\tau_\infty$ is the amount of coordinate $\tau$ elapsed during the brane trajectory from the symmetric point to the boundary, first defined at the beginning of Section \ref{sec:euclidean}. 

By inspecting the brane equation of motion \eqref{eqn:braneeuclidean} it is easy to conclude that, near the asymptotic boundary where $f(r)\approx g(r)\approx h(r)\approx r^2$, we have $\tau(r)\approx \pm(\tau_\infty- 1/r)$. Therefore, near $\tau=\mp\tau_\infty$ the conformal factor $f(\tau)$ in the metric \eqref{eqn:wormholeconformalform} behaves as

\begin{equation}
    f(\tau) \approx \frac{-3/ \Lambda_4}{(\tau_{\infty} \pm \tau)^2}
\end{equation}
where the effective 4D cosmological constant on the brane is $\Lambda_4=T^2-1/L^2<0$ (where we temporarily restored the AdS radius $L$). Therefore, in the 4D effective theory on the brane, the two regions around $\tau=\pm\tau_\infty$ are two (planar) asymptotically AdS${}_4$ regions connected by a planar eternally traversable wormhole. The $\tau$ coordinate is along the wormhole. 

The traversability of the wormhole is guaranteed by the positive minimum of the wormhole conformal factor $f(\tau)$. It can be quantitatively confirmed by a simple numerical check that the proper distance along the $\tau$ direction 
\begin{equation}
    \Delta \ell=\int_{\tau_1}^{\tau_2}\sqrt{f(\tau)+\frac{(r'(\tau))^2}{f(\tau)}}d\tau
\end{equation}
between any two bulk points, with $-\tau_\infty<\tau_1,\tau_2<\tau_\infty$, is finite for $b<\sqrt{3}$, $T<1$. Because the wormhole geometry \eqref{eqn:wormholetraversable} is static, this guarantees traversability. Clearly, timelike geodesics cannot reach the asymptotic boundaries in a finite proper time, but can connect bulk points near the two asymptotic boundaries. On the other hand, as we will see in the next subsection, light rays can travel from one boundary to the other in a finite amount of coordinate time $\zeta$, which is identified with the boundary time.

We remark that, in the 4D effective description, the two boundaries connected by the traversable wormhole are also coupled by auxiliary non-gravitational degrees of freedom on $\mathbb{R}^{2,1}\times I$, see the center line, left panel of Figure \ref{fig:slicingduality}. This ``bath'' accounts for the 5D DOF we integrated out and is a low-energy effective description of the 4D DOF in the microscopic BCFT, whereas the microscopic 3D boundary DOF are represented holographically by the wormhole geometry.\footnote{We remind that this is a simplified description of the role of microscopic DOF in the 4D effective description, see Footnote \ref{footnote:DOF}.} This coupling between the two asymptotic boundaries is what allows the wormhole to be traversable. From an effective 4D bulk point of view, it guarantees the presence of negative energy supporting the wormhole (see Section \ref{subsec:otherconstr} for further comments); from a microscopic point of view, it shows that the 3D holographic DOF living on the asymptotic AdS${}_4$ boundaries are coupled, and the information transfer possible through the wormhole is possible also microscopically.

\subsection{Traversability and no-go arguments}
\label{subsec:nogotheorem}

In the previous subsection we showed the existence of a 4D, planar, nearly-Poincar\'e-symmetric, eternally traversable wormhole on the brane. It was argued in \cite{Freivogel:2019lej} that, in more than two spacetime dimensions, it is generically not possible to build semiclassical eternally traversable wormholes that satisfy Poincar\'e invariance in the boundary directions. 
There are two reasons why our construction is not in tension with the results of \cite{Freivogel:2019lej}. The first one is that the 4D effective description on the brane is only approximate and it involves massive gravity (although the mass of the brane-localized graviton is small), as we have already explained. The second one is that Poincar\'e invariance is also not exact: the wormhole is slightly anisotropic, see equations \eqref{eqn:wormholetraversable}-\eqref{eqn:wormholeconformalform}.

An additional check of the compatibility of our results with \cite{Freivogel:2019lej} is achieved by verifying that in the limit in which the assumption of \cite{Freivogel:2019lej} are satisfied, namely the localized graviton becomes massless and the Poincar\'e symmetry on the brane exact, the wormhole becomes non-traversable. This limit correspond to the $b\to\sqrt{3}$, $T\to 1$ limit, in which the brane is pushed infinitely close to the boundary of the soliton spacetime while preserving a positive boundary width parameter $\tau_0$. We remark that this limit cannot be taken strictly, as we have discussed in Section \ref{sec:MBBsol}, because the 5D bulk solution ceases to exist in the $b\to \sqrt{3}$ limit. This fact already implies that in the limit in which the assumptions of \cite{Freivogel:2019lej} are exactly satisfied, our solution does not exist.

\begin{figure}
    \centering
    \includegraphics[width=0.5\linewidth]{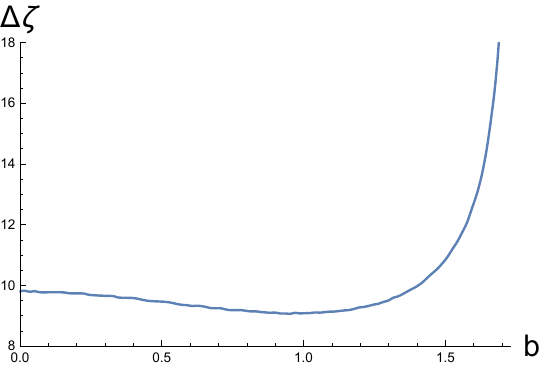}
    \caption{Boundary time $\Delta \zeta$ necessary for a light ray shot in the $\tau$ direction to traverse the wormhole from boundary to boundary as a function of the magnetic field parameter $b$ for $T=0.9999$. For any finite $b$, $\Delta\zeta$ is finite, signaling traversability. In the $b\to \sqrt{3}$ limit (where the assumptions of \cite{Freivogel:2019lej} would be satisfied), the time diverges, signaling loss of traversability. The results for small $b$ should be taken cautiously, because $\tau_0<0$ in that regime.}
    \label{fig:wormholeproptime}
\end{figure}

However, we can also verify the loss of traversability as we approach this limit. To do so, we compute the boundary time $\Delta \zeta$\footnote{Notice that the bulk coordinate time $\zeta$ is the same as the boundary time, which is obtained by conformally rescaling the asymptotic limit of the metric \eqref{eqn:wormholeconformalform} by $f(\tau)$.} necessary for a light ray shot along the wormhole direction $\tau$ from one boundary to reach the other boundary, and check it diverges in the $b\to\sqrt{3}$, $T\to 1$ limit. This is given by
\begin{equation}
    \Delta \zeta = 2\int_{0}^{\tau_{\infty}}\frac{1}{\sqrt{h(\tau)}}\sqrt{f(\tau) + \frac{(r'(\tau))^{2}}{f(\tau)}} d\tau.
\end{equation}
We provide an explicit numerical evaluation of $\Delta \zeta$ as a function of the magnetic field parameter $b$ in Figure~\ref{fig:wormholeproptime} for T=0.9999. Observe that for $0<b<\sqrt{3}$, $\Delta \zeta$ is finite, suggesting traversability. In this regime, the graviton is massive and Poincar\'e symmetry is only approximate, so that our solution evades the no-go arguments discussed in \cite{Freivogel:2019lej}.\footnote{We remark that at low values of $b$ the solution in Figure \ref{fig:wormholeproptime} has $\tau_0<0$ (i.e. the brane self-intersect) and no gravity localization for a portion of the brane trajectory. The results in that region of the plot should then be taken cautiously. Anyway, the low-$b$ region of the plot is irrelevant for our argument.} Notice that $\Delta\zeta$ diverges as $b\to\sqrt{3}$, signaling loss of traversability. We also verified that the same behavior persists for all values of $T$. This confirms that in the limit where the assumptions of \cite{Freivogel:2019lej} are valid, our wormhole also becomes non-traversable.

\subsection{Confinement in the dual theory from Braneworld Wormholes}

\label{subsec:wormholeconfinement}

The magnetically charged AdS${}_5$ soliton cut off by the ETW brane studied in this section is dual to the vacuum state of a 4D BCFT on $\mathbb{R}^{2,1}\times I$, see the first line, left panel of Figure \ref{fig:slicingduality}. This is simply the Lorentzian version of the 4D Euclidean BCFT discussed in Section \ref{sec:euclidean}. It is in a vacuum state because the Euclidean path integral preparing this state is non-compact in the Euclidean time direction $z$.

The behavior of this theory is expected to be qualitatively similar to that of the holographic dual of the soliton spacetime studied in Section \ref{sec:MBBdual}. In fact, the only difference between the two is that the theory lives on $\mathbb{R}^{2,1}\times I$ with conformal boundary conditions imposed at the boundaries of the interval rather than on $\mathbb{R}^{2,1}\times S^1$. In particular, the theory on $\mathbb{R}^{2,1}\times I$ also flows in the IR to a (2+1)-dimensional theory on $\mathbb{R}^{2,1}$ which has a mass gap and is confining.

On one hand, the confining spectrum of massive particles associated with a bulk scalar field could be computed in complete analogy with Section \ref{sec:MBBdual}, after imposing appropriate boundary conditions for the scalar field at the ETW brane (whose presence quantitatively modifies the spectrum). On the other hand, we can also employ the 4D effective braneworld wormhole description to compute the same spectrum, following \cite{Antonini:2022opp}.\footnote{Obtaining a quantitative match between the spectra obtained by the 4D and 5D bulk calculations, although possible, is non-trivial because the mass of a scalar field on the brane is non-trivially related to the mass of a corresponding 5D bulk scalar field, see e.g. \cite{Ghoroku:2002gx}.} In Section \ref{sec:MBBdual}, the 
bulk radial cutoff corresponding the dual theory's IR cutoff and setting the confinement scale was provided by the location where the $S^1$ pinches off in the cigar geometry. From the 4D braneworld point of view, this cutoff is provided by the center of the wormhole, which is a finite proper distance away from any bulk point. In other words, the confinement scale is provided by the wormhole length.

In the following, we quickly review and apply the procedure from \cite{Antonini:2022opp} to obtain the mass spectrum by performing canonical quantization on a free scalar field $\phi$ with mass $m$ within our braneworld wormhole geometry. We are interested in the regime where gravity localization holds, for which $f \approx g \approx h$ and the metric takes the conformal form \eqref{eqn:wormholeconformalform}.
We proceed by solving the Klein-Gordon equation
\begin{equation}
    (\Box{}-\mu^2)\phi = 0
\end{equation}
to find a complete basis of solutions (labeled by $i=0,1,2,...$) of the form
\begin{equation}
    f_{k_{x},k_{y},i}(\zeta,x,y,\tau) = \frac{1}{2\pi \sqrt{2\omega}f(\tau)}e^{-i \omega \zeta}e^{i k_{x}x} e^{i k_{y}y} u_{i}(\tau),
    \label{eq:basis}
\end{equation}
where we chose a convenient normalization and used translational symmetry in the $\zeta,x,y$ directions. Plugging this ansatz into the Klein-Gordon equation yields a 1D Schrödinger equation
\begin{equation}
    -u_{i}''(\tau) + V(\tau) u_{i}(\tau) = \lambda_{i}u_{i}(\tau)
    \label{eqn:schrodinger}
\end{equation}
where the potential $V(\tau)$ is defined as 
\begin{equation}
    V(\tau) =\frac{f''(\tau)}{f(\tau)}+\mu^2 f^2(\tau)
    \label{eq:potential}
\end{equation}
and $\lambda_i\equiv \omega^2-k_x^2-k_y^2$. The asymptotic form of the potential near the two boundaries is
\begin{equation}
    V(\tau) \sim \frac{\alpha}{(\tau_{\infty}\pm \tau)^{2}},
\end{equation}
where $\alpha \equiv 2-3\mu^2/\Lambda_4$.
As usual in asymptotically AdS spacetimes, there are normalizable and non-normalizable solutions, which behave asymptotically as $u(\tau) \sim (\tau_{\infty} \mp \tau)^{\Delta_{+}}$ and $u(\tau) \sim (\tau_{\infty} \mp \tau)^{\Delta_{-}}$, respectively, where $\Delta_{\pm} \equiv (1 \pm \sqrt{1+4 \alpha})/2$. We are interested in solutions which are normalizable at both boundaries.
Normalizability at both boundaries guarantees that the set of solutions is discrete, with discrete eigenvalues $\lambda_i$ for the potential and the lowest eigenvalue strictly positive $\lambda_0>0$ \cite{Antonini:2022opp}.\footnote{This can be understood in analogy with a 1D particle in a box Schr\"odinger problem.} Since we expanded the basis of solutions \eqref{eq:basis} in Fourier modes for $\zeta,x,y$, which are the non-compact boundary coordinates, we can interpret the relation

 \begin{equation}
     \omega = \sqrt{k_{x}^2+k_{y}^{2} + \lambda_{i}}
 \end{equation}
as a dispersion relation for particles in the dual IR confining theory on $\mathbb{R}^{2,1}$.\footnote{We remind that an important subtlety is that this 3D confining theory is \textit{not} given by the pair of 3D holographic DOF living on the two AdS${}_4$ asymptotic boundaries and associated with the braneworld wormhole description. It is instead the theory obtained by considering the full dual 4D BCFT (including the 4D DOF coupling the pair of 3D holographic DOF) and flowing in the IR to a 3D theory.} $m_i=\sqrt{\lambda_i}$ can then be interpreted as a discrete confining spectrum of massive particles (glueballs), with $\lambda_0>0$ confirming the presence of a mass gap.

The spectrum of masses can be computed numerically. For a choice of $\mu=1$ for the mass of the scalar field on the braneworld, $b=\sqrt{3}-0.001$, and $T=0.9999$, we obtain
\begin{equation}
    m_{n}=\{ 2.81,2.87,2.93,2.97,3.04\}
\end{equation}
for the first five masses. 

\begin{figure}
    \centering
    \includegraphics[width=0.5\linewidth]{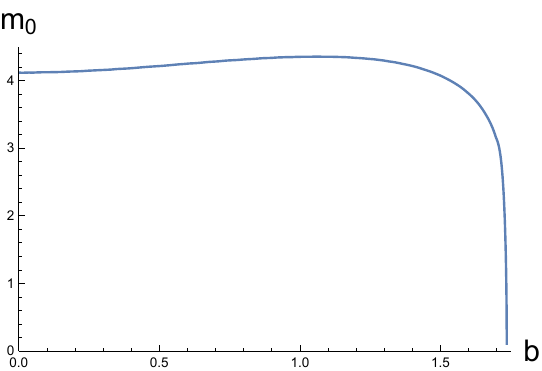}
    \caption{Lowest mass $m_{0} = \sqrt{\lambda_{0}}$ in the confining spectrum of massive particles of the dual IR theory on $\mathbb{R}^{2,1}$ as a function of the magnetic field parameter $b$ for a bulk scalar field with mass $\mu=1$. Observe that as $b \to \sqrt{3}$, $m_{0} \to 0$, which signals deconfinement. This behavior is qualitatively analogous to the one observed in Section~\ref{sec:MBBdual} for the spectrum obtained from a 5D calculation in the soliton spacetime. The small increase of $m_0$ for small $b$ is due to the wormhole length decreasing before increasing again and diverging for $b\to\sqrt{3}$. The results for small $b$ should be taken cautiously, because $\tau_0<0$ in that regime.}
    \label{fig:wormholespectra}
\end{figure}

We also verified that in the limit $b\to\sqrt{3}$ the mass gap goes to zero, see Figure \ref{fig:wormholespectra}, and the spectrum of masses becomes continuous. This behavior, which matches qualitatively the one observed for the AdS${}_5$ soliton spectrum in Section \ref{sec:MBBdual}, can once again be understood in terms of the IR cutoff length scale diverging. On one hand, from a braneworld wormhole point of view, the proper distance between the wormhole's center and any point in the bulk diverges. On the other hand, from a boundary theory point of view, just like in the pure soliton case of Section \ref{sec:MBBdual}, the theory lives on $\mathbb{R}^{3,1}$ in the $b\to\sqrt{3}$ limit. In fact, for fixed tension the boundary width parameter $\tau_0$ diverges in the $b\to\sqrt{3}$ limit. Therefore the boundary interval $I$ becomes infinitely long, the dual 4D theory has no boundary, and it does not flow in the IR to a (2+1)-dimensional confining theory.

\subsection{Bulk reconstruction and slicing duality}
\label{sec:slicing}

Let us briefly comment about the properties of bulk reconstruction. The absence of a horizon in the wormhole geometry (and, from a doubly-holographic perspective, in the soliton geometry) implies that the whole bulk geometry is in the causal wedge of the dual theory. This in turn implies that bulk observables in the braneworld wormhole are easy to reconstruct from the dual theory. In particular, ordinary HKLL reconstruction \cite{Hamilton:2005ju,Hamilton:2006az} is sufficient to represent bulk fields in terms of boundary operators. Geometric observables, such as the complete form of the wormhole conformal factor $f(\tau)$, can also be reconstructed. For example, an explicit procedure to reconstruct $f(\tau)$ from the confining spectra of massive particles studied in the previous subsection was provided in \cite{Antonini:2022opp}.

The fact that observables in the braneworld cosmology of Section \ref{sec:cosmology} are related to observables in the braneworld wormhole by double analytic continuation implies that the former can be reconstructed from the 4D BCFT vacuum dual to the braneworld wormhole picture. For instance, given two spectra of confining particles, we can reconstruct the wormhole conformal factor $f(\tau)$. By analytically continuing $\tau=it$, we then obtain the cosmological scale factor (which is guaranteed to be real by the double reflection symmetry of the Euclidean saddle) in terms of the cosmological conformal time $t$. Similarly, a correlation function in the braneworld cosmology on the $t=z=0$ slice is equal to a correlation function on the $\tau=\zeta=0$ slice of the braneworld wormhole, which can be reconstructed via HKLL. Away from this codimension-2 slice, bulk correlation functions are related by double analytic continuation, but the story should be similar.

\begin{figure}
    \centering
    \includegraphics[width=0.8\textwidth]{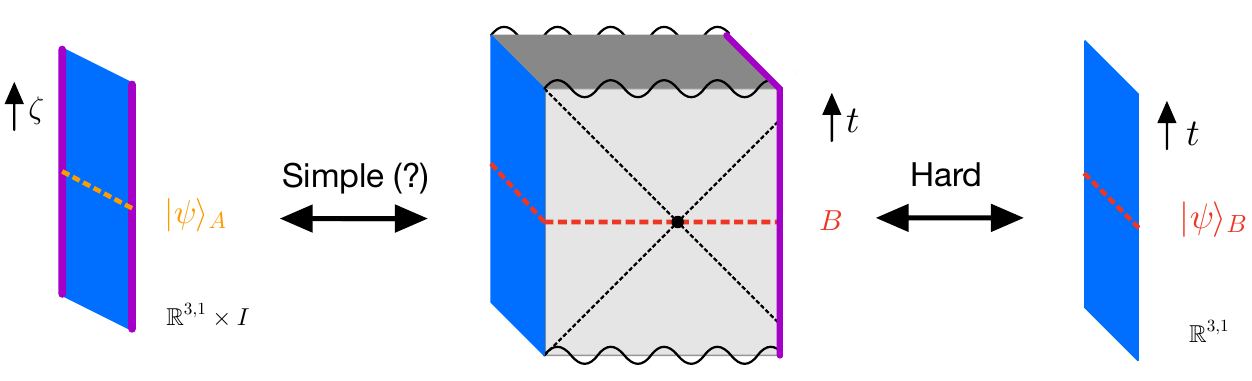}
    \caption{The dictionary between bulk and boundary observables in the MBB braneworld cosmology picture (right) is exponentially complex because the brane lies behind the MBB horizon. The slicing duality provides a different dictionary relating bulk observables in the MBB braneworld cosmology picture to boundary observables in the vacuum $\ket{\psi}_A$ of the confining theory dual to the soliton braneworld wormhole picture. This dictionary is expected to be simpler, because the entire soliton spacetime lies in the causal wedge of its dual theory.}
    \label{fig:dictionaries}
\end{figure}

The procedure outlined above defines a new dictionary between observables in the braneworld cosmology (and more generically in the MBB spacetime cut off by the ETW brane) and observables in the vacuum state of the BCFT dual to the soliton braneworld wormhole picture. Recall that the conventional dictionary between bulk and boundary observables in the MBB braneworld cosmology picture is exponentially complex, see Section \ref{sec:cosmodual}. The slicing duality, first introduced in \cite{VanRaamsdonk:2021qgv,Antonini:2022blk,Antonini:2022ptt} and realized in our setup, therefore provides an alternative, seemingly much simpler dictionary, see Figure \ref{fig:dictionaries}. It remains to be verified whether in general the difficulty is hidden in the analytic continuation procedure. But at least for some observables, such as the scale factor and correlation functions at $t=z=0$, the simplification is clear. We leave further investigation on these points to future work \cite{workinprogress}.

\subsection{Relationship to other traversable wormhole constructions}
\label{subsec:otherconstr}

Constructions of traversable wormholes in four dimensions in magnetically charged setups were studied in \cite{Maldacena:2018gjk,Maldacena:2020sxe,Bintanja:2021xfs}. In \cite{Maldacena:2018gjk}, the wormhole is asymptotically flat and it connects two different subregions of the same asymptotically flat region (see also \cite{Bintanja:2021xfs} for a similar construction of a traversable wormhole connecting two asymptotically higher dimensional AdS boundaries). In their setup, magnetic field lines are closed, going through the wormhole, out of the two mouths, and closing in the ambient flat spacetime. The wormhole is sustained by the negative Casimir energy associated with fermions moving along these field lines. It seems plausible that something similar happens in our setup, with closed magnetic field lines passing through the wormhole and the non-gravitational auxiliary degrees of freedom (with an appropriate boundary condition at the interface at the asymptotic AdS${}_4$ boundary). In fact, magnetic field lines in the braneworld wormhole description are along the $\tau$ direction and ``threading'' the wormhole. Bulk fermions should also be present in a full description of our setup, because, as we have explained in Section \ref{sec:MBBsoliton}, in the dual theory fermions are present, for which SUSY-breaking boundary conditions must be imposed \cite{Witten:1998zw,Horowitz:1998ha}. We leave the investigation of this possibility to future work.

On the other hand, the authors of \cite{Maldacena:2020sxe} built a braneworld wormhole. The 5D Lorentzian bulk spacetime is given by the infinite $\beta$ solution (first built in \cite{DHoker:2009mmn}) of the Einstein's equations \eqref{eq:einsteineq}, with the analytic continuation corresponding to our MBB solution ($\tau=it$). Their ETW brane is static, similar to an AdS Karch-Randall brane \cite{Karch:2000ct}. Therefore this connects two different region of the \textit{same} asymptotic AdS${}_4$ boundary, as it was pointed out in \cite{Maldacena:2020sxe}.
In order to understand why, it is useful to consider the $\mathbb{R}^{3,1}$ boundary of the 5D bulk solution without brane as an infinite radius limit of a spatially $S^3$ boundary. The braneworld wormhole built in \cite{Maldacena:2020sxe} connects two points\footnote{Or, more correctly, codimension-1 surfaces.} on the $S^3$. These are therefore clearly connected also on the AdS${}_4$ boundary geometry (through the two remaining directions of the $S^3$). When taking the large radius limit, the points are still connected on the boundary, and therefore the wormhole should be understood as connecting two regions of the same AdS${}_4$ boundary.

In the setup studied in the present paper, this is not the case. In fact, our braneworld wormhole is embedded in the magnetically charged AdS${}_5$ soliton spacetime, whose boundary is topologically $\mathbb{R}^{2,1}\times S^1$ in the absence of the ETW brane, which, using a similar reasoning as in the previous paragraph, is an infinite radius limit of a spatially $S^2\times S^1$ boundary. Our braneworld wormhole connects two different points on the $S^1$ (which is parametrized by the $\tau$ coordinate). Since on the AdS${}_5$ boundary these points are connected only through the $S^1$---which is the dimension absent in the AdS${}_4$ boundary---the wormhole should be interpreted as connecting two different asymptotic AdS${}_4$ boundaries, each one with a $S^2$ spatial topology, which becomes $\mathbb{R}^2$ in the infinite radius limit.

\section{Discussion}
\label{sec:discussion}
In this paper we have built 4D braneworld cosmologies and 4D eternally traversable braneworld wormholes by embedding ETW branes in 5D magnetically charged spacetimes. The two Lorentzian setups arise from the same Euclidean saddle of the gravitational path integral, in which the ETW brane takes the form of a Maldacena-Maoz wormhole, and are related to each other by double analytic continuation. In Euclidean signature, the holographic dual description is given by a 4D Euclidean BCFT. The Lorentzian dual of the braneworld cosmology setup is given by a pure excited state of a 4D CFT on $\mathbb{R}^{3,1}$. The dual of the braneworld wormhole setup is given by the ground state of a BCFT on $\mathbb{R}^{2,1}\times I$ which confines in the IR (on length scales much larger than the size of the interval $I$).

Several interesting questions remain to be answered. From a braneworld cosmology point of view, it would be interesting to understand what sources of energy density drive the accelerated expansion at early times and in general the cosmological evolution. Similarly, from a braneworld wormhole point of view, it is not clear what is the source of negative energy density sustaining the traversable wormhole. We have advocated for the magnetic field being responsible for both of these effects, but further study is necessary to confirm this expectation. Clearly, making steps towards a better analytic understanding (or possibly a complete analytic solution) of the 5D bulk spacetimes involved in our analysis would lead to the sharpest answers to these questions. In fact, that would allow us to read out the source of negative energy from the braneworld cosmology and braneworld wormhole Einstein's equations, similar to what we reviewed in Section \ref{sec:cosmoaccel} for the AdS-RN case. In the braneworld wormhole picture, it would then be interesting to examine whether the source of negative energy is an enhanced Casimir energy similar to those studied in \cite{May:2021xhz,Swingle:2022vie}, and whether the mechanism generating it is similar to that of asymptotically flat wormholes \cite{Maldacena:2018gjk}.

It would also be interesting to study the quantitative agreement between the holographic entanglement entropy computed using Hartman-Maldacena-like Ryu-Takayanagi surfaces ending on the ETW brane in our doubly holographic cosmological setup, and entanglement islands in cosmology (see Section \ref{sec:islands}). This would also give a more detailed understanding of the encoding of the braneworld cosmology in the dual excited state of the holographic theory on $\mathbb{R}^{3,1}$, via analogy with the encoding of the black hole interior in the Hawking radiation after the Page time. In fact, the reconstruction of observables in the braneworld cosmology from the dual theory is generically understood to be exponentially complex, because the brane sits inside a Python's lunch. This also suggests a non-isometric encoding \cite{Akers:2021fut,Akers:2022qdl} of the cosmological physics in the dual theory, which raises the question of when EFT in the cosmology can be fully trusted. At the qualitative level of this discussion, because the outermost QES of our Python is a classical surface given by the bifurcation surface, one can expect the EFT in the Python, including in the braneworld cosmology, to hold as long as bulk observers are not allowed to perform exponentially (in $1/G$) complex operations \cite{Akers:2022qdl}. A more thorough study of these questions is needed to arrive to a definite answer.

Another interesting direction to explore further is the extent to which the slicing duality between the braneworld cosmology and the braneworld wormhole allows us to define a different, simpler dictionary between observables in the cosmology and observables in the ground state of the microscopic theory dual to the wormhole picture \cite{workinprogress}. Because the dictionary between the braneworld cosmology and its own microscopic dual theory is exponentially complex, if such a simplification occurs, one would then expect the relationship between the two boundary representations of the same bulk observable in the two microscopic Lorentzian theories (the dual of the soliton and the dual of the MBB) to be exponentially complex.

The extension of our setup beyond the simplest constant tension approximation for the brane would also be interesting. For example including a brane-localized scalar field in a potential could allow us to obtain additional phases of accelerated expansion for the cosmological universe, which could match the one currently observed in our own universe \cite{VanRaamsdonk:2023ion}.

Finally, obtaining an explicit microscopic BCFT realization of our setup would also be invaluable. This could be obtained, for example, by studying $\mathcal{N}=4$ SYM in an external magnetic field (which is dual to our 5D bulk setups \cite{DHoker:2009mmn}) on a manifold with a boundary $\mathbb{R}^3\times I$, with conformal boundary conditions imposed at the endpoints of the interval. We remind that the external source coupled to the $U(1)$ subgroup of the R-symmetry already partially breaks supersymmetry. It was advocated in \cite{VanRaamsdonk:2021qgv,Antonini:2022blk} that in (Euclidean) microscopic theories dual to Euclidean wormholes like ours, supersymmetry should be completely broken, but in a specific way. Namely, with the boundary conditions at the two endpoints of the interval separately preserving complementary sets of supersymmetries. The supersymmetry breaking would then be manifest in the IR of the theory, leading to the confining properties we analyzed. This is similar in spirit to the AdS soliton case studied in \cite{Witten:1998zw,Horowitz:1998ha}, where supersymmetry is also broken by boundary conditions (for fermions on the $S^1$). It would be interesting to test this idea by explicitly building a $\mathcal{N}=4$ SYM${}_B$ BCFT with conformal boundary conditions satisfying these properties, and verifying its agreement with the bulk setup of the present paper.

\section*{Acknowledgements}

We would like to thank Maciej Kolanowski for collaboration in the early stages of this work. We also thank Gary Horowitz, Luca Iliesiu, Brett McInnes, Arvin Shahbazi-Moghaddam, Petar Simidzija, Brian Swingle, Mark Van Raamsdonk, and Chris Waddell for useful discussions. We especially thank Eric D'Hoker for several discussions and comments. S.A. is supported by the U.S. Department of Energy through DE-FOA-0002563. This research was supported in part by grant NSF PHY-2309135 to the Kavli Institute for Theoretical Physics (KITP).

\appendix

\section{Numerical magnetic black brane solution}
\label{app:MBBbc}

In this appendix, we explain how the boundary conditions \eqref{eq:MBBbc} needed to find the MBB numerical solution are obtained.

In order to find a numerical solution to the system \eqref{eq:einsteineq}, we must impose appropriate boundary conditions at the horizon, to then numerically find asymptotically AdS${}_5$ solutions. The first boundary condition we have is clearly $f(r_H)=0$. Here we derive the other boundary conditions at the horizon and make some comments which are useful to construct the numerical solution and in the rest of the paper:
\begin{itemize}
    \item Using the freedom to rescale the radial and the time coordinates $r,t$, we can set the derivative of one of the metric functions at the horizon to unity while keeping the ansatz \eqref{eq:MBBmetric} unchanged. A shift of the radial coordinate also allows us to set $r_H=1$. In this paper, we choose to set $g'(1)=1$. This choice, which differs from the $f'(1)=1$ choice of the authors of \cite{DHoker:2009mmn}, is useful to advocate for gravity localization in the braneworld setup studied in Sections \ref{sec:euclidean}, \ref{sec:cosmology}, and \ref{sec:wormhole}. In fact, gravity localization occurs when the ETW brane is deep into the asymptotically AdS region. With our choice of coordinates $g'(1)=1$, such a region corresponds to smaller (and therefore numerically better under control) values of the radial coordinate with respect to the $f'(1)=1$ (or $h'(1)=1$) choices of coordinates. 
    \item A constant rescaling of the planar coordinates $x,y,z$ allows us to set $g(1)=h(1)=1$. In doing this we choose to keep the form of the electromagnetic field strength (and therefore of the system \eqref{eq:einsteineq}) unchanged, $F=b dx\wedge dy$, where we indicated with $b$ the magnetic field parameter in our choice of coordinates. With this boundary condition, we will in general obtain a solution for $g$ and $h$ which behaves asymptotically as $g(r)\sim v(b)r^2$, $h(r)\sim w(b)r^2$, with $v(b)$ and $w(b)$ constants depending only on the value of the magnetic field parameter $b$. After obtaining our solution, we will therefore need to numerically find the value of $v(b),w(b)$ and rescale $\sqrt{v(b)}(x,y)\to (x,y)$ and $\sqrt{w(b)}z\to z$ (i.e. $g\to v(b)g$ and $h\to w(b)h$) to obtain the correct asymptotically AdS${}_5$ solution. The final form of the electromagnetic field strength is therefore $F=b/v(b)dx\wedge dy$, namely the physical magnetic field is given by $B=b/v(b)$.
    \item Finally, we need boundary conditions at the horizon for the derivatives of $f$ and $h$. In the system \eqref{eq:einsteineq} we have three unknown functions and four equations. The first three equations are dynamical second order equations, whereas the fourth one is a first order constraint equation. The latter can be imposed at the horizon and is guaranteed to remain satisfied along a solution of the dynamical equations. Notice that at the horizon the third equation also becomes first order. Combining the third and fourth equation evaluated at the horizon $r=1$, we find the boundary conditions $f'(1)=8(3-b^2)/3$, $W'(1)=(6+b^2)/(12-4b^2)$.
\end{itemize} 
To summarize, the set of boundary conditions we have to impose at the horizon at $r=1$ is (expressed in terms of $V$ and $W$):
\begin{equation}
    f(1)=V(1)=W(1)=0, \quad f'(1)=\frac{8}{3}(3-b^2), \quad V'(1)=\frac{1}{2},\quad W'(1)=\frac{6+b^2}{4(3-b^2)}.
    \label{eq:MBBbcapp}
\end{equation}
We emphasize that these boundary conditions are completely generic. In fact, it is easy to check that any asymptotically AdS${}_5$ solution of the system \eqref{eq:einsteineq} with a horizon at $r=r_H$ can be mapped to an asymptotically AdS${}_5$ solution\footnote{After the appropriate rescaling of $g$ and $h$ by $v(b)$, $w(b)$ as explained above.} satisfying the boundary conditions \eqref{eq:MBBbcapp} by a change of coordinates (rescaling of all coordinates and shift of the radial coordinate) which results in an overall conformal rescaling of the asymptotic metric. The conformal nature of the AdS boundary then guarantees that the two solutions are physically equivalent. This result clearly holds independently of the choice of spatial topology of the asymptotic boundary ($\mathbb{R}^3$ or $\mathbb{T}^3$).

\section{Deriving the Brane Equation in the Euclidean Analysis} 
\label{app:braneeqn}

In this appendix we provide a detailed derivation of equation~\eqref{eqn:braneeuclidean}. The Euclidean metric in 5-dimensions takes the form of equation~\eqref{eqn:euclidean5dmetric}. The induced metric on the brane takes the form $h_{ab} =g_{\mu \nu} e_{a}^{\mu} e_{b}^{\nu}$ such that 
\begin{equation}
    e_{a}^{\mu}=\frac{dx^{\mu}}{dy^{a}}
\end{equation}
where $x^{\mu}$ , ($\mu=\{ \tau,r,x,y,z\}$) and $y^{a}$, ($a=\{ \tau,x,y,z\}$) are the coordinates that describe the 5-dimensional bulk spacetime and the 4-dimensional brane, respectively.

The motion of the planar brane in the bulk spacetime can be parameterized by $r(\tau)$. Thus, we can select $e_{\tau}^{\mu}=(1,r',\vec{0})$. We can further define the unit normal vector as
\begin{equation}
    n_{\mu} = N(-r',1,\vec{0})
\end{equation}
where $r' = dr/d\tau$ and $N$ is the normalization factor such that $n_{\mu}n^{\mu}=1$. Thus,
\begin{equation}
    N=\sqrt{\frac{f(r)}{(r')^2+(f(r))^2}}.
\end{equation}
Given this, we can calculate the extrinsic curvature 
\begin{equation}
    K_{ab}=\nabla_{\nu}n_{\mu}e_{a}^{\mu}e_{b}^{v}.
\end{equation}
We obtain the following results for $K_{ab}$
\begin{align}
    K_{\tau \tau} &= N \left [ \frac{3}{2} \frac{f'(r)}{f(r)} (r')^2 - r''+\frac{1}{2} f(r) f'(r) \right ] \\ 
    K_{x x} & = K_{y y} = \frac{N}{2} f(r) g'(r) \\
    K_{z z} & = \frac{N}{2} f(r) h'(r).
\end{align}
The $\tau \tau$ component of equation~\eqref{eqn:robinbraneeqn} then yields equation~\eqref{eqn:braneeuclidean} which we display here:
\begin{equation}
    \frac{dr}{d\tau} = \pm \frac{f \sqrt{4 fV'W' + 4 f V'^2 + fW'^2 -9T^2}}{3T}.
\end{equation}
In this equation, the $+$ and $-$ sign denotes the contracting and expanding phase, respectively.

\section{Action comparison and phase structure}
\label{app:actioncomparison}

In order to determine whether the Euclidean saddle with a single, connected ETW brane we studied in Section \ref{sec:euclidean} is dominant in the gravitational Euclidean path integral, we must compare its on-shell action with other phases that could potentially contribute. In the absence of an ETW brane, at finite periodicity $\beta$ for the $\tau$ coordinate there is only one known saddle contributing to the Euclidean path integral: the Euclidean MBB \eqref{eq:euclideanMBB}, with boundary manifold $\mathbb{R}^3\times S^{1}$. Analogously, at $\beta=\infty$, where the Euclidean MBB is not well defined (this would correspond to the strict $b=\sqrt{3}$ limit), there is only one saddle contributing to the Euclidean path integral. This is the saddle associated with the zero-temperature solution constructed in \cite{DHoker:2009mmn}, which interpolates between $\text{AdS}_{3} \times \mathbb{R}^{2}$ deep into the bulk and $\text{AdS}_{5}$ asymptotically, with boundary manifold $\mathbb{R}^4$. By analogy with \cite{DHoker:2009mmn}, in this appendix we will refer to our Euclidean MBB as the ``finite temperature'' solution, and to the $\beta=\infty$ saddle as the ``zero temperature'' solution, although this nomenclature makes sense only when $\tau$ is identified with the Euclidean time (i.e. in the MBB setting, but not in the soliton setting).

\begin{figure}
    \centering
    \includegraphics[width=0.75\linewidth]{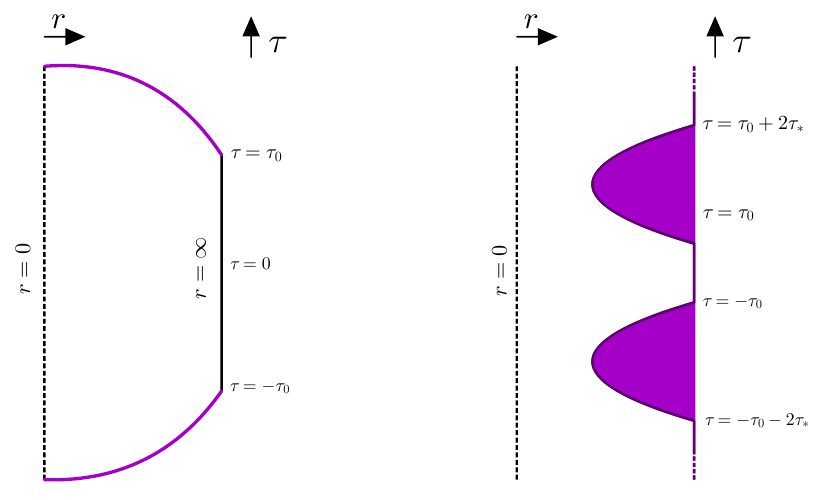}
    \caption{Left: Hypothetical alternative Euclidean saddle with two disconnected ETW branes (in purple) arising from the zero temperature solution of \cite{DHoker:2009mmn}. This is analogous to the pure AdS phase of \cite{Cooper:2018cmb}. This saddle, if it existed, would contribute to the same Euclidean path integral as the one studied in the main sections of this paper and depicted in Figure \ref{fig:eucltrajectory}, and we would have to compare their on-shell actions to determine which saddle is dominant. Right: The saddle depicted on the left does not exist because in the zero temperature solution of our setup and for any $T>1/\sqrt{3}$, the brane does not shrink all the way to $r=0$, but it turns around at $r=r^*$ and expands again to the asymptotic boundary in a finite amount of $\tau$ coordinate. The portions of bulk spacetime cut off by the ETW branes are shaded in purple. Clearly, this saddle has different boundary conditions with respect to the finite temperature one, with the asymptotic boundary being $\mathbb{R}^3\times ((-\infty,-\tau_0-2\tau^*]\cup [-\tau_0,\tau_0]\cup [\tau_0+2\tau^*,\infty))$. Therefore, it does not contribute to the same Euclidean path integral.}
    \label{fig:twobranes}
\end{figure}

When the Euclidean geometry contains ETW branes, one could expect that both of these bulk saddles (appropriately cut off by ETW branes) could contribute to the same Euclidean path integral. In fact, the boundary manifold is now $\mathbb{R}^3\times I$ with $I=[-\tau_0,\tau_0]$, which can arise from adding a codimension-1 boundary to either the $\mathbb{R}^{3}\times S^1$ or the $\mathbb{R}^4$ boundaries. For the finite temperature saddle, as we have seen, there is a single connected ETW brane intersecting the asymptotic boundary at $\tau=-\tau_0$ and $\tau=\tau_0$, see Figure \ref{fig:eucltrajectory}. On the other hand, the zero-temperature scenario could result in two disconnected ETW branes with each one intersecting the boundary at one of the two endpoints of the interval. This would be similar to the pure AdS phase described in \cite{Cooper:2018cmb}, which is given by a pure Euclidean AdS spacetime cut off by two disconnected ETW branes and is an alternative saddle to the Euclidean Schwarzschild-AdS extensively studied in \cite{Cooper:2018cmb}. The left panel of Figure~\ref{fig:twobranes} displays this hypthetical configuration of the ETW brane in the zero temperature solution. 

However, this alternative phase does not exist in our setup. The reason is that the ETW brane in the zero temperature solution does not shrink all the way to $r=0$ in Euclidean signature, but it has a turning point at $r=r^*$ for some finite $\tau-\tau_\infty=\tau^*$, and then it expands again to the asymptotic boundary, see the right panel of Figure \ref{fig:twobranes}. The asymptotic AdS${}_5$ boundary of the resulting Euclidean saddle therefore is not simply given by $\mathbb{R}^3\times I$, but by a disconnected manifold given by $\mathbb{R}^3$ times $(-\infty,-\tau_0-2\tau^*]\cup [-\tau_0,\tau_0]\cup [\tau_0+2\tau^*,\infty)$. This Euclidean geometry clearly has different boundary conditions with respect to our finite temperature saddle examined in the main sections, hence it does not contribute to the same Euclidean path integral.

In order to reach this conclusion, we must examine, following \cite{DHoker:2009mmn}, the zero-temperature solution interpolating between AdS${}_3\times \mathbb{R}^2$ and AdS${}_5$. To obtain this solution, we solve Einstein's equations~\eqref{eq:einsteineq} by assuming the preservation of Lorentz invariance in the $(t,y)$ directions which is equivalent to setting $f = h=  e^{2W}$. The system \eqref{eq:einsteineq} then simplifies to
\begin{equation}
    \begin{aligned}
     &2V''+W''+2(V')^2+(W')^2=0\\
     &V''+2W''+4(W')^2-V'W'=2B^2e^{-4V-2W}\\
     &(V')^2+(W')^2+4V'W'=6e^{-2W}-B^2e^{-4V-2W}.
    \end{aligned}
    \label{eq:einsteineqzero}
\end{equation}
The small $r$ behavior dictates the boundary conditions: $g=e^{2V} \sim B/\sqrt{3}$, $f=h=e^{2W} \sim 3r^2$ at $r\sim 0$. $B$ can be absorbed by an appropriate rescaling of all the coordinates. Thus, the solution is unique and independent of the magnetic field. We must also impose $g'(0)=0$ (i.e. $V'(0)=0)$ to guarantee the smoothness of the solution (this can also be obtained by using the constraint equations  in the third line of the system \eqref{eq:einsteineqzero}). The Einstein's equations can then be numerically integrated to yield a solution with an asymptotic behavior of $e^{2V} = vr^{2}$ and $e^{2W} = r^{2} $. $v$ can be absorbed by a redefinition of $V$ to yield an asymptotically $\text{AdS}_{5}$ solution. This background solution must then be cut off using two ETW branes, one anchored at the boundary at $\tau=-\tau_0$ and the other one at $\tau=\tau_0$. In particular, the brane trajectory in this solution solves the exact same equation \eqref{eqn:braneeuclidean} we have derived for the finite temperature solution, after identifying $f=h=e^{2W}$, namely
\begin{equation}
    \frac{dr}{d\tau}=\pm\frac{f}{3T}\sqrt{f'\frac{g'}{g}+f\left(\frac{g'}{g}\right)^2+\frac{(f')^2}{4f}-9T^2},
    \label{eq:zerotempbrane}
\end{equation}
where we used $V'=g'/(2g)$, $W'=f'/(2f)$. 

\begin{figure}
    \centering
    \includegraphics[width=0.5\textwidth]{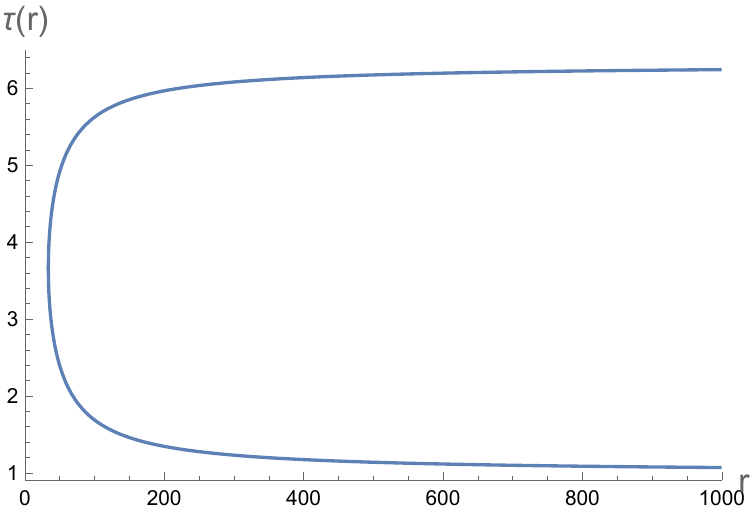}
    \caption{Solution $\tau(r)$ of equation \eqref{eq:zerotempbrane} with $b=1$, $T=0.9999$ for a brane anchored at $\tau=\tau_0=1$. As expected (see right panel of Figure \ref{fig:twobranes}), the brane turns around at $r=r^*=33.7$ and expands again to the asymptotic boundary in a finite amount of $\tau$ coordinate. This saddle therefore does not contribute to the same Euclidean path integral as the finite temperature solution of our interest.}
    \label{fig:branetrajzero}
\end{figure}

Let us focus on the brane anchored at the boundary at $\tau=\tau_0$. In order to obtain a smooth solution like the one depicted in the left panel of Figure \ref{fig:twobranes}, we would need $dr/d\tau$ to diverge at $r=0$ and be real for any $r>0$. From the boundary conditions for $f$, $g$ and their derivatives reported above, it is clear that the argument of the square root in equation \eqref{eq:zerotempbrane} is negative at $r=0$ for any $1/\sqrt{3}<T<1$ (which includes the regime of our interest $T\lesssim 1$).\footnote{The key difference between our setup and the pure AdS one studied in \cite{Cooper:2018cmb} is that, due to the AdS${}_3\times \mathbb{R}^2$ topology of our spacetime deep into the bulk, the first two terms in the square root in equation \eqref{eq:zerotempbrane} vanish. The planar nature of our solution also implies that the overall $f$ factor vanishes at $r=0$, unlike the pure AdS case of \cite{Cooper:2018cmb}.} In fact, $dr/d\tau=0$ at $r=r^*>0$ in this case, and we obtain the saddle depicted in the right panel of Figure \ref{fig:twobranes} instead, with a brane anchored at the boundary at $\tau=\tau_0$ \textit{and} $\tau=\tau_0+2\tau^*$. We plot in Figure \ref{fig:branetrajzero} a numerical solution of this equation of motion for a brane anchored at $\tau=\tau_0=1$ for $b=1$, $T=0.9999$, displaying the behavior expected from the discussion above. On the other hand, $dr/d\tau$ is positive for $r>0$ and vanishes at $r=0$ for any $0<T<1/\sqrt{3}$. Therefore, the solution is not smooth (there is a kink for the brane at $r=0$). In this case, the brane reaches $r=0$ for $\tau\to\infty$. In either case, a smooth Euclidean saddle contributing to the same Euclidean path integral as the finite temperature solution of our interest does not exist. We can therefore conclude that the solution with a single ETW brane studied in the main text is the only (known) saddle of the Euclidean path integral and therefore it dominates the gravitational ensemble. 

We remark that the zero temperature setup cut off by ETW branes studied in this appendix and depicted in the right panel of Figure \ref{fig:twobranes} describes 5D Euclidean wormholes connecting different (disconnected and decoupled) asymptotic AdS${}_5$ boundaries. These disconnected boundaries are given by the union of two half-hyperplanes ($\mathbb{R}^3\times (-\infty,-\tau_0-2\tau^*]$ and $\mathbb{R}^3\times [\tau_0+2\tau^*,\infty)$) and a strip ($\mathbb{R}^3\times [-\tau_0,\tau_0]$). It would be interesting to study these Euclidean wormhole solutions and their possible Lorentzian analytic continuations in more detail. We leave this endeavour to future work.

\addtocontents{toc}{\protect\enlargethispage{2\baselineskip}}
\section{Equation of state and deceleration parameter}
\label{app:accel}

In this appendix, we constrain the equation of state of the species driving the cosmological accelerated expansion. For simplicity, we will work under the (well-justified, as we have seen) approximation of an isotropic cosmology. Recall the perfect fluid equation of state
\begin{equation}
    w=\frac{P}{\rho},
\end{equation}
where $P$ is the pressure and $\rho$ is the energy density. The deceleration parameter $q$ can be written as \cite{Weinberg:2008zzc}
\begin{equation}
    q = \frac{1}{2} \sum_{i}^{N} \Omega_{i}(1+3 w_{i})
    \label{eq:decelpar}
\end{equation}
where $\Omega_{i}= \rho_{i}/\rho_{C}<1$ (with $\rho_{C}$ the critical density) are the density parameters at a given moment of time, and we considered the contribution from $N$ species to the deceleration parameter.
We can use this relation to extract a bound in terms of the deceleration parameter on the value of the parameter $w_X$ for the species driving the accelerated expansion. During the accelerated phase, the species $X$ is the dominant one, and a collection of other subdominant species can be present, whose role counteracts acceleration. Examples of this include matter, radiation, and a negative cosmological constant.

We denote the contribution of the collection of such species to equation \eqref{eq:decelpar} as $C>0$. 
We can then write

\begin{equation}
    w_{X}= \frac{2}{3}\left (\frac{q-C}{\Omega_{X}} \right  )-\frac{1}{3}.
\end{equation}
We must now distinguish two possible cases: $0<\Omega_X<1$ and $-1<\Omega_X<0$, namely a positive or negative energy density for the species driving the acceleration. For $0<\Omega_X<1$, we arrive at the bound
\begin{equation}
    w_X<\frac{2q-1}{3}.
\end{equation}
The right hand side can then be evaluated numerically to upper bound $w_X$ (see Figure \ref{fig:cosmoaccel} for a numerical evaluation of the deceleration parameter $q$ as a function of cosmic time). Since $q<0$ when the species $X$ is dominant, $w_X<-1/3$ as expected from a positive energy density species supporting acceleration. On the other hand, for negative energy density $-1<\Omega_X<0$ we obtain
\begin{equation}
    w_X>\frac{2|q|-1}{3}.
\end{equation}
Again, the right hand side can be evaluated numerically to lower bound $w_X$. Notice that $w_X>-1/3$, as expected by a negative energy density supporting acceleration. 

We remark that the latter case with negative energy density is likely the correct one. Recall that the energy density for a given species dilutes as the universe expands as $\rho\propto a^{-3(1+w)}$, with $a$ the isotropic scale factor. For $\Omega_X>0$ and $w_X<-1/3$, the energy density of the species $X$ would dilute very slowly when the universe expands (slower than radiation, matter, etc.), and should therefore be dominant at late times. Given that the accelerated expansion occurs at early times, it is likely that $\Omega_X<0$ and $w_X$ is large, possibly larger than ordinary matter and radiation (i.e. $w_X>1/3$). This guarantees that the species $X$ is dominant at early times but it dilutes very fast as the universe expands. This is also analogous to the AdS-RN case, for which the negative energy density stiff matter term appearing in equation \eqref{eq:friedmann} is dominant at early times and has equation of state $w=1$. We therefore expect the species driving the accelerated expansion to be a negative energy source in the cosmology, proportional to the magnetic field parameter $b$.

\bibliographystyle{jhep}
\bibliography{magneticbranes}
\end{document}